\documentclass[nofootinbib,aps,twocolumn,showpacs]{revtex4}
\usepackage{graphicx}
\usepackage{amsfonts}
\usepackage{amssymb}
\usepackage{amsbsy}
\usepackage{amsmath}
\usepackage{mathrsfs}
\usepackage{latexsym}
\usepackage{natbib}
\usepackage{bm}
\usepackage{subfigure} 
\usepackage{color}
\usepackage{wasysym}
\usepackage{bigints}
\usepackage{bbold}
\usepackage{esvect}
\usepackage{float}
\usepackage{booktabs}

\def\kp{k_{p}}

\begin{document}

\title{Radiative process of two entangled uniformly accelerated atoms in a 
thermal bath: a possible case of anti-Unruh event}

\author{Subhajit Barman}
\email{subhajit.b@iitg.ac.in}

\author{Bibhas Ranjan Majhi}
\email{bibhas.majhi@iitg.ac.in}

\affiliation{Department of Physics, Indian Institute of Technology Guwahati, 
Guwahati 781039, Assam, India}
 

\date{\today}

\begin{abstract}

We study the radiative process of two entangled two-level atoms uniformly 
accelerated in a thermal bath, coupled to a massless scalar field. First, by 
using the positive frequency Wightman function from the Minkowski modes with a 
Rindler transformation we provide the transition probabilities for the 
transitions from maximally entangled symmetric and anti-symmetric Bell states to 
the collective excited or ground state in $(1+1)$ and $(1+3)$ dimensions. We 
observe a possible case of {\it anti-Unruh-like} event in these transition 
probabilities, though the $(1+1)$ and $(1+3)$ dimensional results are not 
completely equivalent. We infer that thermal bath plays a major role in the 
occurrence of the anti-Unruh-like effect, as it is also present in the 
transition probabilities corresponding to a single detector in this case. 
Second, we have considered the Green's functions in terms of the Rindler modes 
with the vacuum of Unruh modes for estimating the same. Here the anti-Unruh 
effect appears only for the transition from the anti-symmetric state to the 
collective excited or ground state. It is noticed that here the $(1+1)$ and 
$(1+3)$ dimensional results are equivalent, and for a single detector, we do not 
observe any anti-Unruh effect. This suggests that the entanglement between the 
states of the atoms is the main cause for the observed anti-Unruh effect in this 
case. In going through the investigation, we find that the transition 
probability for a single detector case is symmetric under the interchange 
between the thermal bath's temperature and the Unruh temperature for Rindler 
mode analysis; whereas this is not the case for Minkowski mode. We further 
comment on whether this observation may shed light on the analogy between an 
accelerated observer and a real thermal bath. An elaborate investigation for the 
classifications of our observed anti-Unruh effects, i.e., either {\it weak} or 
{\it strong} anti-Unruh effect, is also thoroughly demonstrated.

\end{abstract}

\maketitle

\section{Introduction}\label{Introduction}

Quantum entanglement is one of the most distinguishing features that 
differentiates quantum physics from the classical. The fact that an entangled 
state of a collective system cannot be separated into the product states of the 
subsystems, acts as the essence of entanglement. It ensures that measurements of 
a physical observable on entangled particles are not independent of each other. 
The existence of quantum entanglement is experimentally verified in systems with 
photons, electrons, etc., see \cite{PhysRevA.87.053822, Hensen:2015ccp}. 
Moreover, the application of quantum entanglement in quantum communications, 
cryptography, and computing \cite{Tittel:1998ja, Salart-2008} has made it a more 
active and desirable field to venture further.

Furthermore, the realization and application of entanglement in flat and curved 
spacetimes through the usage of quantum field theory is considered to be the 
most enthralling recent outcomes, and there has been a growing interest in 
studying these relativistic quantum entanglement effects in recent times, see 
\cite{FuentesSchuller:2004xp, Reznik:2002fz, Lin:2010zzb, 
Ball:2005xa, Cliche:2009fma, MartinMartinez:2012sg, Salton:2014jaa, 
Martin-Martinez:2015qwa, Cai:2018xuo, Menezes:2017oeb, Menezes:2017rby, 
Zhou:2017axh}.
Another interesting phenomena is harvesting vacuum entanglement 
\cite{Reznik:2002fz, Salton:2014jaa, Henderson:2017yuv, Henderson:2020ucx, 
Stritzelberger:2020hde}, i.e., quantum fields can be source for entanglement for 
atoms interacting with it. The degradation of entanglement due to uncontrolled 
coupling to the external field, resulting from the influence of the field-atom 
interaction, is also an actual physical problem in a realistic experimental 
situation. Then it becomes imperative to understand the reasons of these 
degradation, so that sincere predictions can be provided.
All these reasonings motivated the developments on studying the transition rates 
between different states of entangled atoms in different trajectories, which are 
thriving with many new ideas and possibilities, see 
\cite{Rodriguez-Camargo:2016fbq, Menezes:2015iva, Hu:2015lda, Rizzuto:2016ijj, 
Arias:2015moa, Costa:2020aqa, Zhou:2020oqa, Cai:2019pnw, Lima:2019mbt, 
Liu:2018zod, Cai:2017jan, Zhou:2017fmu, Menezes:2015veo, Flores-Hidalgo:2015urj, 
Menezes:2015uaa, Zhou:2016urt}. In this purpose the concept of two-level atomic 
Unruh-deWitt detectors are essential. These point-like atomic detectors, whose 
internal states are coupled to the external field, were conceptualized to 
understand the Unruh effect \cite{Unruh:1976db, Unruh:1983ms} 
{\footnote{The Unruh-deWitt detector setup has been used to investigate whether 
a freely falling observer can detect particles in Boulware vacuum, both for 
black hole \cite{Scully:2017utk, Chakraborty:2019ltu} as well as 
Friedmann-Lama\^{i}tre-Robertson-Walker \cite{Chakraborty:2019ltu} spacetimes. 
Moreover, Unruh effect has also been used to verify the quantum memory of the 
spacetime \cite{Majhi:2020pps, Compere:2019rof}.}}. To be specific, the Unruh 
effect proclaims that the Minkowski vacuum as perceived by a uniformly 
accelerated observer will present a Planckian distribution of particles, with 
the temperature proportional to the acceleration of the observer. There are 
plenty of works predicting the possibility of generating entangled states in 
systems of these two-level atoms interacting with bosonic and other fields 
\cite{Plenio:1998wq, Ficek2003, Tana_2004}, further enriching the plausibility 
of these experiments.
It is to be noted that there are articles discussing the transition 
probabilities of entangled atoms in different scenarios for the static or 
accelerated observers. Like, in \cite{Arias:2015moa} the radiative process of 
the static atoms are discussed in the presence of mirrors. While in 
\cite{Rodriguez-Camargo:2016fbq} the finite time effects of acceleration are 
analyzed. These studies motivates one to study the similar radiative process for 
entangled atoms accelerated in thermal bath, which is not there in literature 
up to our knowledge.

In this work we are going to study the radiative process of two entangled 
accelerated atoms, coupled to a background scalar field, in a thermal bath. The 
transition probability corresponding to single accelerated observers in thermal 
bath are estimated in \cite{Costa:1994yx, Kolekar:2013hra}, considering 
two-level Unruh-deWitt detectors (where in \cite{Hodgkinson:2014iua} the 
scenario for a rotating detector in thermal bath is considered). In the 
derivation of \cite{ Kolekar:2013hra} the Green's function, which is essential 
for the calculation, is constructed from the Minkowski modes with a Rindler 
coordinate transformation. These Green's functions are not time translation 
invariant and one cannot provide the notion of transition probability per unit 
time out of them. However, one can get the idea about the transition probability 
for certain field mode frequency from these estimations. We have considered this 
particular procedure to study the transition probability of entangled atoms 
accelerated in thermal bath in $(1+1)$ and $(1+3)$ dimensions.

On the other hand, we have also constructed the Green's functions out of the 
Rindler modes with the Unruh operators, i.e., with vacuum for Unruh modes(which 
is the Minkowski vacuum here), corresponding to accelerated observers in thermal 
bath, in both $(1+1)$ and $(1+3)$ dimensions. These Green's functions are time 
translational invariant and one can get the notion of transition probability per 
unit time out of them. In both of the cases with the Minkowski and Rindler modes 
we considered studying the transition probabilities for the transitions form the 
symmetric and anti-symmetric Bell states to the collective excited or ground 
state. We observed that these transition probabilities decreases with increasing 
detector acceleration in certain cases, the so called anti-Unruh(-like) 
effect{\footnote{We mention that one cannot provide the notion of transition 
probability per unit time here for the case with the Minkowski modes as the 
Wightman function is not time translational invariant. On the other hand, the 
Unruh or anti-Unruh effect notion is always associated with a transition rate. 
Therefore, for Minkowski modes, we term them as Unruh-like and anti-Unruh-like 
effects to make a distinction from the usual notion. Whereas, for Rindler modes, 
as we will see later, one can provide transition rate expressions. Therefore, 
for this later discussion we reserve the phrases Unruh and anti-Unruh.}} 
\cite{Brenna:2015fga, Garay:2016cpf} (this will be discussed later elaborately). 
We have further provided a thorough study about this effect in this work and 
tried to understand the source of this effect in our case. For the case with the 
Minkowski modes we inferred that the presence of the thermal bath, not the 
entanglement between the detectors, has a significant contribution in the 
occurrence of this anti-Unruh-like effect. On the other hand, for the case with 
the Rindler modes, where the anti-Unruh effect arises only for the transition 
from the anti-symmetric Bell state to the collective excited or ground state, 
the investigations suggest that the entanglement has the major role in the 
occurrence of the anti-Unruh phenomenon. Interestingly, while for the Minkowski 
mode case features of the results in $(1+1)$ and $(1+3)$ dimensions are not 
completely identical with each other, in the Rindler mode case the features are 
common in both dimensions. The consideration of these particular Rindler modes 
then provides further insights into the picture.

On the other hand, it is well known that the Minkowski vacuum fluctuations as 
seen by an uniformly accelerated observer has striking resemblance with that of 
the thermal fluctuations of a static observer in thermal bath. In this regard, 
there are many works in the direction of understanding the distinguishability 
and indistinguishability between them, see \cite{Kolekar:2013hra, 
Kolekar:2013xua, Kolekar:2013aka, Adhikari:2017gyb, Das:2019aii, 
Chowdhury:2019set, Lima:2020czr}. In \cite{Adhikari:2017gyb} the analogy between 
the two was pointed out by showing the satisfaction of the fluctuation 
dissipation theorem by the force due to radiation as measured by the accelerated 
frame (see \cite{Das:2019aii} for this analogy in  de-Sitter, 
Friedmann-Lama\^{i}tre-Robertson-Walker background and \cite{Banerjee:2019tbr} 
for the consideration of anomalous stress tensor). Furthermore, in 
\cite{Kolekar:2013hra, Kolekar:2013xua, Kolekar:2013aka} the 
indistinguishability is proclaimed in certain scenarios. In particular, two 
different observers one accelerated in thermal bath and another with double 
acceleration, i.e., in the Rindler-Rindler frame, are shown to be analogous by 
studying the spectrum of observed particles. Here Bogoliubov transformation is 
utilized to obtain the spectrum of particles seen by a Rindler-Rindler observer 
in the Minkowski vacuum state and the detector response is studied to get 
spectrum of particles for an observer accelerated in a thermal bath. In 
\cite{Chowdhury:2019set} some significant dintinguishabilities between the 
observer static in thermal bath and the one with uniform acceleration are 
provided by studying different components of the renormalized stress energy 
tensor, and comparing them for Rindler-Rindler to Thermal-Rindler cases. All 
these analysis suggest the absence of a straight resolution out of this issue. 
We expected our analysis to shed some light on this matter too. In particular, 
we observed that in the calculation of obtaining the transition probabilities 
corresponding to the two-atom system there are also quantities that resemble the 
situation signifying the case of a single two-level detector accelerated in a 
thermal bath interacting with a scalar field. Namely we shall be denoting these 
quantities by the transition coefficients $\mathcal{F}_{11}$ for the Minkowski 
mode case and $R_{11}$ for the Rindler mode case. We observed that the quantity 
$\mathcal{F}_{11}$ is not symmetric under the interchange between the 
temperature of the thermal bath and the Unruh temperature. However, for the case 
with the Rindler modes with the vacuum of Unruh modes $R_{11}$ is symmetric 
under the same interchange. Thus suggesting this particular case may be the 
ideal representation for an accelerated observer, where the analogy with a 
thermal bath is prominent.

In Sec. \ref{sec:radiative_process} we begin with a brief discussion of our 
model set-up, the two entangled two-level atoms coupled with the vacuum scalar 
field. In this section by perturbatively expanding the time evolution operator 
up to first order in the coupling constant the expression of the transition 
amplitude is provided. From the expression of these transition amplitudes and 
subsequently from the transition probabilities the role of the Green's functions 
corresponding to the detectors will be evident. Next, in Sec. 
\ref{sec:Thermal-GreenFn-Mink} the the expressions of the Green's functions 
corresponding to uniformly accelerated observers in thermal bath are given 
considering the Minkowski modes. In Sec. \ref{sec:Thermal-GreenFn-Unruh} the 
expressions of Green's functions for the same systems, considering the Rindler 
modes (with Unruh creation and annihilation operators), are given. In Sec. 
\ref{sec:Transition-prob-Mink} we have estimated the transition probabilities 
for the transitions from the entangled states to the collective excited state in 
the two-atom system considering the Green's functions of Sec. 
\ref{sec:Thermal-GreenFn-Mink}. Subsequently, in Sec. 
\ref{sec:Transition-prob-Unruh} we have considered the Green's functions from 
Sec. \ref{sec:Thermal-GreenFn-Unruh} and estimated the transition probabilities 
for the same transitions. In Sec. \ref{sec:Study-Anti-Unruh} we have studied the 
anti-Unruh(-like) effect resulting in the transition probabilities from the both 
cases considering the Minkowski and Rindler modes. We have concluded this 
article with a discussion of our results in Sec. \ref{sec:discussion}.

\section{Radiative process of two entangled atoms: a model 
set-up}\label{sec:radiative_process}

We begin our analysis elucidating on the radiative process of two entangled 
Unruh-DeWitt detectors. This model has been taken up earlier in several 
situations \cite{Arias:2015moa, Rodriguez-Camargo:2016fbq, Costa:2020aqa}. Since 
we need this, a brief review of it will be presented here in order to make the 
discussion self-sufficient. Also this will help us to have a clear picture of 
the notations which we will introduce in order to define different quantities in 
the analysis. 

\begin{figure}[h]
\centering
 \includegraphics[width=0.7\linewidth]{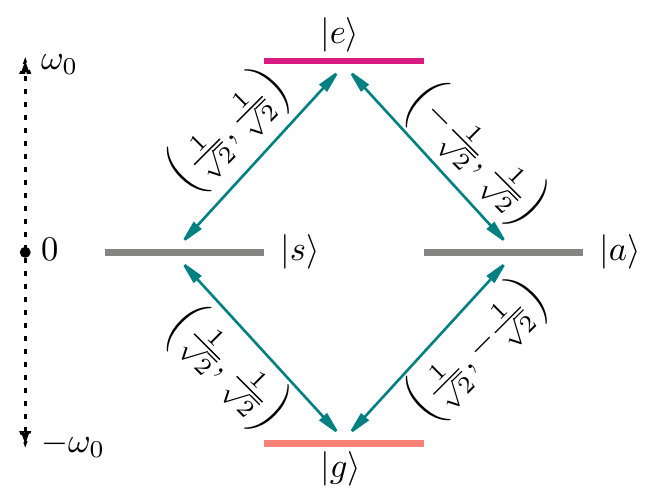}
 \caption{The energy levels corresponding to the eigenstates of the collective 
two-level two-atom system is depicted in this diagram. This figure has been 
taken from \cite{Costa:2020aqa}. The contributions from the monopole moment for 
each transition are also noted.}
 \label{fig:Energy-levels}
\end{figure}

The detectors are composed of point like two-level atoms, which are interacting 
with a massless, minimally coupled scalar field $\Phi(X)$ through monopole 
interaction. The Hamiltonian of this system of two two-level detectors 
interacting with the scalar field is expressed as
\begin{equation}\label{eq:Hamiltonian-total}
 H = H_{A}+H_{F}+H_{int}~,
\end{equation}
where, $H_{A}$ denotes the atomic Hamiltonian free of any interaction, $H_{F}$ 
is the free scalar field Hamiltonian, and $H_{int}$ is the interaction between 
the atoms and the scalar field. As provided by \emph{Dicke} \cite{Dicke:1954, 
Rodriguez-Camargo:2016fbq} one may express the two-atom Hamiltonian 
corresponding to the proper time as
\begin{equation}\label{eq:Hamiltonian-atoms}
 H_{A} = \omega_{0} \left[
S_{1}^{z}\otimes\mathbb{1}_{2}~\frac{d\tau_{1}}{d\tau}+ 
\mathbb{1}_{1}\otimes S_{2}^{z}~\frac{d\tau_{2}}{d\tau} \right]~,
\end{equation}
where, $S_{j}^{z}=(1/2)\left(|e_{j}\rangle \langle e_{j}|-|g_{j}\rangle \langle 
g_{j}|\right)$ denotes the energy operator, with $|g_{j}\rangle$ and 
$|e_{j}\rangle$ respectively representing the ground and excited states of the 
$j^{th}$ atom with $j=1,2$ here. In Eq. (\ref{eq:Hamiltonian-atoms}), 
$\mathbb{1}_{j}$ denote identity matrices, and $\omega_{0}$ the transition 
frequency corresponding to the collective two atom system. In particular for two 
identical atomic detectors the two-atom system has energy eigenvalues and the 
corresponding eigenstates, see \cite{Arias:2015moa}, as
\begin{eqnarray}
 E_{e} &=& \omega_{0}~,~~|e\rangle = |e_{1}\rangle |e_{2}\rangle~;\nonumber\\
 E_{s} &=& 0~,~~|s\rangle = \frac{1}{\sqrt{2}}(|e_{1}\rangle 
|g_{2}\rangle+|g_{1}\rangle |e_{2}\rangle)~;\nonumber\\
 E_{a} &=& 0~,~~|a\rangle = \frac{1}{\sqrt{2}}(|e_{1}\rangle 
|g_{2}\rangle-|g_{1}\rangle |e_{2}\rangle)~;\nonumber\\
 E_{g} &=& -\omega_{0}~,~~|g\rangle = |g_{1}\rangle |g_{2}\rangle~;
\end{eqnarray}
where, $|g\rangle$ and $|e\rangle$ respectively correspond to the ground and 
excited states of the collective system and $|s\rangle$, $|a\rangle$ denote the 
symmetric and anti-symmetric maximally entangled Bell states. A pictorial 
representation of them is shown in Fig. \ref{fig:Energy-levels}. 
\noindent
In (\ref{eq:Hamiltonian-atoms}), $\tau_1$ and $\tau_2$ are proper times 
corresponding to the frames attached with the first and second atomic detectors, 
respectively while $\tau$ denotes that for our observer who will measure the 
transition amplitudes.

In Minkowski spacetime the free Hamiltonian of the massless scalar field is
\begin{equation}\label{eq:Hamiltonian-field}
 H_{F} = \frac{1}{2} \int d^3X\left[ \left(\dot 
\Phi(X)\right)^2+\left|\bm{\nabla}\Phi(X)\right|^2 \right]~,
\end{equation}
where, the overhead dot denotes differentiation with respect to the time 
coordinate, and $\bm{\nabla}$ denotes the vector differential operator. The 
monopole interaction Hamiltonian is given by 
\begin{equation}\label{eq:Hamiltonian-int}
\scalebox{0.95}{$ H_{int}(\tau) = 
\sum_{j=1}^{2}\mu_{j}\kappa_{j}(\tau_{j}(\tau))m^{j}(\tau_{j} 
(\tau))\Phi\left(X_ { j}(\tau_{j}(\tau))\right) \frac{d\tau_{j}(\tau)}{d\tau} 
~,$}
\end{equation}
where, $\mu_{j}$ denote the individual coupling between the detectors and the 
scalar field. On the other hand, $\kappa_{j}(\tau_{j}(\tau))$ and 
$m^{j}(\tau_{j}(\tau))$ respectively denote the switching function, and the 
detector monopole operators. For identical atomic detectors the 
coupling constants between different detectors and the scalar field can be 
assumed to be the same, i.e., $\mu_{1}=\mu_{2}=\mu$. With this consideration 
the time evolution operator can be expressed as 
\begin{eqnarray}\label{eq:TimeEvolution-int}
 U &=& \mathcal{T}~ 
\exp\left\{-i\int_{-\infty}^{\infty}d\tau\mu\right. \nonumber\\
&& \left.\left.\bigg[ 
\kappa_{1}(\tau_{1}(\tau))m^{1}(\tau_{1}(\tau)) \Phi\left(X_{
1}(\tau_{1}(\tau))\right) \frac{d\tau_{1}(\tau)}{d\tau}
\right.\right.\nonumber\\ 
&& +  \left.\left.  
\kappa_{2}(\tau_{2}(\tau))m^{2}(\tau_{2}(\tau)) \Phi\left(X_{
2}(\tau_{2}(\tau))\right) 
\frac{d\tau_{2}(\tau)}{d\tau}\right]\right\}~,\nonumber\\
\end{eqnarray}
where $\mathcal{T}$ signifies that time ordering is done. 

We consider 
$|\omega\rangle$ to be some collective initial state of the two detector system, 
and $|\Omega\rangle$ some collective final state. It is also considered that the 
initial state $|\omega\rangle$ is prepared in the Minkowski vacuum state 
$|0_{M}\rangle$ of the scalar field, and the final state $|\Omega\rangle$ 
evolves to some field state $|\Theta\rangle$. Then the transition amplitude from 
the state $|\omega;0_{M}\rangle$ to $|\Omega;\Theta\rangle$, at the first order 
perturbation level of the coupling constant $\mu$, will be
\begin{eqnarray}\label{eq:Transition-amp}
 \mathcal{A}_{|\omega;0_{M}\rangle\to|\Omega;\Theta\rangle} &=& 
\langle\Omega;\Theta|{U}|\omega;0_{M}\rangle\nonumber\\
~&\approx& 
-i\mu\langle\Omega;\Theta|\int_{-\infty}^{\infty}d\tau 
\left.\bigg[ \kappa_{1}m^{1} \Phi\left(X_{
1}\right)\right.\times\nonumber\\ 
&&  \frac{d\tau_{1}}{d\tau} + \left.  
\kappa_{2}m^{2} \Phi\left(X_{
2}\right) 
\frac{d\tau_{2}}{d\tau}\right]|\omega;0_{M}
\rangle~.
\end{eqnarray}
From this transition amplitude one can obtain 
the transition probability for all possible field states 
$\left\{|\Theta\rangle\right\}$ as
\begin{eqnarray}\label{eq:Transition-prob}
 \Gamma_{|\omega\rangle\to|\Omega\rangle} &=& 
\sum_{\left\{|\Theta\rangle\right\}} \mathcal{A}_{|\omega;0_{M}
\rangle\to|\Omega;\Theta\rangle}^{*} \mathcal 
{A}_{|\omega;0_{M}\rangle\to|\Omega;\Theta\rangle}\nonumber\\
~&\approx& 
\mu^2 \sum_{j,l=1}^{2} m_{\Omega\omega}^{j*} m_{\Omega\omega}^{l}~ 
F_{jl}\left(\Delta {E}\right) 
~,
\end{eqnarray}
where $\Delta {E}=E_{\Omega}-E_{\omega}$, and $m_{\Omega\omega}^{j} = 
\langle\Omega|m^{j}(0)|\omega\rangle$. The monopole moment operator is defined 
as 
\begin{equation}
m^{j}(0)=|e_{j}\rangle \langle g_{j}|+|g_{j}\rangle \langle e_{j}|~,
\end{equation} 
and one can utilize this expression to acquire the contributions due to 
individual transitions through the monopole moments in the expression of the  
transition probability. In particular one can find out 
$m_{se}^{1}=m_{se}^{2}=1/\sqrt{2}$, $m_{ae}^{1}=-m_{ae}^{2}=-1/\sqrt{2}$, which 
respectively denote the contributions from the monopole moments due to the 
transitions from the symmetric and anti-symmetric states to the collective 
excited state. On the other hand, the contributions due to the transitions from 
the symmetric and anti-symmetric states to the collective ground state are 
$m_{sg}^{1}=m_{sg}^{2}=1/\sqrt{2}$, and $m_{ag}^{2}=-m_{ag}^{1}=-1/\sqrt{2}$. It 
can be observed that the transition from the collective excited to the ground 
state or reverse is not possible as in that case $m_{eg}^{j}=0=m_{ge}^{j}$. One 
can look into Fig. \ref{fig:Energy-levels}, taken from \cite{Costa:2020aqa}, for 
a diagramatic representation of the energy levels and for the monopole operator 
expectation values corresponding to different transitions.

Let us now shift our attention to the transition coefficients in Eq. 
(\ref{eq:Transition-prob}), the explicit form of which is given by 
\begin{eqnarray}\label{eq:Transition-coeff}
 F_{jl}\left(\Delta {E}\right) &=& 
\int_{-\infty}^{\infty}d\tau d\tau'~ 
e^{-i\left(\tau_{j}(\tau)-\tau_{l}(\tau')\right)\Delta {E}} 
\nonumber\\
~&& G_{jl}^{+}\left(\tau_{j}(\tau),\tau_{l}(\tau')\right) \frac{ 
d\tau_ { j }}{d\tau} \frac{d\tau_{l}}{d\tau'}\kappa_{j} 
\kappa_{l}~,
\end{eqnarray}
where, the positive frequency Wightman function is defined as 
\begin{equation}\label{eq:Two-point-fn-gen}
 G_{jl}^{+}\left(\tau_{j}(\tau),\tau_{l}(\tau')\right) = \langle 
0_{M}|\Phi\left[X_{j}(\tau_{j}(\tau))\right]\Phi\left[X_{l}(\tau_{l}
(\tau'))\right ]|0_{M}
\rangle~.
\end{equation}
Later we shall evaluate these transition coefficients specifically considering 
the first detector to be the place where our observer is located, i.e., then $\tau$ 
shall become $\tau_{1}$. It will result in a factor of 
$d\tau_{2}(\tau_{1})/d\tau_{1}$ when one considers the contribution of the 
second detector, see \cite{Rodriguez-Camargo:2016fbq, Costa:2020aqa}. Now we 
shall proceed to evaluate the transition probabilities for the detectors, 
uniformly accelerating with different accelerations, in a thermal bath.  For 
this the explicit expression for the Wightman function is required which we will 
discuss in the next couple of sections. Since the field mode functions with 
respect to uniformly accelerated frame can be represented by using Minkowski as 
well as Rindler decomposition, below both will be discussed. We also mention 
that in our subsequent analysis we are going to consider the switching function 
$\kappa_{j}=1$.

\section{Thermal Wightman function corresponding to Minkowski 
mode}\label{sec:Thermal-GreenFn-Mink}

In this section we first consider a quantum statistical system of finite 
temperature and evaluate the Wightman function corresponding to positive 
frequency mode of free {\it massless real scalar field} with respect to 
Minkowski coordinates. We call these modes as Minkowski modes and with respect 
to them the vacuum is the usual Minkowski vacuum. Then we discuss about the 
Rindler spacetime, which corresponds to a uniformly accelerated object. We complete the 
section with consideration of the accelerated detectors in a thermal bath and by 
constructing the required form of Wightman functions, necessary for the 
evaluation of the transition probability in these trajectories.

\subsection{An observer in a Thermal bath}\label{TR_spacetime1}

We take an observer to be in equilibrium with a thermal bath characterized by 
the parameter $\beta=1/(k_{B}T)$, with $k_{B}$ being the \emph{Boltzmann 
constant} and $T$ the temperature of the thermal bath. In this background we 
further consider a massless minimally coupled scalar field 
$\Phi(X)=\Phi(T,\mathbf{X})$. Then one can obtain the thermal Green's (Wightman) function 
by taking Gibbs ensemble average of the operator $\Phi(X_{2})\Phi(X_{1})$ as 
\begin{eqnarray}\label{eq:Greens-fn-thermal1}
 G^+_{\beta}(X_{2};X_{1}) &=& \langle \Phi(X_{2})\Phi(X_{1}) 
\rangle_{\beta}\nonumber\\
 ~&=& \frac{1}{Z}~\textrm{Tr}\left[e^{-\beta H} \Phi(X_{2})\Phi(X_{1})\right] ~,
\end{eqnarray}
where, $X_{1}$ and $X_{2}$ are two events in the spacetime and  $Z = 
\textrm{Tr}[\exp(-\beta H)]$ denotes the partition function. Here $H$ is the 
Hamiltonian of free massless scalar field (earlier we denoted this as $H_F$ in 
(\ref{eq:Hamiltonian-field})). In Minkowski spacetime, using the standard Fock 
quantization, the scalar field can be expressed in terms of the annihilation and 
creation operators $a_{n}$,  $a^{\dagger}_{n}$ and the positive and negative 
frequency mode functions as 
\begin{eqnarray}\label{eq:Field-Minkowski}
 \Phi(X) = \sum_{k} 
\frac{f_{k}(\mathbf{X})}{\sqrt{2\omega_{k}}}\left(a_{k}e^{-i\omega_{k}T}+a^{
\dagger}_{k} e^ { i\omega_{k}T}\right)~.
\end{eqnarray}
In Fourier domain the scalar field Hamiltonian acts as a sum of infinitely many 
simple Harmonic oscillators. Then one can consider 
$H_{k}=a^{\dagger}_{k}a_{k}\omega_{k}$ as the Hamiltonian operator corresponding 
to the $k^{th}$ excitation, and express the partition function to be 
$Z=\textrm{Tr}\left[e^{-\beta H}\right]=\left(1-e^{-\beta \omega_{k}}\right)^{-1}$. 
Furthermore, the thermal Wightman function 
\cite{Mijic:1993wm, Weldon:2000pe, Chowdhury:2019set} from Eq. 
(\ref{eq:Greens-fn-thermal1}) can be expressed as 
\begin{eqnarray}\label{eq:Greens-fn-thermal2}
 G_{\beta}^{+}(X_{2};X_{1}) &=& \sum_{k} 
\frac{f_{k}(\mathbf{X}_{2})f_{k}^{*}(\mathbf{X}_{1})}{2\omega_{k}}  \left[
\frac{e^{i\omega_{k}\Delta 
T}}{e^{\beta\omega_{k}}-1}\right.\nonumber\\
~&&~~~~~~~~~~ \left. +~~\frac{e^{-i\omega_{k}\Delta T}}{ 1-e^ { - 
\beta\omega_{k}}}\right ] ,
\end{eqnarray}
where, $\Delta T=T_{2}-T_{1}$, and $T_{2}\ge T_{1}$ is considered. From this 
expression the notion of positive frequency Green's function is implicit. 

In a $(1+1)$ dimensional spacetime the mode functions are 
$f_{k}(\mathbf{X})=(1/\sqrt{2\pi})e^{ikX}$, and taking the discrete to 
continuous momentum limit one can express the thermal Green's function as
\begin{eqnarray}
 G_{\beta}^{+}(X_{2};X_{1}) &=& \int \frac{dk}{4\pi\omega_{k}}
 \left[ \frac{e^{ik\Delta X+i\omega_{k}\Delta 
T}}{e^{\beta\omega_{k}}-1}\right.\nonumber\\
~&&~~~~~~~~~ \left. +~~\frac{e^{ik\Delta X-i\omega_{k}\Delta 
T}}{1-e^{- 
\beta\omega_{k}}}\right ]~.
~\label{eq:Two-point-fn-thermal1}
\end{eqnarray}
One can explicitly perform the momentum integration in the expression of this 
Green's function, see \cite{Chowdhury:2019set} (we give this in Appendix 
\ref{Apn:Diff-GreenFn-analogy}; see Eq. (\ref{eq:Greens-fn-thermal3})).

Similarly, in $(1+3)$ dimensions 
the mode functions are $f_{n}(\mathbf{X})=(1/\sqrt{(2\pi)^3})e^{i 
\mathbf{k}.\mathbf{X}}$, and taking the discrete to continuous momentum limit 
the thermal Green's function becomes
\begin{eqnarray}
 G_{\beta}^{+}(X_{2};X_{1}) &=& \int \frac{d^3k}{(2\pi)^3 2\omega_{k}}
 \left[ \frac{e^{i\mathbf{k}.\Delta \mathbf{X}+i\omega_{k}\Delta 
T}}{e^{\beta\omega_{k}}-1}\right.\nonumber\\
~&& \left. ~~~~~~~~~~+~\frac{e^{i\mathbf{k}.\Delta 
\mathbf{X}-i\omega_{k}\Delta 
T}}{1-e^{- 
\beta\omega_{k}}}\right ]~.
\label{eq:Greens-fn-thermal-1p3}
\end{eqnarray}
Here also the momentum integration can be explicitly carried out to provide a 
position space representation of the $(1+3)$ dimensional thermal Green's 
function, see \cite{Chowdhury:2019set} (see Eq. (\ref{eq:Greens-fn-thermal4}) of 
Appendix \ref{Apn:Diff-GreenFn-analogy}). Below we shall express this in the 
frame of an uniformly accelerated observer. Since the Wightman functions are 
scalar quantities, one needs to just use coordinate transformations in order to 
obtain the same in accelerated frame.

\subsection{An accelerated observer in a thermal bath}\label{TR_spacetime2}

The coordinates of a uniformly accelerated object are confined to specific 
regions in Minkowski spacetime. These specific regions constitute the Rindler 
wedges \cite{Crispino:2007eb} in a Minkowski spacetime. Like Minkowski spacetime 
these Rindler wedges also make up for static globally hyperbolic spacetimes. The 
motion of a uniformly accelerated observer is studied considering these Rindler 
frames.

Let us first consider a $(1+1)$-dimensional Minkowski spacetime denoted 
by the coordinates $(T,X)$, with the line element
\begin{equation}\label{eq:Minkowski-metric}
 ds^2 = -dT^2+dX^2~.
\end{equation}
The transformation to the coordinates $(\eta,\xi)$ in the right Rindler wedge 
(RRW), i.e., the region $|T|<X$ in the Minkowski spacetime, is
\begin{eqnarray}\label{eq:Rindler1-trans}
 T &=& \frac{e^{a\xi}}{a} \sinh{a\eta}\nonumber\\
 X &=& \frac{e^{a\xi}}{a} \cosh{a\eta}~.
\end{eqnarray}
In a similar manner one can also define a coordinate transformation suitable to 
the left Rindler wedge (LRW), confined in a region $|T|<-X$ of the Minkowski 
spacetime. The line-elements corresponding to both of the right and left 
Rindler wedges are expressed as 
\begin{equation}\label{eq:Rindler1-metric}
 ds^2 = e^{2a\xi}\left[-d\eta^2+d\xi^2\right]~.
\end{equation}
It should be noted that the generalization of this $(1+1)$ dimensional Rindler 
transformation (\ref{eq:Rindler1-trans}) and (\ref{eq:Rindler1-metric}) to 
$(1+3)$ dimensions, when the observer is accelerating along $X$-direction, is 
quite simple. The Minkowski $Y$ and $Z$ coordinates remain the same for an 
observer accelerated in the $X$ direction. 

In Rindler spacetime the proper time can be estimated to be $\tau=e^{a\xi}\eta$ 
and proper acceleration $b=ae^{-a\xi}$. One can express the coordinate 
transformation from Eq. (\ref{eq:Rindler1-metric}) in terms of the proper 
time and proper acceleration as
\begin{eqnarray}\label{eq:Rindler1-trans2}
 T &=& \frac{1}{b} \sinh{b\tau}\nonumber\\
 X &=& \frac{1}{b} \cosh{b\tau}~.
\end{eqnarray}
As one considers $\xi=0$, then $\eta$ and $a$ respectively denote the proper 
time and acceleration of an accelerated observer. 

Now we are going to consider 
accelerated observers in thermal background. The accelerated observers will be 
described by the Rindler coordinates. We begin by looking into the Green's 
function in this case. It can be seen that one can just use the Rindler 
transformation from Eq. (\ref{eq:Rindler1-trans2}) and put the expressions of 
$\Delta T$, $\Delta X$ in Eq. (\ref{eq:Greens-fn-thermal3}) and Eq. 
(\ref{eq:Greens-fn-thermal4}) to get the Green's function in $(1+1)$ and $(1+3)$ 
dimensions for accelerated observers in thermal background. However, up to our 
knowledge, these forms of Green's functions, as {\it not time translation 
invariant} in accelerated frame, are not suitable to compute the transition 
coefficients of Eq. (\ref{eq:Transition-coeff}). To circumvent this issue the 
Green's functions (\ref{eq:Two-point-fn-thermal1}) and 
(\ref{eq:Greens-fn-thermal-1p3}), where the momentum integration are not yet 
carried out, are considered (see e.g. \cite{Kolekar:2013hra}, where this trick has been used). 

In a $(1+1)$-dimensional thermal background one can express the thermal 
two-point function (\ref{eq:Two-point-fn-thermal1}) in a different form with the 
consideration $\omega_{k}=|k|$ for a massless scalar field, which will be 
convenient for our calculations, as
\begin{eqnarray}\label{eq:Two-point-fn-thermal2}
 G_{\beta}^{+}(X_{j};X_{l}) &=& \int_{0}^{\infty} 
\frac{d\omega_{k}}{4\pi \omega_{k}} 
\left[\tfrac{e^{i\omega_{k}(\Delta T_{jl}-\Delta X_{jl})}+e^{i\omega_{k}(\Delta 
T_{jl} +\Delta 
X_{jl})}}{e^{\beta\omega_{k}}-1} \right.\nonumber\\
&& + \left. \tfrac{e^{-i\omega_{k}(\Delta T_{jl}-\Delta 
X_{jl})}+e^{-i\omega_{k}(\Delta 
T_{jl}+\Delta X_{jl})}}{1-e^{-\beta\omega_{k}}}\right].
\end{eqnarray}
From Eq. (\ref{eq:Rindler1-trans2}) one can obtain
\begin{eqnarray}
 \Delta T_{jl}-\Delta X_{jl} &=& 
-\frac{1}{b_{j}}e^{-b_{j}\tau_{j}}+\frac{1}{b_{l}}e^{-b_{l}
\tau_{l}}\nonumber\\
 \Delta T_{jl}+\Delta X_{jl} &=& 
\frac{1}{b_{j}}e^{b_{j}\tau_{j}}-\frac{1}{b_{l}}e^{b_{l}
\tau_{l}}~,\label{eq:TR-DT&DX}
\end{eqnarray}
where, $\Delta T_{jl}=T_{j,2}-T_{l,1}$ and $\Delta X_{jl} = X_{j,2}-X_{l,1}$. 
Here $j$, $l$ are the notations corresponding to different detectors and the 
subscript $1$, $2$ denote different spacetime points. On the other hand, one 
can also consider the $(1+3)$ dimensional thermal Green's function as given in 
Eq. (\ref{eq:Greens-fn-thermal-1p3}) and get
\begin{eqnarray}
 G_{\beta}^{+}(X_{j};X_{l}) &=& 
\int_{0}^{\pi}\sin{\theta}d\theta \int_{0}^{\infty} 
\tfrac{\omega_{k}d\omega_{k}}{2(2\pi)^2}
 \left[ \tfrac{e^{i\omega_{k}(\Delta X_{jl} \cos{\theta}+\Delta 
T_{jl})}}{e^{\beta\omega_{k}}-1}\right.\nonumber\\
~&& \left. ~~~~~~~~~~+~\tfrac{e^{i\omega_{k}(\Delta 
X_{jl}\cos{\theta}-\Delta 
T_{jl})}}{1-e^{- 
\beta\omega_{k}}}\right ]~,
\label{eq:Greens-fn-TR-4D}
\end{eqnarray}
where, $\Delta T_{jl}$ and $\Delta X_{jl}$ are given by the previously mentioned 
expressions, and we have taken $\Delta Y_{jl}=0=\Delta Z_{jl}$ as the detectors 
are considered to be moving on the same $X-T$ plane. In particular, using the 
coordinate transformation from Eq. (\ref{eq:Rindler1-trans2}) one can obtain
\begin{eqnarray}
 X_{j} \cos{\theta}+T_{j} &=& 
\frac{1}{2b_{j}} \left(\delta_{1}~ e^{b_{j}\tau_{j}}-\delta_{2}~ 
e^{-b_{j}\tau_{j}}\right)\nonumber\\
 X_{j} \cos{\theta}-T_{j} &=& 
\frac{1}{2b_{j}} \left(-\delta_{2}~ e^{b_{j}\tau_{j}}+\delta_{1}~ 
e^{-b_{j}\tau_{j}}\right)~,\label{eq:TR-DT&DX-3D}
\end{eqnarray}
where, $\delta_{1}=1+\cos{\theta}$ and $\delta_{2}=1-\cos{\theta}$. One can put 
these expressions in Eq. (\ref{eq:Greens-fn-TR-4D}) to get the Green's function 
corresponding to accelerated observers in $(1+3)$ dimensional thermal bath. 
These expressions (Eq. (\ref{eq:Two-point-fn-thermal2}) and 
(\ref{eq:Greens-fn-TR-4D})) will be used for our later purpose. It should be 
mentioned that the $(1+1)$ dimensional Green's function of Eq. 
(\ref{eq:Two-point-fn-thermal2}) after the transformations (\ref{eq:TR-DT&DX}) 
and the $(1+3)$ dimensional Green's function of Eq. (\ref{eq:Greens-fn-TR-4D}) 
with the transformations (\ref{eq:TR-DT&DX-3D}) do not remain time translation 
invariant. In the next section we are going to discuss about the Green's 
function for accelerating observers in thermal bath constructed considering the 
Rindler modes.

\section{Thermal Wightman function corresponding to Rindler 
modes}\label{sec:Thermal-GreenFn-Unruh}

The field can also be decomposed with respect to modes, defined in Rindler 
coordinates. The Rindler vacuum is the vacuum for these Rindler modes. Below we will find the thermal positive 
frequency Wightman function corresponding these modes in vacuum of the Unruh modes which is here Minkowski vacuum.

\subsection{$(1+1)$ dimensions}

Let us start our discussion in $(1+1)$ dimensions. In terms of the Rindler 
coordinates the equation of motion for a minimally coupled, massless free scalar 
field $\Phi$ is $\Box\Phi=e^{-2a\xi}(-\partial^2_{\eta}\Phi+\partial^2_{\xi} 
\Phi)=0$. The solution of this equation suggests a set of modes each in the 
right and left Rindler wedge as \cite{book:Birrell, book:carroll}
\begin{eqnarray}\label{eq:Rindler-modes}
 ^{R}u_{k} &=& \frac{1}{\sqrt{4\pi\omega}} 
e^{ik\xi-i\omega\eta}~~~\textup{in~RRW}\nonumber\\ 
~&=& 0~~~~~~~~~~~~~~~~~~~~\textup{in~LRW}\nonumber\\
 ^{L}u_{k} &=& \frac{1}{\sqrt{4\pi\omega}} 
e^{ik\xi+i\omega\eta}~~~\textup{in~LRW}\nonumber\\
~&=& 0~~~~~~~~~~~~~~~~~~~~\textup{in~RRW}.
\end{eqnarray}
In terms of these Rindler modes the scalar field can be expressed, see 
\cite{book:Birrell}, 
as
\begin{equation}\label{eq:Field-Rindler-1p1}
 \scalebox{0.85}{$\Phi(X) = $} \sum_{k=-\infty}^{\infty} 
\scalebox{0.85}{$\left[b^{R}_{k}~ ^{R}u_{k} + b^{R^{\dagger}}_{k}~ 
^{R}u_{k}^{*}+b^{L}_{k}~ ^{L}u_{k} + b^{L^{\dagger}}_{k}~ ^{L}u_{k}^{*}\right]$} 
,
\end{equation}
where, superscript $L$ and $R$ denote modes and the operators corresponding to 
the left and the right Rindler wedges respectively. The operators correspond to 
Rindler vacuum $|0_{R}\rangle$, i.e. $b_{k}^{R}|0_{R}\rangle =0 
=b_{k}^{L}|0_{R}\rangle$. In particular, in the right Rindler wedge where the 
field modes $ ^{L}u_{k}=0$ the scalar field takes the form
\begin{equation}\label{eq:Field-Rindler-1p1-RRW}
 \Phi^{R}(X) = \sum_{k=-\infty}^{\infty} 
\left[b^{R}_{k}~ ^{R}u_{k} + 
b^{R^{\dagger}}_{k}~ ^{R}u_{k}^{*}\right] ~.
\end{equation}
We shall use this scalar field decomposition to obtain the Green's function with 
respect to the Minkowski vacuum for an accelerated observer described in the 
RRW. However, it should be noted that the operators $b^{R}_{k}$ and 
$b^{R^{\dagger}}_{k}$ in Eq. (\ref{eq:Field-Rindler-1p1-RRW}) correspond to the 
Rindler vacuum $|0_{R}\rangle$. To circumvent this issue and to obtain the 
desired Green's function we seek help of a prescription provided by Unruh 
\cite{Unruh:1976db}. In the following discussions we are going to delineate this 
prescription namely the Unruh modes and the operators.\vspace{0.15cm}

As defined in Eq. (\ref{eq:Rindler-modes}) the field modes are separately 
non-vanishing in the two different Rindler wedges of the Minkowski spacetime. 
Unruh in $1976$ provided a prescription out of these different modes which are 
valid in the whole region of the Minkowski spacetime. They are obtained from the 
combination of the Rindler modes $^{R}u_{k} + e^{-\pi\omega/a}~ ^{L}u^{*}_{-k}$ 
and $^{R}u^{*}_{-k} + e^{\pi\omega/a}~ ^{L}u_{k}$. In terms of these modes the 
scalar field can be expressed as \cite{book:Birrell}
\begin{eqnarray}\label{eq:Field-Unruh}
 \Phi(X) &=& \sum_{k=-\infty}^{\infty} 
\frac{1}{\sqrt{2\sinh{\frac{\pi\omega}{a}}}} 
\left[d^{1}_{k}\left(e^{\frac{\pi\omega}{2a}}~ ^{R}u_{k} + 
e^{-\frac{\pi\omega}{2a}}~ ^{L}u^{*}_{-k}\right)\right.\nonumber\\
~&& ~~~+~ \left. d^{2}_{k}\left(e^{-\frac{\pi\omega}{2a}}~ ^{R}u^{*}_{-k} + 
e^{\frac{\pi\omega}{2a}}~ ^{L}u_{k}\right) \right] + h.c.~.
\end{eqnarray}
Here $h.c.$ stands for Hermitian conjugate. It is observed that the Unruh modes 
have the positive frequency analyticity property corresponding to the Minkowski 
time, same as the Minkowski modes. Then the annihilation operators from the two 
sets of Unruh operators $(d^{1}_{k},d^{1^{\dagger}}_{k})$ and 
$(d^{2}_{k},d^{2^{\dagger}}_{k})$ annihilate the Minkowski vacuum
\begin{equation}\label{eq:UM-annihilation-op-M0}
 d^{1}_{k}|0_{M}\rangle=d^{2}_{k}|0_{M}\rangle=0~.
\end{equation}
Because of this particular feature of Eq. (\ref{eq:UM-annihilation-op-M0}) it is 
now quite less cumbersome to obtain any Minkowski state expectation value. This 
can be achieved by transforming Rindler operators in terms of these Unruh 
operators, see \cite{book:Birrell}, using the relations
\begin{eqnarray}\label{eq:bR-1p1-RRW}
b_{k}^{L} &=& \frac{1}{\sqrt{2\sinh{\frac{\pi\omega}{a}}}} 
\left[e^{\frac{\pi\omega}{2a}} d_{k}^{2}+e^{-\frac{\pi\omega}{2a}} 
d_{-k}^{1^{\dagger}}\right]\nonumber\\
 b_{k}^{R} &=& \frac{1}{\sqrt{2\sinh{\frac{\pi\omega}{a}}}} 
\left[e^{\frac{\pi\omega}{2a}} d_{k}^{1}+e^{-\frac{\pi\omega}{2a}} 
d_{-k}^{2^{\dagger}}\right]~,
\end{eqnarray}
which is similar to the Bogoliubov transformation. Then putting this 
transformation in Eq. (\ref{eq:Field-Rindler-1p1-RRW}) one can get the 
expression of the field in the RRW in terms of the Unruh operators as
\begin{eqnarray}\label{eq:RRW-field-Unruh}
 \Phi^{R}(X) &=& \sum_{k=-\infty}^{\infty} 
\frac{1}{\sqrt{2\sinh{\frac{\pi\omega}{a}}}} 
\left[d^{1}_{k}~e^{\frac{\pi\omega}{2a}}~ ^{R}u_{k} \right.\nonumber\\
~&& ~~~+~ \left. d^{2}_{k}~e^{-\frac{\pi\omega}{2a}}~ ^{R}u^{*}_{-k} \right] + 
h.c.~.
\end{eqnarray}
This expression of the scalar field in RRW is the same as the representation of 
Eq. (\ref{eq:Field-Unruh}) with $ ^{L}u_{k}$ taken to be zero.

First let us evaluate the Green's function for a single accelerated detector, 
without any thermal bath, considering the scalar field in RRW, i.e., the field 
decomposition from Eq. (\ref{eq:Field-Rindler-1p1-RRW}) with Unruh operators. 
The field modes are expressed by Eq. (\ref{eq:RRW-field-Unruh}). Then one can 
obtain the Green's function, with $\omega=\omega_{k}=|k|$ considering the 
massless scalar field, for an accelerated observer with respect to the Rindler 
modes as
\begin{eqnarray}\label{eq:Greens-fn-Unruh}
 G^{+}_{R}\left(\Delta\xi,\Delta\eta\right) &=& \int_{-\infty}^{\infty} 
\frac{dk}{4\pi\omega_{k}} 
\left[\frac{e^{ik\Delta\xi-i\omega_{k}\Delta\eta}}{1-e^{\frac{-2\pi\omega_{k}
} { a }}}\right. \nonumber\\
~&& \left.~~~~~ +~~ 
\frac{e^{ik\Delta\xi+i\omega_{k}\Delta\eta}}{e^{\frac{2\pi\omega_{k}}{ a 
}}-1}\right]~.
\end{eqnarray}
In the above vacuum has been chosen to be that of Unruh mode, which is Minkowski vacuum here.
For a derivation of this expression one can consider looking into Appendix 
\ref{Apn:Unruh-Greens-fn}. Comparing this expression (\ref{eq:Greens-fn-Unruh}) 
with that of the thermal Green's function (\ref{eq:Two-point-fn-thermal1}), we 
notice that they are exactly the same with $\beta$ is now identified as 
$2\pi/a$, and $\Delta\xi$ and $\Delta\eta$ resemble $\Delta T$ and $\Delta X$. 
Therefore, one can clearly proclaim that an accelerated observer with the 
Rindler modes in RRW mimics the thermal background.\vspace{0.1cm}

On the other hand, one can obtain the Green's function corresponding to 
accelerated observers, with respect to Rindler modes, in a thermal bath in a 
similar manner. We consider the Hamiltonian corresponding to the $k^{th}$ 
excitation to be  $H_{k} = (d^{1^{\dagger}}_{k} d^{1}_{k}+ 
d^{2^{\dagger}}_{k}d^{2}_{k})\omega_{k}$ (as the vacuum under study is that of Unruh mode) to evaluate the Green's function, 
defined by Eq. (\ref{eq:Greens-fn-thermal1}) in a thermal background. This turns 
out to be
\begin{eqnarray}\label{eq:Greens-fn-TU}
&& G_{\beta_R}^{+}\left(\Delta\xi_{jl},\Delta\eta_{jl}\right)\nonumber\\  
&&~~~~~~= \int_{-\infty}^{\infty} 
\frac{dk}{8\pi\omega_{k}\sqrt{\sinh{\frac{\pi\omega_{k}}{a_{j}}}\sinh{
\frac{\pi\omega_{k}}{a_{l}}}}}\times\nonumber\\ 
&&~~~~~~\left[\frac{1}{1-e^{-\beta\omega_{k}
}}\left\{e^{ik\Delta\xi_{jl}-i\omega_{k}\Delta\eta_{jl}} 
~e^{\frac{\pi\omega_{k}}{2}\left(\frac{1}{a_{j}}+\frac{1}{a_{l}}\right)}
\right.\right. \nonumber\\
~&&~~~~~~ \left. ~~+~ 
e^{ik\Delta\xi_{jl}+i\omega_{k}\Delta\eta_{jl}}~
e^{-\frac{\pi\omega_{k}}{2}\left(\frac{1}{a_{j}}+\frac{1}{a_{l}}\right)}
\right\}\nonumber\\
&&~~~~~~ +  \frac{1}{e^{ \beta\omega_{k}}-1}
\left\{e^{-ik\Delta\xi_{jl}+i\omega_{k}\Delta\eta_{jl}} 
~e^{\frac{\pi\omega_{k}}{2}\left(\frac{1}{a_{j}}+\frac{1}{a_{l}}\right)}
\right.\nonumber\\ 
~&&~~~~~~ \left.\left.~~ +~ e^{-ik\Delta\xi_{jl}-i\omega_{k}\Delta\eta_{jl}}~ 
e^{-\frac{\pi\omega_{k}}{2}\left(\frac{1}{a_{j}}+\frac{1}{a_{l}}\right)} 
\right\}\right ] , 
\end{eqnarray}
where, $\Delta\xi_{jl}=\xi_{j,2}-\xi_{l,1}$ and 
$\Delta\eta_{jl}=\eta_{j,2}-\eta_{l,1}$. For a derivation of this expression one 
can look into Appendix \ref{Apn:TU-Greens-fn}. It should be noted that Eq. 
(\ref{eq:Greens-fn-Unruh}) corresponds to the Green's function of a single 
accelerated observer expressed in terms of the Rindler modes. Whereas, Eq. 
(\ref{eq:Greens-fn-TU}) corresponds to accelerated observers, generally 
considered to be of different accelerations, in thermal bath expressed in terms 
of the Rindler modes. In particular, when $a_{j}=a_{l}=a $ and $\beta\to\infty$ 
we get back the expression of Eq. (\ref{eq:Greens-fn-Unruh}). Furthermore, one 
may consider $\xi_{j}=0$ then the proper accelerations of the observers are 
$b_{j}=a_{j}$ and the proper times $\tau_{j}=\eta_{j}$. Then one may also notice 
the clear difference of Eq. (\ref{eq:Greens-fn-TU}) from its counter part Eq. 
(\ref{eq:Two-point-fn-thermal2}) after the transformation (\ref{eq:TR-DT&DX}) 
corresponding to the Minkowski mode. The Green's function, coming from Rindler 
mode, is time translation invariant, whereas for the Minkowski modes it was not. 
We are going to use Eq. (\ref{eq:Greens-fn-TU}) also to obtain the transition 
probabilities later.

\subsection{$(1+3)$ dimensions}

In this subsection we talk about Rindler mode decomposition of the massless real 
scalar field in $(1+3)$ dimensions. The positive frequency mode solutions of the 
scalar field equation of motion $\Box\Phi=0$ in $(1+3)$ dimensions in terms of 
the Rindler coordinates in the right and the left Rindler wedges are
\begin{eqnarray}\label{eq:Rindler-modes-3D}
 ^{R}u_{\omega,\kp} &=& 
\frac{1}{2\pi^2}\sqrt{\frac{\sinh{\left(\frac{\pi\omega}{a}\right)}}{a}}~ 
\mathcal{K}\left[\frac{i\omega}{a},\frac{|\kp| e^{a\xi}}{a}\right]\nonumber\\
~&& ~~~~~~\times~~ 
e^{-i\omega\eta+i\vec{\kp}.\vec{x}}~~~~~~~~\textup{in~RRW}\nonumber\\ 
~&=& 0~~~~~~~~~~~~~~~~~~~~~~~~~~~~~~~~~\textup{in~LRW}\nonumber\\
 ^{L}u_{\omega,\kp} &=& 
\frac{1}{2\pi^2}\sqrt{\frac{\sinh{\left(\frac{\pi\omega}{a}\right)}}{a}}~ 
\mathcal{K}\left[\frac{i\omega}{a},\frac{|\kp| e^{a\xi}}{a}\right]\nonumber\\
 ~&&~~~~~~\times~~
e^{i\omega\eta+i\vec{\kp}.\vec{x}}~~~~~~~~~~\textup{in~LRW}\nonumber\\
~&=& 0~~~~~~~~~~~~~~~~~~~~~~~~~~~~~~~~~\textup{in~RRW}~,
\end{eqnarray}
where, $\mathcal{K}\left[n,z\right]$ denotes the modified Bessel function of the 
second kind of order $n$, and $\vec{x}$ is perpendicular to the direction of 
acceleration, i.e., in the $Y-Z$ plane, see \cite{Compere:2019rof, 
Crispino:2007eb, Higuchi:2017gcd}. Here $\vec{\kp}$ denotes the transverse wave 
vector in the $Y-Z$ plane. Like the $(1+1)$ dimensional case here also one can 
decompose the scalar field confined to the right Rindler wedge as
\begin{equation}\label{eq:Field-Rindler-1p3-RRW}
 \Phi^{R}(X) = \sum_{\omega=0}^{\infty}\sum_{\kp=-\infty}^{\infty}
\scalebox{0.95}{$\left[b^{R}_{\omega,\kp}~ ^{R}u_{\omega,\kp} + 
b^{R^{\dagger}}_{\omega,\kp}~ ^{R}u_{\omega,\kp}^{*}\right]$} ~.
\end{equation}
where, the operators correspond to Rindler vacuum $|0_{R}\rangle$, i.e. 
$b_{\omega,\kp}^{R}|0_{R}\rangle =0$. Here also our main aim is to obtain the 
Green's function using the field decomposition of Eq. 
(\ref{eq:Field-Rindler-1p3-RRW}) with respect to the Minkowski vacuum, which 
corresponds to an accelerating observer. In this regard, similar to the $(1+1)$ 
dimensional case one can utilize the concepts of the Unruh modes and operators. 
In particular, one can obtain the modes $^{R}u_{\omega,\kp} + e^{-\pi\omega/a}~ 
^{L}u^{*}_{\omega,-\kp}$ and $^{R}u^{*}_{\omega,-\kp} + e^{\pi\omega/a}~ 
^{L}u_{\omega,\kp}$ in terms of the Rindler modes which are valid in the whole 
Minkowski spacetime and have the positive frequency analyticity property with 
respect to the Minkowski time. In terms of these modes the scalar field is
\begin{eqnarray}\label{eq:Field-Unruh-3D}
 \Phi(X) &=& \sum_{\omega=0}^{\infty}\sum_{\kp=-\infty}^{\infty} 
\frac{1}{\sqrt{2\sinh{\frac{\pi\omega}{a}}}} \times\nonumber\\
~&& \left[d^{1}_{\omega,\kp}\left(e^{\frac{\pi\omega}{2a}}~ ^{R}u_{\omega,\kp} 
+ e^{-\frac{\pi\omega}{2a}}~ 
^{L}u^{*}_{\omega,-\kp}\right) + \right.\nonumber\\
~&& \left. d^{2}_{\omega,\kp}\left(e^{-\frac{\pi\omega}{2a}}~ 
^{R}u^{*}_{\omega,-\kp} + 
e^{\frac{\pi\omega}{2a}}~ ^{L}u_{\omega,\kp}\right) \right] + h.c.~.\nonumber\\
\end{eqnarray}
Here also the lowering operators from the two sets of Unruh annihilation and 
creation operators $(d^{1}_{\omega,\kp},d^{1^{\dagger}}_{\omega,\kp})$ and 
$(d^{2}_{\omega,\kp},d^{2^{\dagger}}_{\omega,\kp})$, annihilate the Minkowski 
vacuum
\begin{equation}\label{eq:UM-annihilation-op-M0-3D}
 d^{1}_{\omega,\kp}|0_{M}\rangle=d^{2}_{\omega,\kp}|0_{M}\rangle=0~.
\end{equation}
Then it will be helpful to express the scalar field in the RRW from Eq. 
(\ref{eq:Field-Rindler-1p3-RRW}), using a relation analogous to Eq. 
(\ref{eq:bR-1p1-RRW}), in form 
\begin{eqnarray}\label{eq:RRW-field-Unruh-1p3}
 \Phi^{R}(X) = \sum_{\omega=0}^{\infty}\sum_{\kp=-\infty}^{\infty} 
\frac{1}{\sqrt{2\sinh{\frac{\pi\omega}{a}}}} \times 
~~~~~~~~~~~~~~~~~~~~~&&\nonumber\\
\left[d^{1}_{\omega,\kp}e^{\frac{\pi\omega}{2a}}~ ^{R}u_{\omega,\kp} 
 + d^{2}_{\omega,\kp}e^{-\frac{\pi\omega}{2a}}~ 
^{R}u^{*}_{\omega,-\kp} \right] + h.c.~,&&\nonumber\\
\end{eqnarray}
in terms of the Unruh operators. It should be mentioned that like the $(1+1)$ 
dimensional case here also this field decomposition is exactly same as the Unruh 
field decomposition from Eq. (\ref{eq:Field-Unruh-3D}) with the Rindler mode 
corresponding to the left Rindler wedge $^{L}u_{\omega,\kp}$ considered to be 
zero. Like before we evaluate the Green's function for accelerated detectors 
considering the field decomposition of Eq. (\ref{eq:RRW-field-Unruh-1p3}), where 
the accelerating observer is considered to be in the right Rindler wedge. Then 
the Green's function for an accelerated observer without thermal bath with 
respect to the Rindler modes is obtained as
\begin{equation}\label{eq:Greens-fn-Unruh-3D}
 \scalebox{0.96}{$G^{+}_{R}\left(\Delta\xi,\Delta\eta\right)$} = 
\int_{0}^{\infty}
\scalebox{1}{$\frac{\omega d\omega}{(2\pi)^2} 
\left[\frac{e^{-i\omega\Delta\eta}}{1-e^{\frac{-2\pi\omega
} { a }}} + 
\frac{e^{i\omega\Delta\eta}}{e^{\frac{2\pi\omega}{ a 
}}-1}\right]  e^{-2a\xi_{0}}$}~.
\end{equation}
In deriving this expression in $(1+3)$ dimensions we have considered that the 
accelerated detector to be positioned at a fixed Rindler $\xi$ coordinate, 
$\xi=\xi_{0}$.  

In a similar manner, one can obtain the Green's function 
corresponding to accelerated observers, with respect to Rindler modes in a 
thermal bath in $(1+3)$ dimensions. We use the RRW field decomposition from Eq. 
(\ref{eq:RRW-field-Unruh-1p3}) and take the Hamiltonian to be $H_{\omega,\kp} = 
(d^{1^{\dagger}}_{\omega,\kp} d^{1}_{\omega,\kp} + d^{2^{\dagger}}_{\omega,\kp} 
d^{2}_{\omega,\kp})\omega$ to evaluate the Green's function, defined by Eq. 
(\ref{eq:Greens-fn-thermal1}) in a thermal background. This Green's function is
\begin{eqnarray}\label{eq:Greens-fn-TU-3D}
&& G_{\beta_R}^{+}\left(\Delta\eta_{jl}\right)\nonumber\\  
&&= \int_{0}^{\infty}d\omega~\int 
\frac{ d^2\kp}{(2\pi)^4} \frac{2}{\sqrt{a_{j}a_{l}}}\nonumber\\ 
&&\left[\frac{e^{-i\omega\Delta\eta_{jl}} 
~e^{\frac{\pi\omega}{2}\left(\frac{1}{a_{j}}+\frac{1}{a_{l}}\right)}
 + e^{i\omega\Delta\eta_{jl}}~
e^{-\frac{\pi\omega}{2}\left(\frac{1}{a_{j}}+\frac{1}{a_{l}}\right)}}{1-e^{
-\beta\omega
}}\right.\nonumber\\
&& + \left.  \frac{e^{i\omega\Delta\eta_{jl}} 
~e^{\frac{\pi\omega}{2}\left(\frac{1}{a_{j}}+\frac{1}{a_{l}}\right)}
 + e^{-i\omega\Delta\eta_{jl}}~ 
e^{-\frac{\pi\omega}{2}\left(\frac{1}{a_{j}}+\frac{1}{a_{l}}\right)}}{e^{ 
\beta\omega}-1}
\right ] \nonumber\\
~&&~~~~~~\mathcal{K}\left[\frac{i\omega}{a_{j}},\frac{|\kp| 
e^{a_{j}\xi_{j}}}{a_{j}}
\right]\mathcal{K}\left[\frac{i\omega}{a_{l}},\frac{|\kp| 
e^{a_{l}\xi_{l}}}{a_{l}}
\right]~,
\end{eqnarray}
where, $\Delta\eta_{jl}=\eta_{j,2}-\eta_{l,1}$ and $\xi_{j}$ is the fixed 
Rindler spatial coordinate corresponding to the $j^{th}$ detector. Note again that the above one is time translational invariant. We are going 
to utilize these above mentioned considerations to obtain the transition 
probabilities of Eq. (\ref{eq:Transition-prob}) for two accelerated atoms 
immersed in a thermal bath considering the Rindler modes in $(1+3)$ dimensions.

\section{Transition probability for accelerated atoms in thermal bath with 
respect to Minkowski modes}\label{sec:Transition-prob-Mink}

In this section we are going to estimate the transition probability from Eq. 
(\ref{eq:Transition-prob}) for two entangled atoms accelerated in a thermal 
bath. We shall consider here the $(1+1)$ and $(1+3)$ dimensional Green's 
functions (\ref{eq:Two-point-fn-thermal2}) and (\ref{eq:Greens-fn-TR-4D}) with 
the coordinate transformations (\ref{eq:TR-DT&DX}) and (\ref{eq:TR-DT&DX-3D}), 
which correspond to accelerated detectors in thermal background with respect to 
the Minkowski modes. Our observer is considered to be co-moving with the first 
accelerated detector, and we perform all of our evaluations with respect to this 
first frame. The proper time of the first atom $\tau=\tau_{1}$ will be used to 
carry out the integration in Eq. (\ref{eq:Transition-coeff}) to evaluate the 
transition coefficients. We consider for both of the atoms the Rindler parameter 
$a$ to be the same. Then for different atoms with different constant $\xi$ the 
proper accelerations $b_{1}$ and $b_{2}$ will be different with different proper 
times. This provides a spatial separation between them in the $X$-direction. Now 
if one considers both of the observers with respect to the same Rindler time 
$\eta$ one can obtain a relation between the proper times as 
\begin{equation}
\tau_{2}=\frac{b_{1}}{b_{2}}\tau_{1}~.
\label{B1}
\end{equation}
See \cite{Rodriguez-Camargo:2016fbq} and Appendix \ref{Apn:rel-proper-time} for 
details. We shall use this relation to evaluate the the transition coefficients.

\subsection{$(1+1)$ dimensions}

To evaluate the transition coefficients in $(1+1)$ dimensions for two entangled 
atoms accelerated in a thermal bath considering the Minkowski modes we consider 
the Green's function (\ref{eq:Two-point-fn-thermal2}) and put it in Eq. 
(\ref{eq:Transition-coeff}) with the coordinate transformation 
(\ref{eq:TR-DT&DX}). Let us denote $\alpha_{j} = (b_{1}/b_{2})^{\delta_{2,j}}$, 
where $\delta_{i,j}$ denotes the Kronecker delta with detector's indices $i$ and 
$j$. Then the proper time of the $j^{th}$ atom, by Eq. (\ref{B1}), is given by 
$\tau_{j}=\alpha_{j}\tau_{1}$. One can then express the coefficient functions 
$F_{jl}\left(\Delta {E}\right)$ as
\begin{eqnarray}\label{eq:Transition-coeff-TR-F11}
F_{jl}\left(\Delta {E}
\right) &=& 
\int_{0}^{\infty} 
\frac{d\omega_{k}}{4\pi \omega_{k}}
~\alpha_{j}\alpha_{l}\nonumber\\
~&&
\left[\tfrac{\mathcal{I}_{1}(
b_{j})~ \mathcal{I}^{*}_{1}(
b_{l}) + \mathcal{I}_{2}(
b_{j})~ \mathcal{I}^{*}_{2}(
b_{l})}{e^{\beta\omega_{k}}-1}\right.\nonumber\\ 
&+& 
\left.\tfrac{\mathcal{I}_{3}(
b_{j})~ \mathcal{I}^{*}_{3}(
b_{l})+\mathcal{I}_{4}(
b_{j})~ \mathcal{I}^{*}_{4}(
b_{l})}{1-e^{-\beta\omega_{k}}}\right] 
,\nonumber\\
\end{eqnarray}
where 
\begin{eqnarray}\label{eq:Transition-coeff-TR-F11-2}
 \mathcal{I}_{1}(b_{j}) &=& 
\int_{-\infty}^{\infty} d\tau_{1} 
e^{-i\Delta {E} 
\tau_{j}} 
\exp{\left(-\tfrac{i\omega_{k}}{b_{j}}e^{-b_{j}\tau_{j}}\right)}
\nonumber\\
 \mathcal{I}_{2}(b_{j}) &=& 
 \int_{-\infty}^{\infty} d\tau_{1} 
e^{-i\Delta {E} 
\tau_{j}} 
\exp{\left(\tfrac{i\omega_{k}}{b_{j}}e^{b_{j}\tau_{j}}\right)}
\nonumber\\
 \mathcal{I}_{3}(b_{j}) &=& 
 \int_{-\infty}^{\infty} d\tau_{1} 
e^{-i\Delta {E} \tau_{j}} 
\exp{\left(\tfrac{i\omega_{k}}{b_{j}}e^{-b_{j}\tau_{j}}\right)}
\nonumber\\
 \mathcal{I}_{4}(b_{j}) &=& 
\int_{-\infty}^{\infty} d\tau_{1} 
e^{-i\Delta {E} 
\tau_{j}} 
\exp{\left(-\tfrac{i\omega_{k}}{b_{j}}e^{b_{j}\tau_{j}}\right)}
~.\nonumber\\
\end{eqnarray}
To simplify these integral expressions one can make change of variables 
$e^{-b_{j}\tau_{j}}=y$ and $e^{b_{j}\tau_{j}}=z$, and we also use of the 
relation $\tau_{j}=\alpha_{j}\tau_{1}$ between the proper times 
corresponding to two differently accelerated observers. Then the integrals 
become
\begin{eqnarray}\label{eq:Transition-coeff-TR-F11-3}
 \mathcal{I}_{1}(b_{j}) &=& 
\tfrac{1}{b_{j}\alpha_{j}} 
\int_{0}^{\infty} 
dz~ 
z^{-1+\frac{i\Delta {E}}{b_{j}}} 
e^{-\frac{i\omega_{k}}{b_{j}}z}\nonumber\\
&=& \tfrac{1}{b_{j}\alpha_{j}} 
\left(\frac{\omega_{k}}{b_{j}}\right)^{-\frac{i\Delta {E}}{
b_{j}
}} 
e^{\frac{\pi\Delta {E}}{2b_{j}}}\Gamma\left(\tfrac{i\Delta {E}}{
b_{j}}\right)\nonumber\\
&=& \mathcal{I}^{*}_{2}(b_{j})\nonumber\\
 \mathcal{I}_{3}(b_{j}) &=& 
\tfrac{1}{b_{j}\alpha_{j}} 
\int_{0}^{\infty} dz~ 
z^{-1+\frac{i\Delta {E}}{b_{j}}} 
e^{\frac{i\omega_{k}}{b_{j}}z}\nonumber\\
&=& \tfrac{1}{b_{j}\alpha_{j}} 
\left(\frac{\omega_{k}}{b_{j}}\right)^{-\frac{i\Delta {E}}{
b_{j}
}} e^{-\frac{\pi\Delta {E}}{2b_{j}}} 
\Gamma\left(\tfrac{i\Delta {E}}{b_{j}}\right) \nonumber\\
&=& \mathcal{I}^{*}_{4}(b_{j})~.
\end{eqnarray}
To perform the above integration we have used the formula \begin{equation} 
\int_0^\infty dx ~x^{s-1}e^{-bx} = e^{-s\ln b}~\Gamma(s)~, \end{equation} with 
the conditions Re$(b)>0$ and Re$(s)>0$. To ensure the convergence of our 
integrals the standard prescription has been adopted here (see e.g. 
\cite{book:PadmanabhanGrav} for details). Using these results one can express 
the coefficient functions in Eq. (\ref{eq:Transition-coeff-TR-F11}) as 
$F_{jl}\left(\Delta {E}\right) = \int_{0}^{\infty} 
d\omega_{k}~\mathcal{F}_{jl}\left(\Delta {E},\omega_{k}\right)$. Here 
$\mathcal{F}_{jl}\left(\Delta {E},\omega_{k}\right)$ denote the transition 
coefficients corresponding to each mode with wave number $k$. These transition 
coefficients are given by
\begin{eqnarray}\label{eq:Transition-coeff-TR-Fjl}
 && \mathcal{F}_{jl}\left(\Delta {E},\omega_{k}\right) =
\tfrac{Re\left[\mathcal{C}_{1}(b_{j},b_{l})\right]}{
2\pi 
\omega_{k} b_{j}b_{l}}~
 \times \nonumber\\
&&~~~~~~~~~~~~~~~
\left[\tfrac{e^{\frac{\pi\Delta {E}}{2}\left(\frac{
1}{b_{j}}+\frac{1}{b_{l}
}\right)}}{e^{\beta\omega_{
k}}-1} +\tfrac{e^{-\frac{\pi\Delta {E}}{2}\left(\frac{
1}{b_{j}}+\frac{1}{b_{l}
}\right)}}{1-e^{-\beta\omega_{
k}}} 
\right],
\end{eqnarray}
where
\begin{eqnarray}\label{eq:Expression-C1}
 \mathcal{C}_{1}(b_{j},b_{l}) &=&
\left(\tfrac{\omega_{k}}{b_{j}}\right)^{-\frac{i\Delta {E}}{
b_{j}}} 
\left(\tfrac{\omega_{k}}{b_{l}}\right)^{\frac{i\Delta {E}}{
b_{l}}}\nonumber\\
~&&~~~~~ \Gamma\left(\tfrac{i\Delta {E}}{b_{j}}\right)
\Gamma\left(-\tfrac{i\Delta {E}}{ b_{l}}\right)~.
\end{eqnarray}
From (\ref{eq:Transition-coeff-TR-Fjl}) and 
(\ref{eq:Expression-C1}) one can obtain the expression of the transition 
coefficient $\mathcal{F}_{11}(\Delta {E},\omega_{k})$ as
\begin{eqnarray}\label{eq:Transition-coeff-TR-F11-4}
 \mathcal{F}_{11}\left(\Delta {E},\omega_{k}\right) &=& 
\frac{1}{\omega_{k}\Delta {E} b_{1}} 
\left[\frac{1}{e^{\beta\omega_{k}}-1}\frac{1}{1-e^{\frac{-2\pi\Delta {E}}{
b_{1} }}} \right.\nonumber\\
&&~~ +~~ \left. 
\frac{1}{1-e^{-\beta\omega_{k}}}\frac{1}{e^{\frac{2\pi\Delta {E}}{
b_{1} }}-1}\right]~,
\end{eqnarray}
where, we have used the Gamma function identity $\Gamma(i z)\Gamma(-i 
z)=\pi/(z\sinh{\pi z})$. Also Eq. (\ref{eq:Transition-coeff-TR-Fjl}) and 
(\ref{eq:Expression-C1}) shows that for equal acceleration $b_{2}=b_{1}$ all the 
transition coefficients $\mathcal{F}_{jl}\left(\Delta {E},\omega_{k}\right)$ 
become equal to $\mathcal{F}_{11}\left(\Delta {E},\omega_{k}\right)$. It should 
be mentioned that for a single detector, accelerated in thermal background, 
$\mathcal{F}_{11}\left(\Delta {E},\omega_{k}\right)$ corresponds to the required 
transition probability. The same has been obtained earlier in 
\cite{Kolekar:2013hra}. Note that it is not symmetric under the exchange 
$\beta\leftrightarrow (2\pi)/b_1$ due to the over all multiplicative factor 
$(1/b_1)$. This originates in our calculation as it is based on the Minkowski 
mode. Later we will notice that Rindler modes do not give rise to this 
asymmetric property.

\begin{figure}[h]
\centering
 \includegraphics[width=0.8\linewidth]{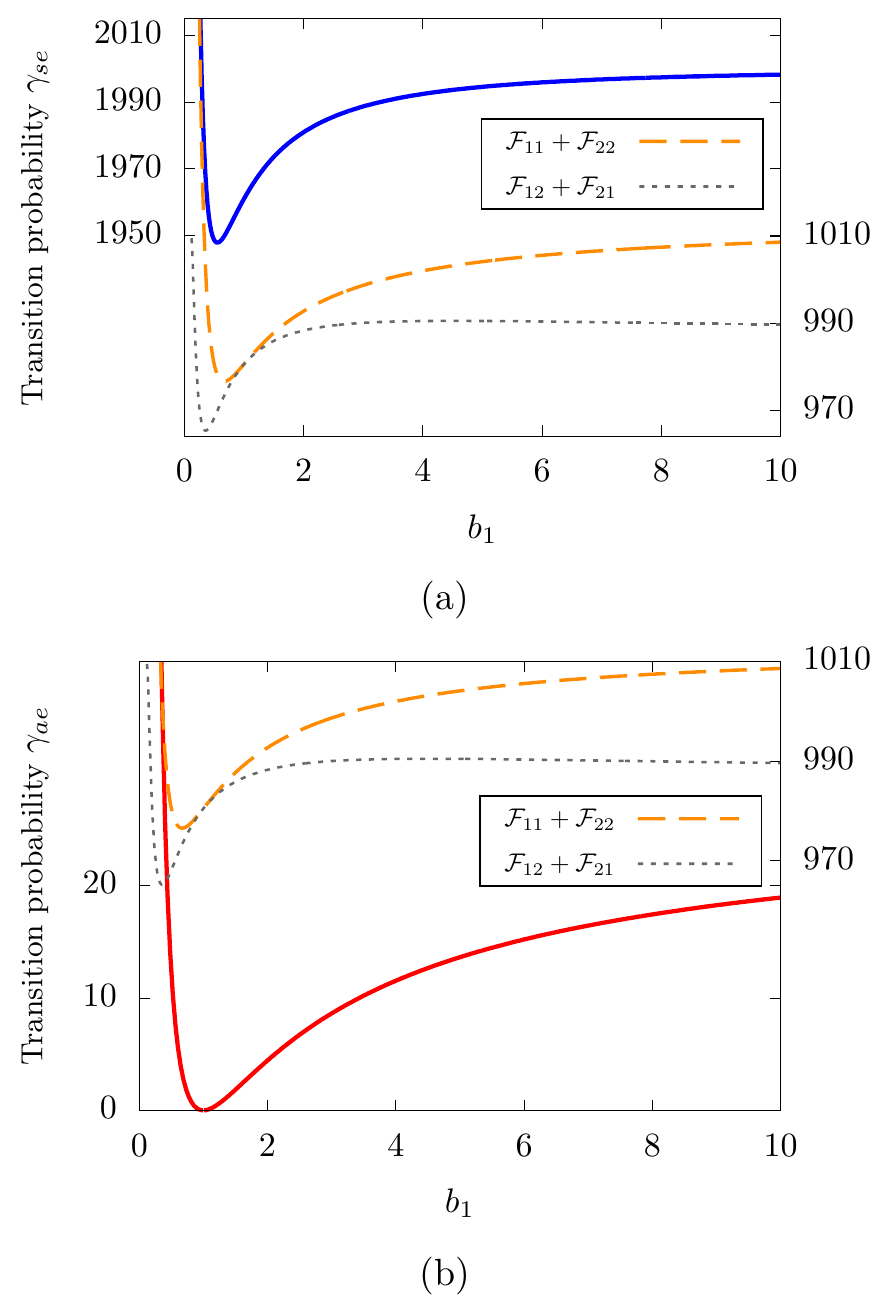}
 \caption{$(1+1)$ dimensions: (a) The transition probability from the symmetric 
state to the collective excited state denoted by the solid blue line. (b) The 
transition probability from the anti-symmetric state to the collective excited 
state denoted by the solid red line. In both of the plots the sum of the 
transition coefficients $(\mathcal{F}_{11}+\mathcal{F}_{22})$ for the 
transitions from the symmetric or anti-symmetric states to the excited state are 
denoted by orange dashed lines. The sum of the transition coefficients 
$(\mathcal{F}_{12} + \mathcal{F}_{21})$ for the transitions from the symmetric 
or anti-symmetric states to the excited state are denoted by gray dotted lines. 
In all of the above cases $b_{2}=1$ is kept fixed and $b_{1}$ is varied. The 
value of the other parameters are $\Delta {E}=0.1$, $\omega_{k}=0.1$ and 
$\beta=2\pi$.}
 \label{fig:TsgTR-P1}
\end{figure}

Now we shall evaluate the transition probability between different states 
of the two-atom system using Eq. (\ref{eq:Transition-prob}). For instance the 
transition probability from the symmetric entangled state $|s\rangle$ to the 
collective excited state $|e\rangle$ comes out to be $\Gamma_{se} = \int_{0}^{\infty} 
d\omega_{k}~\gamma_{se}$, where the expression of $\gamma_{se}$ is given by,
\begin{eqnarray}\label{eq:Tran-prob-se}
 \gamma_{se} &=& \frac{\mu^2}{2} 
\left[\{\mathcal{F}_{11}(\omega_{0},\omega_{k})+\mathcal{F}_{22}(\omega_{0},
\omega_{k})\}
\right.\nonumber\\
~&& ~~\left.~ +~ \{\mathcal{F}_{12}
(\omega_{0},\omega_{k} )+\mathcal{F}_{21}(\omega_{0}),\omega_{k}\}\right]~.
\end{eqnarray}
Whereas the same between the anti-symmetric state 
$|a\rangle$ to the excited state $|e\rangle$ is provided by
\begin{eqnarray}\label{eq:Tran-prob-ae}
 \gamma_{ae} &=& \frac{\mu^2}{2} 
\left[\{\mathcal{F}_{11}(\omega_{0},\omega_{k})+\mathcal{F}_{22}(\omega_{0},
\omega_{k})\}
\right.\nonumber\\
~&& ~~\left.~ -~ \{\mathcal{F}_{12}
(\omega_ { 0 },\omega_{k} )+\mathcal{F}_ { 21 } (\omega_{0},\omega_{k}) 
\}\right]~.
\end{eqnarray}
It should be noted that in both of the above cases the change in energy level of 
the collective system is $\Delta {E}=\omega_{0}-0=\omega_{0}$. Furthermore, for 
the transitions from the symmetric and anti-symmetric states to the ground state 
one has $\Delta {E}=0-(-\omega_{0})=\omega_{0}$. Then one can find out the 
transition probabilities from the  symmetric and anti-symmetric states to the 
ground state, also provided by the $\gamma_{se}$ and $\gamma_{ae}$, 
respectively. Also note that for equal proper acceleration of the two atoms the 
transition probability from the anti-symmetric state to the collective excited 
state or to the collective ground state becomes zero. The transition probability 
corresponding to the transition from the symmetric and anti-symmetric entangled 
states to the collective excited state as a function of $b_1$ are depicted in 
Fig. \ref{fig:TsgTR-P1}. 
\noindent
It should be noted that in this figure we have plotted 
the quantities from Eq. (\ref{eq:Tran-prob-se}) and (\ref{eq:Tran-prob-ae}) per 
unit $\mu^2/2$. Furthermore, in the subsequent studies also we shall take the 
same consideration. Here since the observer is attached with the first atom, we 
show the variation with respect to first atom's proper acceleration $b_1$.
For the symmetric case, the entanglement between the states of atoms, acts as 
enhancement in the transition probability (the cross terms, e.g. 
$\mathcal{F}_{12}$ and $\mathcal{F}_{21}$ are added), whereas in the 
anti-symmetric situation entanglement provides decrease of transition 
probability. Interesting point to be noted from the figure that in both of the 
cases the transition coefficients at first tend to decrease with increasing 
acceleration $b_{1}$ of the first observer, giving rise to a possible case of 
the anti-Unruh-like effect, and then increases with it. Later, before 
providing our concluding remarks, we shall elaborately 
discuss about this phenomena of anti-Unruh-like effect in this scenario.

\subsection{$(1+3)$ dimensions:}

To estimate the transition coefficients between different states of a collective 
system of two entangled atoms accelerated in a thermal bath in $(1+3)$ 
dimensions one can put the expression of the Green's function 
(\ref{eq:Greens-fn-TR-4D}) in Eq. (\ref{eq:Transition-coeff}).  With this and 
the substitution of coordinate transformations (\ref{eq:TR-DT&DX-3D}) one 
obtains
\begin{eqnarray}\label{eq:TranCoeffTR-Fij3D}
&& F_{jl}\left(\Delta {E}\right) = 
\int_{0}^{\infty} 
d\omega_{k} 
\int_{0}^{\pi}\frac{\omega_{k}\sin{\theta}}{2(2\pi)^2}~ d\theta ~
\alpha_{j}\alpha_{l}
\nonumber\\
&&~\times \left[\frac{\mathcal{I}_{1_{3D}}(b_{j})~ 
\mathcal{I}^{*}_{1_{3D}}( 
b_{l})}{e^{\beta\omega_{k}}-1} \right.
+ \left. \frac{\mathcal{I}_{2_{3D}}( b_{j})~ 
\mathcal{I}^{*}_{2_{3D}}( b_{l})}{1-e^{-\beta\omega_{k}}}\right]~.
\end{eqnarray}
The integrals $\mathcal{I}_{1_{3D}}(b_{j})$ can be evaluated to be 
\begin{eqnarray}\label{eq:TranCoeff-int1}
\mathcal{I}_{1_{3D}}(b_{j}) &=& \int_{-\infty}^{\infty} d\tau_{1}~ 
e^{-i\tau_{j}\Delta {E}}~e^{i\omega_{k}(X_{j}\cos{\theta}+T_{j})}\nonumber\\ 
~&=&\frac{2~ e^{\frac{\pi\Delta {E}}{2b_{j}}}}{b_{j}\alpha_{j}} 
\left(\frac{\delta_{1}}{\delta_{1}}\right)^{\frac{i\Delta {E}}{2b_{j} 
}}\mathcal{K}\left[\tfrac{i\Delta {E}}{b_{j}},\tfrac{\omega_{k} 
\sqrt{\delta_{1} \delta_{2}}}{b_{j}} \right],\nonumber\\
\end{eqnarray}
and integrals $\mathcal{I}_{2_{3D}}(b_{j})$ as
\begin{eqnarray}\label{eq::TranCoeff-int2}
\mathcal{I}_{1_{3D}}(b_{j}) &=& \int_{-\infty}^{\infty} d\tau_{1}~ 
e^{-i\tau_{j}\Delta {E}}~e^{i\omega_{k}(X_{j}\cos{\theta}-T_{j})}\nonumber\\ 
~&=&\frac{2e^{-\frac{\pi\Delta {E}}{2b_{j}}}}{b_{j}\alpha_{j}} 
\left(\frac{\delta_{1}}{\delta_{2}}\right)^{-\frac{i\Delta {E}}{2b_{j} 
}}\mathcal{K}\left[\tfrac{i\Delta {E}}{b_{j}},\tfrac{\omega_{k} 
\sqrt{\delta_{1} \delta_{2}}}{b_{j}} \right],\nonumber\\
\end{eqnarray}
where, $\delta_{1}=1+\cos{\theta}$, $\delta_{2}=1-\cos{\theta}$, and 
$\mathcal{K}\left[n,z\right]$ denotes the modified Bessel function of the second 
kind of order $n$. Then one express the coefficient functions like before as 
$F_{jl}\left(\Delta {E}\right) = \int_{0}^{\infty} d\omega_{k}~ \mathcal{F}_{jl} 
\left(\Delta {E},\omega_{k}\right)$, where
\begin{eqnarray}\label{eq:TranCoeffTR-Fij3D-2}
&& \mathcal{F}_{jl}\left(\Delta {E},\omega_{k}\right) = 
\int_{0}^{\pi} \sin{\theta}~d\theta~\frac{\omega_{k}}{2\pi^2b_{j}b_{l}} 
~\mathcal{C}_{2}(\theta,b_{j},b_{l})\nonumber\\
&&~~~~~~~~~~~~~~~~~~~~ \left[\frac{e^{\frac{\pi}{2}\left(\frac{
\Delta {E}}{b_{j}}+\frac{\Delta {E}}{b_{l}
}\right)}}{e^{\beta\omega_{
k}}-1} \left(\tfrac{\delta_{1}}{\delta_{2}}\right)^{\frac{i\Delta {E}}{2
}\left(\frac{1}{b_{j}}-\frac{1}{b_{l}}\right)} \right.\nonumber\\
&&~~~~~~~~~~ + \left.
\frac{e^{-\frac{\pi}{2}\left(\frac{
\Delta {E}}{b_{j}}+\frac{\Delta {E}}{b_{l}
}\right)}}{1-e^{-\beta\omega_{
k}}}
\left(\tfrac{\delta_{1}}{\delta_{2}}\right)^{-\frac{i\Delta {E}}{2
}\left(\frac{1}{b_{j}}-\frac{1}{b_{l}}\right)}
\right].
\end{eqnarray}
and the quantity $\mathcal{C}_{2}(\theta,b_{j},b_{l})$ is given by
\begin{eqnarray}\label{eq:Expression-C2}
 \mathcal{C}_{2}(\theta,b_{j},b_{l}) &=&
\mathcal{K}\left[\tfrac{i\Delta {E}}{b_{j}},\tfrac{\omega_{k} \sqrt{\delta_{ 
1 } \delta_{2}}}{b_{j}} \right] \left(\mathcal{K}\left[\tfrac{i\Delta 
{E}}{b_{l}},\tfrac{\omega_{k} \sqrt{\delta_{1} \delta_{2}}}{b_{l}} 
\right]\right)^{*}.\nonumber\\
\end{eqnarray}
We found it to be suitable to perform this $\theta$ integration numerically and 
then to plot the resulting transition probabilities. However, for equal proper 
accelerations one can perform this $\theta$ integral analytically to express 
the transition coefficient $\mathcal{F}_{11} \left(\Delta {E}, 
\omega_{k}\right)$ as
\begin{eqnarray}\label{eq:Transition-coeff-TR-F11-3D}
&& \mathcal{F}_{11}\left(\Delta {E},\omega_{k}\right) =
\frac{\omega_{k}}{4\pi b_{1}^2} 
\left[\frac{1}{e^{\beta\omega_{k}}-1}\frac{1}{1-e^{\frac{-2\pi\Delta {E}}{
b_{1} }}} \right.\nonumber\\
&&~~~~ ~~~~~~~~~~ + \left. 
\frac{1}{1-e^{-\beta\omega_{k}}}\frac{1}{e^{\frac{2\pi\Delta {E}}{
b_{1} }}-1}\right]~\mathcal{C}_{3}(b_{1})~,
\end{eqnarray}
where the expression of $\mathcal{C}_{3}(b_{1})$ is given by
\begin{eqnarray}
&& \mathcal{C}_{3}(b_{1}) = \frac{b_{1}}{\Delta {E}}~ 
_2F_3\left(\tfrac{1}{2},1;\tfrac{3}{2},1-\tfrac{i \Delta {E}}{b_{1}},\tfrac{i 
\Delta {E}}{b_{1}}+1;\tfrac{\omega_{k}^2}{b_{1}^2}\right)\nonumber\\
~&& + \scalebox{0.9}{$ 2\cosh \left(\frac{\pi  \Delta 
{E}}{b_{1}}\right)~Im\left[ \Gamma 
\left(-\frac{2 i \Delta {E}}{b_{1}}-1\right) 
\left(\frac{\omega_{k}}{b_{1}}\right)^{\frac{2 i \Delta {E}}{b_{1}}}\right.$}
\nonumber\\
~&& \left.~_1F_2\left(\tfrac{i \Delta {E}}{b_{1}}+\tfrac{1}{2};\tfrac{i 
\Delta {E}}{b_{1}}+\tfrac{3}{2},\tfrac{2 i \Delta {E}}{b_{1}}+1;\tfrac{
\omega_{k}^2}{b_{1}^2}\right)\right]~.
\end{eqnarray}
Here $_pF_q\left(m;n;z\right)$ denotes the generalized hypergeometric function. 
Note again that the above expression is not symmetric under the interchange 
$\beta\leftrightarrow (2\pi)/b_1$ and as we mentioned earlier, it is due to our 
choice of mode which is Minkowski here. 

\begin{figure} [h]
\centering
 \includegraphics[width=0.8\linewidth]{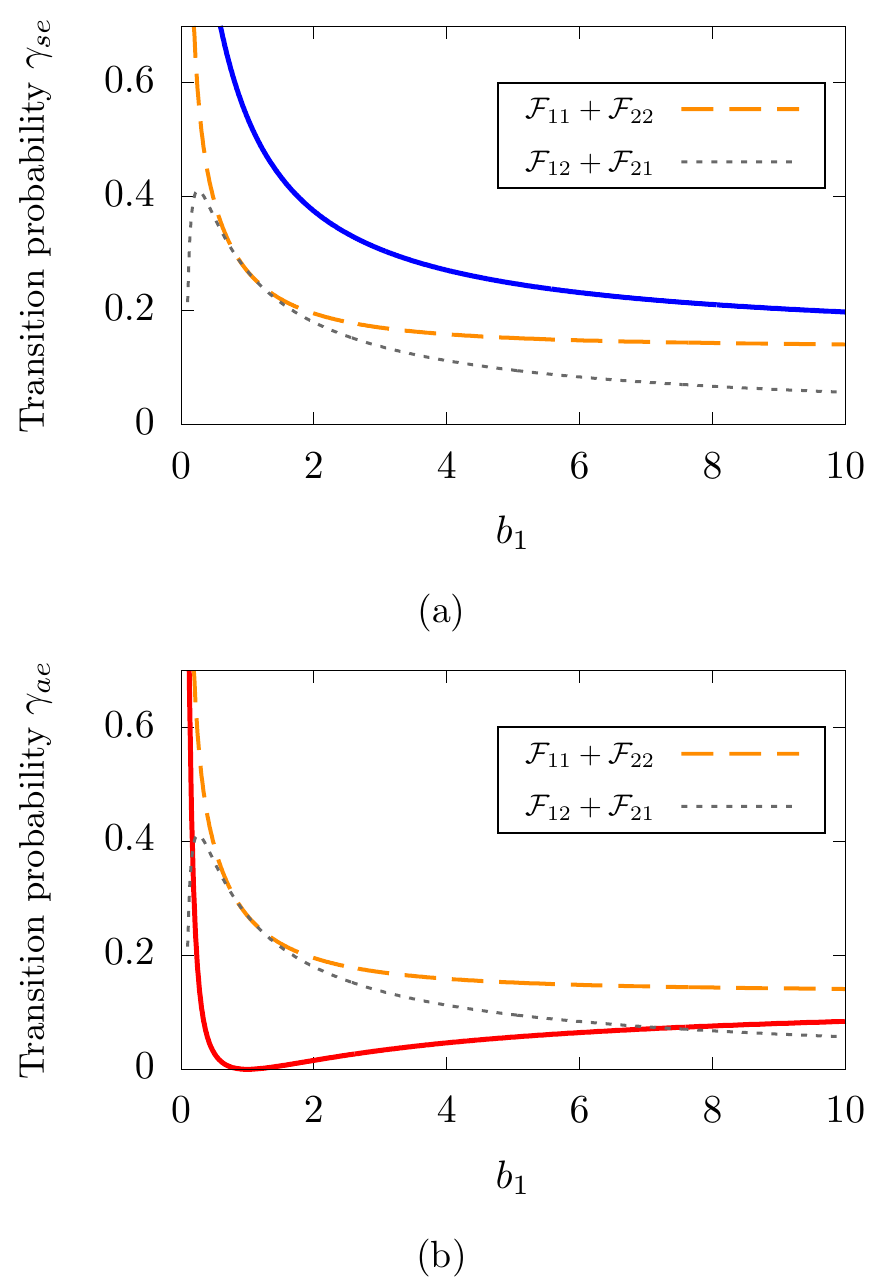}
 \caption{$(1+3)$ dimensions: (a) The transition probability from the symmetric 
state to the collective excited state denoted by the solid blue line. (b) The 
transition probability from the anti-symmetric state to the collective excited 
state denoted by the solid red line. In both of the cases $b_{2}=1$ is kept 
fixed and $b_{1}$ is varied. The orange dashed and gray dotted lines denote the 
contributions from $(\mathcal{F}_{11}+\mathcal{F}_{22})$ and 
$(\mathcal{F}_{12}+\mathcal{F}_{21})$ respectively. We have considered the 
values of the other parameters to be $\Delta {E}=0.1$, $\omega_{k}=0.1$ and 
$\beta=2\pi$.}
 \label{fig:TsgTR-P3D}
\end{figure}

One can get the transition probability for the transitions from the symmetric 
and anti-symmetric states to the excited and ground states in this case as well. 
The expressions will be given by (\ref{eq:Tran-prob-se}) and 
(\ref{eq:Tran-prob-ae}) again, but in this case $\mathcal{F}_{jl}$ are 
determined by (\ref{eq:TranCoeffTR-Fij3D-2}). The Transition probabilities are 
depicted in Fig. \ref{fig:TsgTR-P3D}. 
\noindent
Note again that the entanglement provides enhancement in transition for the 
symmetric case whereas decrement to that for anti-symmetric situation. Here also 
the anti-Unruh-like effects are evident and we shall properly discuss about it 
later in this paper. But there is a distinct difference in the features of 
transition coefficients compared to $(1+1)$ dimensional case. Here $\gamma_{se}$ 
shows anti-Unruh-like phenomenon for all values of $b_1$ while this is not true 
in $(1+1)$ dimensions in same parameter range. It should also be noted that all 
of the transition coefficients become equal to $\mathcal{F}_{11} \left(\Delta 
{E}, \omega_{k}\right)$ in the equal acceleration case $b_{2}=b_{1}$ and the 
transition probabilities from the anti-symmetric state to the collective excited 
or ground states become zero.

\section{Transition probability for accelerated atoms with the Rindler modes in 
a Thermal background}\label{sec:Transition-prob-Unruh}

The scalar field in terms of the Rindler modes in RRW are meant to resemble the 
thermal characteristics with respect to Minkowski vacuum for an accelerated observer. From the $(1+1)$ and $(1+3)$ 
dimensional field decompositions (\ref{eq:Field-Unruh}) and 
(\ref{eq:Field-Unruh-3D}), and the Green's functions (\ref{eq:Greens-fn-Unruh}), 
(\ref{eq:Greens-fn-TU}) and (\ref{eq:Greens-fn-Unruh-3D}), 
(\ref{eq:Greens-fn-TU-3D}) it can be observed that they are all represented in 
terms of the Rindler parameters $a_{j}$ and the Rindler coordinates $\eta_{j}$, 
$\xi_{j}$. To evaluate the transition coefficients using these Green's functions 
one has to move to the proper time and proper acceleration which will be quite 
convoluted in this case. On the other hand, one can approach this issue in a 
more straight forward manner, where the convenient way is to keep these 
quantities in terms of $a_{j}$, $\eta_{j}$, and $\xi_{j}$ as they are. Then 
consider the same system of two accelerated observers in a different manner so 
that the transformation from these coordinates to the proper time and 
acceleration follows easily. In this regard we consider the parameters are 
different $a_{1}\neq a_{2}$ corresponding to the two observers. The proper 
accelerations and proper times are given by $b_{j}=a_{j}e^{-a_{j}\xi_{j}}$ and 
$\tau_{j}=\eta_{j}e^{a_{j}\xi_{j}}$, respectively. Now as one considers 
$\xi_{j}=0$ the proper accelerations and proper times become $b_{j}=a_{j}$ and 
$\tau_{j}=\eta_{j}$. Then for observers with equal Rindler time, the proper 
times are also equal (a discussion on this has been presented in Appendix 
\ref{Apn:rel-proper-time}). We have seen from our previous analysis that the 
exact form of the proportionality constant in the relation between the proper 
times has no role in the transition probabilities. Therefore, it should not be 
absurd to take the previous consideration of equal proper time.

We are going to utilize these above mentioned considerations to obtain the 
transition probabilities of Eq. (\ref{eq:Transition-prob}) for two accelerated 
atoms immersed in a thermal bath considering the Rindler modes. Then one can use 
$\Delta\xi_{jl}=0$ and $\Delta\eta_{jl} = \Delta\tau_{jl} = 
\tau_{j,2}-\tau_{l,1}$ in the Green's function of Eq. (\ref{eq:Greens-fn-TU}) 
and (\ref{eq:Greens-fn-TU-3D}). It should be noted that these Green's functions 
here are time translation invariant. Then one may perform the integration in Eq. 
(\ref{eq:Transition-coeff}) by switching to the coordinates, 
$u_{jl}=\tau_{j,2}-\tau_{l,1}$ and $v_{jl}=\tau_{j,2}+\tau_{l,1}$. After 
dividing the transition coefficients by, $(\mu^{2} 
\int_{-\infty}^{\infty}dv_{jl})$ one can get the response functions, which 
signify the transition probabilities per unit time, as
\begin{equation}\label{eq:response-fn-TU}
 R_{jl}(\Delta {E}) = \int_{-\infty}^{\infty} 
du_{jl}~ e^{-iu_{jl} \Delta {E}}~ G^{+}_{\beta_{R}}(u_{jl})~.
\end{equation}
In the subsequent analysis we shall use the above one here to find the transition probabilities. 

\subsection{$(1+1)$ dimensions}

In $(1+1)$ dimensions using the Green's function from Eq. 
(\ref{eq:Greens-fn-TU}) we evaluate these response functions 
(\ref{eq:response-fn-TU}) and get
\begin{eqnarray}\label{eq:response-TU}
 && R_{jl}(\Delta {E}) = 
\int_{0}^{\infty}\frac{d\omega_{k}}{
4\omega_{k}\sqrt{\sinh{ \frac { \pi\omega_{k} } {a_{j}} } \sinh {
\frac{\pi\omega_{k}}{a_{l}}}}}\nonumber\\ 
&\times&\left[\delta(\omega_{k}-\Delta 
{E})\left\{\frac{e^{-\frac{\pi\omega_{k}}{2} \left(\frac { 1 } { a_ { j } }
+\frac
{ 1 } { a_{l} }
\right) } }{1-e^{-\beta\omega_{k} }} +  
\frac{e^{\frac{\pi\omega_{k}}{2}\left(\frac{1}{a_{j}}+\frac{1}{a_{l}}\right)}}
{ e^ { 
\beta\omega_{k}}-1}\right\}\right.\nonumber\\
&+& \left. \delta(\omega_{k}+\Delta 
{E})\left\{\frac{e^{\frac{\pi\omega_{k}}{2} \left(\frac { 1 } { a_ { j } } 
+\frac { 1 } { a_{l} } \right) } }{1-e^{-\beta\omega_{k} }} +  
\frac{e^{-\frac{\pi\omega_{k}}{2}\left(\frac{1}{a_{j}}+\frac{1}{a_{l}}\right)}} 
{ e^ { \beta\omega_{k}}-1}\right\} \right]~.\nonumber\\
\end{eqnarray}
Here $\delta(x-a)$ denotes the \emph{Dirac delta distribution}. In both of our 
considered transitions from the symmetric and anti-symmetric entangled states to 
the collective excited or the ground state the transition energy $\Delta {E}>0$. 
In the previous equation $\omega_{k}$ could only take positive values. Then from 
Eq. (\ref{eq:response-TU}) considering $\Delta {E}>0$ we get
\begin{eqnarray}\label{eq:response-TU-greater}
 && R_{jl}(\Delta {E}) = 
\frac{1}{
4\Delta {E}\sqrt{\sinh{ \frac { \pi\Delta {E} } {a_{j}} } \sinh {
\frac{\pi\Delta {E}}{a_{l}}}}}\nonumber\\ 
&&~~~~~~\times~\left[\frac{e^{-\frac{\pi\Delta {E}}{2}\left(\frac{1}{a_{j}}
+\frac
{ 1 } { a_{l} }
\right) } }{1-e^{-\beta\Delta {E} }} +  
\frac{e^{\frac{\pi\Delta {E}}{2}\left(\frac{1}{a_{j}}+\frac{1}{a_{l}}\right)}}
{ e^ { 
\beta\Delta {E}}-1}\right]~.
\end{eqnarray}
For $j=l=1$ we get the expression of the response function 
$R_{11}(\Delta {E})$ to be
\begin{eqnarray}\label{eq:response-TU-greater11}
 R_{11}(\Delta {E}) &=& \frac{1}{2\Delta {E}} 
\left[\frac{1}{e^{\beta\Delta {E}}-1}\frac{1}{1-e^{\frac{-2\pi\Delta {E} }{ 
a_{1} }}} \right.\nonumber\\ &&~~~~ +~~ \left. 
\frac{1}{1-e^{-\beta\Delta {E}}}\frac{1}{e^{\frac{2\pi\Delta {E}}{ a_{1} 
}}-1}\right]~.
\end{eqnarray}
Note that, contrary to Minkowski mode analysis, here $R_{11}$, which is regarded 
as the single detector's response function, is symmetric under the exchange 
$\beta\leftrightarrow (2\pi)/a_1$. This probably provides a justification in 
replacing thermal bath by a uniformly accelerated observer  with respect to 
Rindler mode (with Unruh operators), rather than Minkowski mode.

\begin{figure}[h]
\centering
 \includegraphics[width=0.8\linewidth]{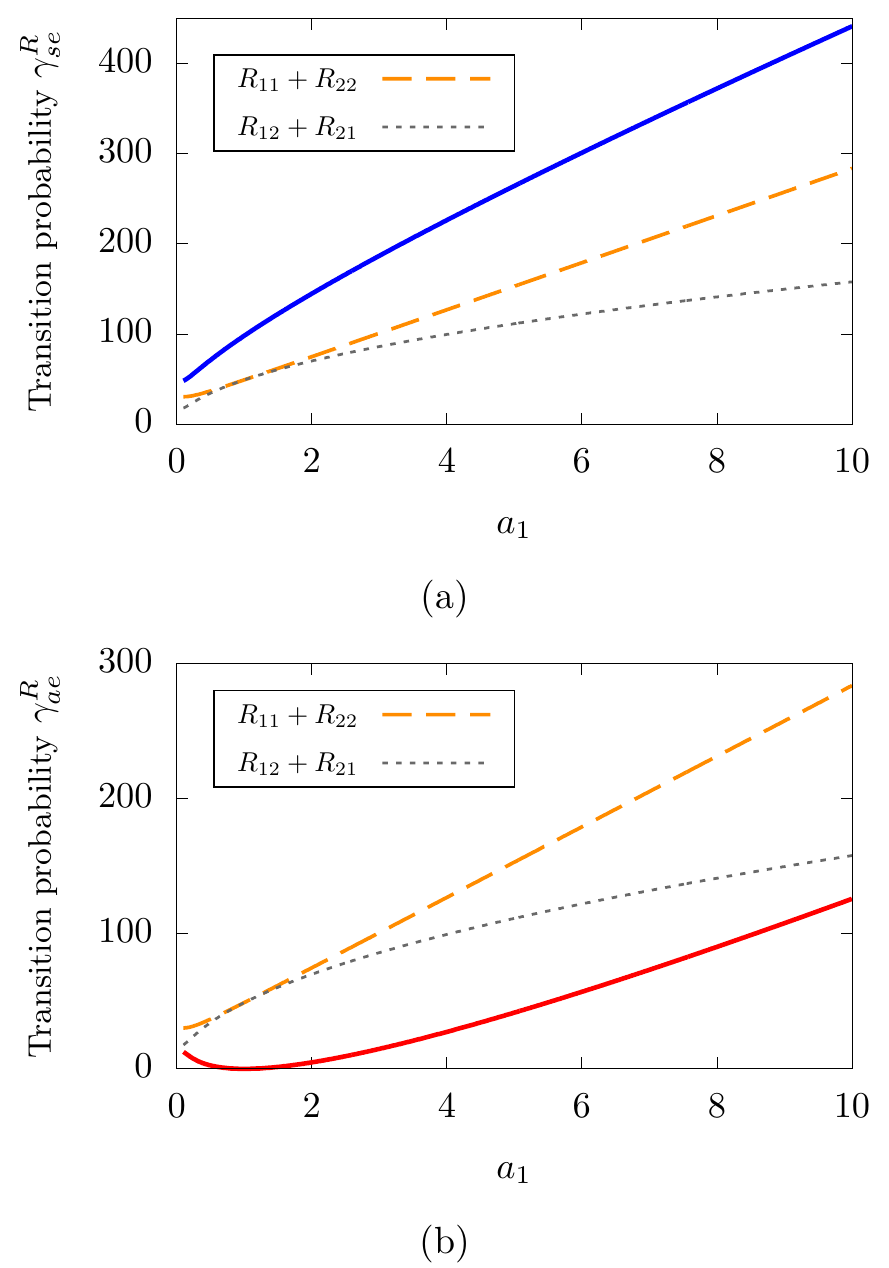}
 \caption{$(1+1)$ dimensions: (a) The transition probability from the symmetric 
state to the collective excited state for two observers with the Rindler modes, 
immersed in a thermal bath. The transition probability is denoted by solid blue 
line. (b) The transition probability from the anti-symmetric state to the 
collective excited state for two observers with the Rindler modes, immersed in 
a 
thermal bath, denoted by solid red line. In both of the cases $a_{2}=1$ is kept 
fixed and $a_{1}$ is varied. The other parameters are $\Delta {E}=0.1$, and 
$\beta=2\pi$. The orange dashed and gray dotted lines denote the contributions 
from $(R_{11}+R_{22})$ and $(R_{12}+R_{21})$ respectively.} \label{fig:Tsg-TU-1}
\end{figure}

With the help of these response functions one can obtain the transition 
probabilities per unit time $\gamma_{se}^{R}$ and $\gamma_{ae}^{R}$, 
between different atomic states in a similar manner as done in Eq. 
(\ref{eq:Tran-prob-se}) and (\ref{eq:Tran-prob-ae}), as
\begin{eqnarray}\label{eq:Tran-prob-sae-TU}
 \gamma_{se}^{R} &=&  
\left[\{R_{11}(\omega_{0},\omega_{k})+R_{22}(\omega_{0},
\omega_{k})\}
\right.\nonumber\\
~&& ~~\left.~ +~ \{R_{12}
(\omega_{0},\omega_{k} )+R_{21}(\omega_{0}),\omega_{k}\}\right]~,\nonumber\\
~\textup{and}&&\nonumber\\
 \gamma_{ae}^{R} &=& \left[\{R_{11}(\omega_{0},\omega_{k})+R_{22}(\omega_{0},
\omega_{k})\}
\right.\nonumber\\
~&& ~~\left.~ -~ \{R_{12}
(\omega_ { 0 },\omega_{k} )+R_ { 21 } (\omega_{0},\omega_{k}) 
\}\right]~.
\end{eqnarray}
These have been depicted in  Fig. \ref{fig:Tsg-TU-1}. 
\noindent
It should be noted that we 
have ignored the $1/2$ factor coming from the expectation value of the monopole 
moments, as it will not affect any qualitative prediction. It should also be 
mentioned that these transition probabilities (\ref{eq:Tran-prob-sae-TU}) are 
qualitatively different from the previous ones (\ref{eq:Tran-prob-se}) and 
(\ref{eq:Tran-prob-ae}). In the previous case the transition probability 
corresponded to certain mode frequency whereas here it corresponds to unit time.

From this figure, one can observe that the transition probability for the 
transition from the symmetric state to the collective excited state has no 
anti-Unruh effect for the considered values of the fixed parameters. For the 
transition from the anti-symmetric state to the collective excited state, there 
is a visible occurrence of the anti-Unruh effect for the same considered values 
of the fixed parameters.  We will again take up this issue in the next section.

\subsection{$(1+3)$ dimensions}

In a similar fashion as done in the $(1+1)$ dimensional case we consider the 
Green's function from Eq. (\ref{eq:Greens-fn-TU-3D}), which corresponds to two 
accelerated atoms in a thermal bath described with respect to the Rindler 
modes. 
Using this Green's function we evaluate the response functions of Eq. 
(\ref{eq:response-fn-TU}) and get
\begin{eqnarray}\label{eq:response-TU-3D}
 && R_{jl}(\Delta {E})~~ = 
\int_{0}^{\infty}d\omega~\int
\frac{ d^2\kp}{(2\pi)^3} \frac{2}{\sqrt{a_{j}a_{l}}}\nonumber\\ 
&& \times\left[\delta(\omega-\Delta 
{E})\left\{\frac{e^{-\frac{\pi\omega}{2} \left(\frac { 1 } { a_ { j } }
+\frac
{ 1 } { a_{l} }
\right) } }{1-e^{-\beta\omega }} +  
\frac{e^{\frac{\pi\omega}{2}\left(\frac{1}{a_{j}}+\frac{1}{a_{l}}\right)}}
{ e^ { 
\beta\omega}-1}\right\}\right.\nonumber\\
&& + \left. \delta(\omega+\Delta 
{E})\left\{\frac{e^{\frac{\pi\omega}{2} \left(\frac { 1 } { a_ { j } } 
+\frac { 1 } { a_{l} } \right) } }{1-e^{-\beta\omega }} +  
\frac{e^{-\frac{\pi\omega}{2}\left(\frac{1}{a_{j}}+\frac{1}{a_{l}}\right)}} 
{ e^ { \beta\omega}-1}\right\} \right]\nonumber\\
~&&~~~~~~~~~~~~~~~\mathcal{K}\left[\frac{i\omega}{a_{j}},\frac{|\kp|}{a_{j}}
\right]\mathcal{K}\left[\frac{i\omega}{a_{l}},\frac{|\kp|}{a_{l}}\right]~,
\end{eqnarray}
where, we have considered $\xi_{j}=0$ for both of the observers. Here 
$\delta(x-a)$ denotes the \emph{Dirac delta distribution}. From this Eq. 
(\ref{eq:response-TU-3D}) considering $\Delta {E}>0$ we get
\begin{eqnarray}\label{eq:response-TU-greater-3D}
  R_{jl}(\Delta {E}) = 
\tfrac{2}{(2\pi)^2\sqrt{a_{j}a_{l}}}\left[\tfrac{e^{-\frac{\pi\Delta 
{E}}{2}\left(\frac{1}{a_{j}}
+\frac{1}{a_{l}}
\right)}}{1-e^{-\beta\Delta {E} }} + 
\tfrac{e^{\frac{\pi\Delta {E}}{2}\left(\frac{1}{a_{j}}+\frac{1}{a_{l}}\right)}}
{ e^{\beta\Delta {E}}-1}\right] &&\nonumber\\
~~~~~~~~\times\int_{0}^{\infty} 
\kp d\kp~\mathcal{K}\left[\tfrac{i\Delta 
{E}}{a_{j}},\tfrac{|\kp|}{a_{j}}\right]\mathcal{K}\left[\tfrac{i\Delta 
{E}}{a_{l}},\tfrac{|\kp|}{a_{l}}\right].~~~&&
\end{eqnarray}
The expression of the response function $R_{11}(\Delta {E})$ is obtained from 
this equation with $j=l=1$ as
\begin{eqnarray}\label{eq:response-TU-greater11-3D}
 R_{11}(\Delta {E}) &=& \frac{\Delta {E}}{2\pi} 
\left[\frac{1}{e^{\beta\Delta {E}}-1}\frac{1}{1-e^{\frac{-2\pi\Delta {E} }{ 
a_{1} }}} \right.\nonumber\\ &&~~~~ +~~ \left. 
\frac{1}{1-e^{-\beta\Delta {E}}}\frac{1}{e^{\frac{2\pi\Delta {E}}{ a_{1} 
}}-1}\right]~,
\end{eqnarray}
which signifies the contribution of a single detector accelerated in a thermal 
bath described in terms of the Rindler modes. This also exhibits the 
$\beta\leftrightarrow (2\pi)/a_1$ symmetry.

\begin{figure}[h]
	\centering
	\includegraphics[width=0.8\linewidth]{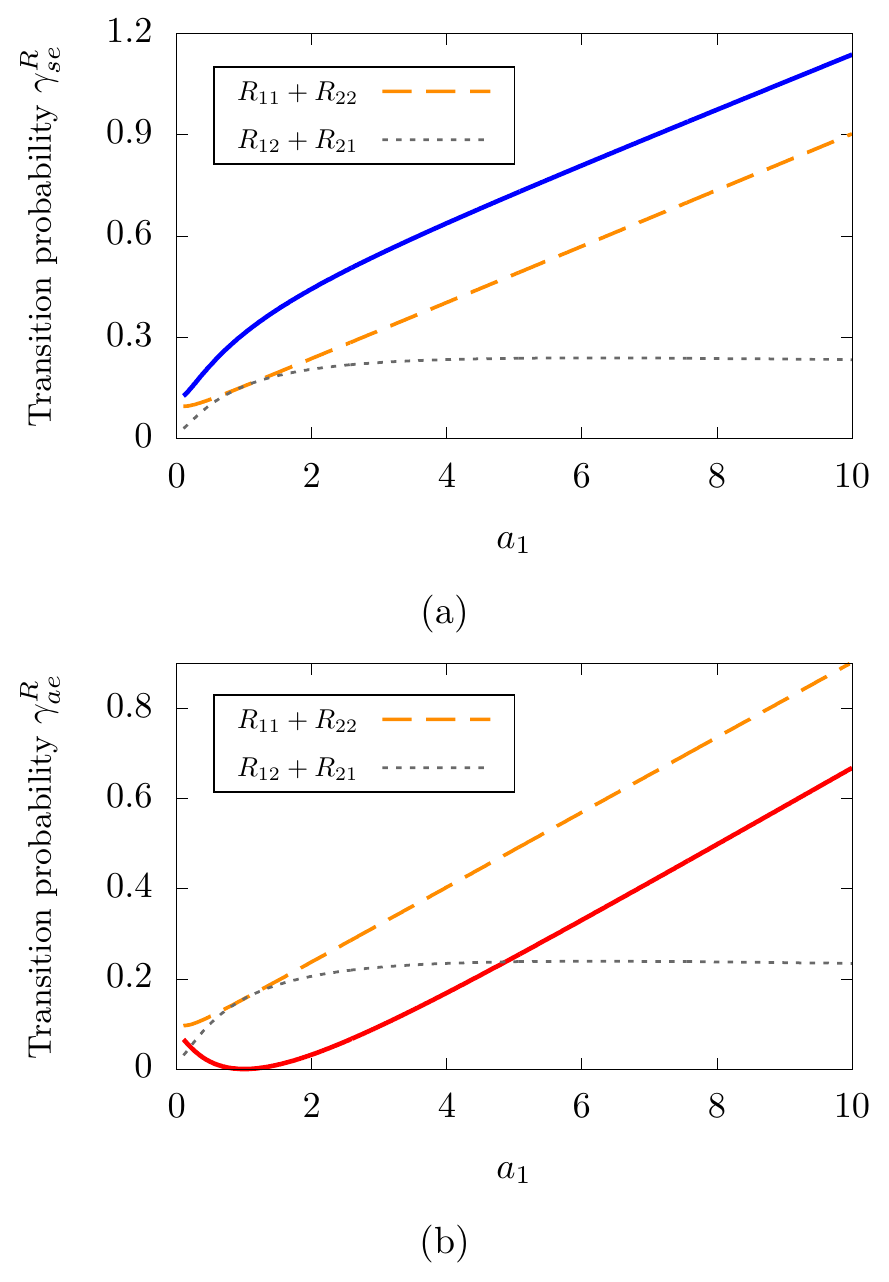} 
	\caption{$(1+3)$ dimensions: (a) The transition probability from the 
symmetric state to the collective excited state denoted by the solid blue line. 
(b) The transition probability from the anti-symmetric state to the collective 
excited state denoted by the solid red line. In both of the cases the Rindler 
modes are considered. In both of the cases $a_{2}=1$ is kept fixed and $a_{1}$ 
is varied. The value of the other parameters are $\Delta {E}=0.1$, and 
$\beta=2\pi$.} \label{fig:TsgTU-P3D}
\end{figure}

With the help of the response functions of Eq. (\ref{eq:response-TU-greater-3D}) 
and using  Eq. (\ref{eq:Tran-prob-sae-TU}), one can obtain the transition 
probabilities per unit time $\gamma_{se}^{R}$ and $\gamma_{ae}^{R}$, between 
different atomic states in $(1+3)$ dimensions. These have been depicted in  Fig. 
\ref{fig:TsgTU-P3D}.
\noindent
Note that the features of the transition probabilities are identical to $(1+1)$ 
dimensional case. For both dimensions, $\gamma^{R}_{se}$ shows no anti-Unruh 
effect while $\gamma^{R}_{ae}$ contains anti-Unruh phenomenon. This similarity 
was not there in Minkowski mode analysis. Furthermore, the expressions of the 
response functions (\ref{eq:response-TU-greater11}) and 
(\ref{eq:response-TU-greater11-3D}) corresponding to a single detector with 
Rindler modes correctly provides those for an accelerated 
detector \cite{book:Birrell} in a zero temperature (i.e., $\beta\to\infty$ 
limit) Minkowski background. Whereas that is not apparent in the case with the 
Minkowski modes (see Eq. (\ref{eq:Transition-coeff-TR-F11-4}) and 
(\ref{eq:Transition-coeff-TR-F11-3D})). Such 
consistency probably indicates a preference of choosing Rindler modes (with 
Unruh operators) over Minkowski ones in mimicking thermal behaviour by 
accelerated observer. But to be concrete, further investigations are needed. 

\section{Detailed investigation of observed anti-Unruh(-like) 
phenomenon}\label{sec:Study-Anti-Unruh}

In our previous discussions we observed that several transition probabilities 
are decreasing with the increase of acceleration of the observer within a 
particular range. It is usually, as observed earlier in \cite{Brenna:2015fga, 
Garay:2016cpf} for a different situation, called as anti-Unruh effect. In this 
section, the observed anti-Unruh(-like) effect for our model in the previous 
sections, will be further investigated in the light of required conditions for 
the same. In particular, we observed that for accelerated detectors in thermal 
bath considering the Minkowski modes there are anti-Unruh-like effects in 
$(1+1)$ and $(1+3)$ dimensions for both transitions from the symmetric and 
anti-symmetric states to the collective excited state. On the other hand, 
considering the same setup in terms of the Rindler modes with Unruh operators in 
$(1+1)$ and $(1+3)$ dimensions we observed that there is no anti-Unruh effect 
for the transition from the symmetric state to the collective excited state for 
the similar set of fixed parameters as the Minkowski modes. However, we observed 
that there is anti-Unruh effect for the transition from the anti-symmetric state 
to the collective excited state for the same values of the fixed parameters. 
Below we study them thoroughly and  balustrade these facts by verifying the 
consistency with the required anti-Unruh conditions.

\subsection{Anti-Unruh effect: the conditions}
Let us first briefly summarise the conditions of anti-Unruh effect.
In article \cite{Brenna:2015fga} it was first shown that for short times 
the transition probability of an accelerated particle detector decreases with 
increasing acceleration, a phenomena better known as the anti-Unruh effect from 
then. In subsequent article \cite{Garay:2016cpf} by the same authors the 
existence of the anti-Unruh effect for infinite time was also confirmed. The 
statement of the anti-Unruh effect goes like, ``\emph{a uniformly accelerated 
particle detector coupled to the vacuum can cool down as its acceleration 
increases.}" To mathematically realize the existence of the anti-Unruh effect, 
done in \cite{Garay:2016cpf}, there are two particular conditions which have 
to be satisfied. These conditions are first mentioned below 
corresponding to our system of analysis.\vspace{0.1cm}

\subsubsection{Weak anti-Unruh effect} 
In our case the weak anti-Unruh effect is defined by the condition when the 
transition coefficients $(\mathcal{F}_{jl})$, transition probabilities 
$(\gamma_{\omega\Omega})$ or response functions $(R_{jl})$ decrease with 
increasing acceleration of the atoms with all other parameters of the system 
fixed \cite{Garay:2016cpf}, i.e.,
\begin{eqnarray} 
\partial_{b_{1}}\mathcal{F}_{jl}<0~; \,\,\
\partial_{b_{1}}\gamma_{\omega\Omega}<0~; \,\,\,\
\partial_{a_{1}}R_{jl}<0~.
\label{eq:Anti-Unruh-weak}
\end{eqnarray}
Note that here we have also considered taking the differentiation of the 
transition coefficients and response functions rather than only the transition 
probabilities, because these particular coefficients $\mathcal{F}_{11}$ and 
$R_{11}$ signify the transition probability of the single accelerated 
detector in thermal background. We also specify that the differentiation is 
taken with respect to the proper acceleration of the first detector as we have 
performed all of our calculations with respect to this particular 
frame.

\subsubsection{Strong anti-Unruh effect}
 To talk about this condition a definition of excitation to de-excitation ratio 
(EDR) for the transition coefficients, transition probabilities and response 
functions is needed. The EDR corresponding to the transition coefficient 
$\mathcal{F}_{jl}(\Delta {E})$, the transition probability 
$\gamma_{\Omega\omega}(\Delta {E})$ and the response function 
$\mathcal{R}_{jl}(\Delta {E})$ are defined as \cite{Garay:2016cpf}
\begin{eqnarray}
 \mathcal{R}_{\mathcal{F}}(\Delta {E})=\frac{\mathcal{F}_{jl} (\Delta {E} 
)}{\mathcal{F}_{jl}(-\Delta {E})}~;\nonumber\\
 \mathcal{R}_{\gamma}(\Delta {E})=\frac{\gamma_{_{\Omega\omega}} (\Delta {E} 
)}{\gamma_{_{\Omega\omega}}(-\Delta {E})}~;\nonumber\\
 \mathcal{R}_{R}(\Delta {E})=\frac{R_{jl} (\Delta {E} 
)}{R_{jl}(-\Delta {E})}~.
\end{eqnarray}
The corresponding EDR inverse temperature can be defined as \cite{Garay:2016cpf}
\begin{equation}
 \mathcal{B}_{_{EDR}}=-\frac{1}{\Delta {E}}\ln{(\mathcal{R})}~.
\end{equation}
The condition for strong anti-Unruh effect is met when the EDR temperature  
decreases with increasing detector acceleration. The mathematical representation 
of the strong anti-Unruh effect is characterized by the condition of 
\begin{eqnarray}
\partial_{b_{1}}\mathcal{B}_{_{EDR}}(\Delta {E},b_{2},\omega_{k})>0~.
\label{eq:Anti-Unruh-strong}
\end{eqnarray}

Below we shall check whether our observed phenomena are consistent with these 
mentioned conditions. This will not only provide a verification of our aforesaid 
claim, but also provide a classification of the anti-Unruh(-like) phenomenon. 
Before proceeding further, it may be noted that satisfaction of strong 
anti-Unruh condition implies automatic satisfaction of weak condition; while the 
reverse is may not be true (see \cite{Garay:2016cpf} for details). The strong 
anti-Unruh effect always refers to the occurrence of the weak anti-Unruh effect 
unless the conditions
\begin{eqnarray}
  \partial_{b_{1}}\mathbb{F}(-\Delta {E})>0~,~~
  \textup{and}~
  \partial_{b_{1}}\mathbb{F}(\Delta {E})>0~;
  \label{eq:Anti-Unruh-strongToweak}
\end{eqnarray}
are satisfied simultaneously, see \cite{Garay:2016cpf}. Here $\mathbb{F}(\Delta 
{E})$ can be considered to be any of the $\mathcal{F}_{jl}(\Delta {E})$, 
$\gamma_{\omega\Omega}(\Delta {E})$ or $R_{jl}(\Delta {E})$, while for 
$R_{jl}(\Delta {E})$ the derivative is taken with respect to $a_{1}$. The 
occurrence of strong anti-Unruh effect, when one of these conditions or both of 
them are violated signifies the definite satisfaction of the weak condition. On 
the other hand, when both of these conditions (\ref{eq:Anti-Unruh-strongToweak}) 
are simultaneously satisfied, the satisfaction of the strong condition will not 
imply the satisfaction of the weak condition and in that case one cannot comment 
about the nature of this phenomena. We shall check both the conditions from Eq. 
(\ref{eq:Anti-Unruh-weak}) and (\ref{eq:Anti-Unruh-strong}), and accordingly 
categorise the phenomenon as either weak or strong anti-Unruh effect.

\subsection{Case I: Minkowski mode}\label{eq:AU-Mink}
\subsubsection{$(1+1)$-dimensions}

We have seen from our analysis that in the system of two entangled atoms 
accelerated in a thermal background, considering the $(1+1)$ dimensions and 
Minkowski modes, as one estimates the transition probability between different 
states one first gets a decreasing probability with increasing proper 
acceleration $b_{1}$, signifying anti-Unruh-like effect. Then one gets an 
increasing transition probability with increasing acceleration $b_{1}$ 
signifying the Unruh-like effect. These phenomena can be understood from Fig. 
\ref{fig:TsgTR-P1}. Here we are going to check whether the transition 
coefficients and transition probabilities satisfies the weak and strong 
conditions of the so called anti-Unruh effect. After analyzing the results 
we shall like to predict some possible source of origin behind it. 

\begin{figure}[h]
\centering
 \includegraphics[width=0.8\linewidth]{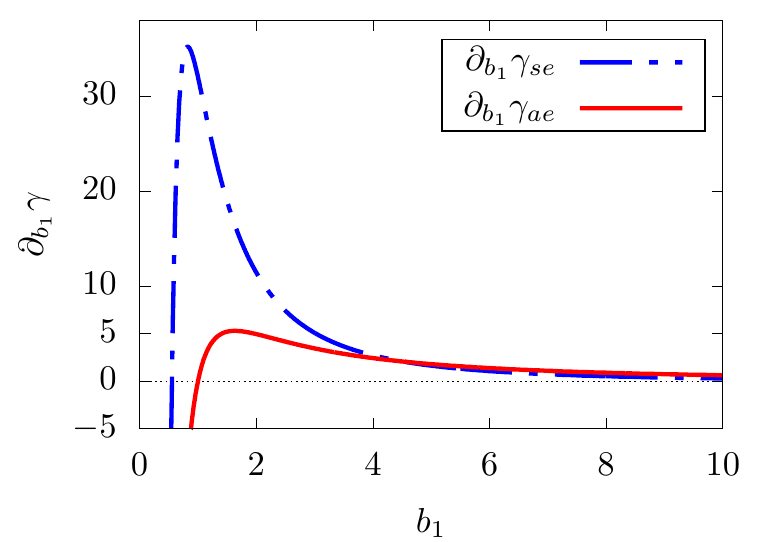}
 \caption{ The quantity $\partial_{b_{1}}\gamma_{se}$ is plotted with respect to 
varying $b_{1}$ for accelerated atoms in thermal background in $(1+1)$ 
dimensions, depicted by dash-dotted blue line. The quantity 
$\partial_{b_{1}}\gamma_{ae}$ is plotted with respect to varying $b_{1}$ for 
accelerated atoms in thermal background in $(1+1)$ dimensions, depicted by red 
line.}
 \label{fig:Anti-Unruh-TsaeW}
\end{figure}

First, to check the weak condition for certain values of the parameters 
$\omega_{k}=0.1$, $\Delta {E}=0.1$, and $b_{2}=1$ we plot 
$\partial_{b_1}\gamma_{\omega\Omega}$ VS $b_1$ in Fig. 
\ref{fig:Anti-Unruh-TsaeW} corresponding to transitions from the symmetric and 
anti-symmetric states to the collective excited state (here we take the same 
values of parameters as taken in Fig. \ref{fig:TsgTR-P1} in order to have a 
proper comparison).
\noindent
These plots are meant to provide confirmation in support of the existence of the 
weak anti-Unruh effect when the functions have negative values. We observe 
that this is the case in the lower regimes of the proper acceleration $b_{1}$ -- 
in the case of $\gamma_{se}$ for $b_{1}$ less than $0.5$ and in the case of 
$\gamma_{ae}$ for $b_{1}$ less than $1$. Note that these are the values of $b_1$ 
up to which the transition probabilities were decreasing (see Fig. 
\ref{fig:TsgTR-P1}).

\begin{figure}[h]
\centering
 \includegraphics[width=0.8\linewidth]{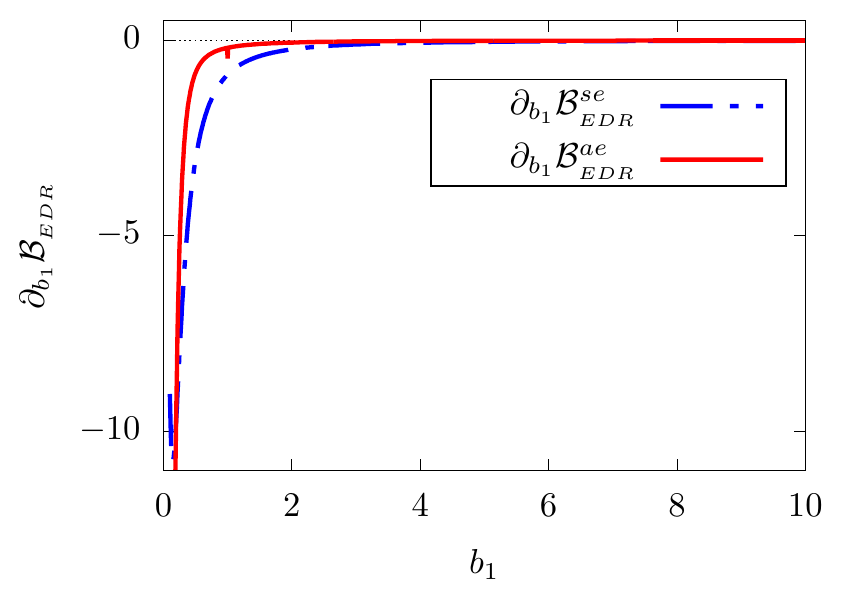}
 \caption{ The quantity $\partial_{b_{1}}\beta_{\gamma_{se}}$ is plotted with 
respect to varying $b_{1}$, depicted by dash-dotted blue line. The quantity 
$\partial_{b_{1}}\beta_{\gamma_{ae}}$ is plotted with respect to varying 
$b_{1}$, depicted by red line.} 
\label{fig:Anti-Unruh-TsaeS}
\end{figure}

Now to check whether this is complied with strong condition, in Fig. 
\ref{fig:Anti-Unruh-TsaeS} 
%
\noindent
we have plotted the differentiation of EDR inverse temperature with respect to 
$b_{1}$ for the same two transitions. Note that for both the cases 
$\partial_{b_1}\mathcal{B}_{_{EDR}}$ are always negative throughout the range 
of $b_1$ and so it does not satisfy the strong condition. Hence here we have 
weak anti-Unruh effect.

\begin{figure}[h]
\centering
 \includegraphics[width=0.75\linewidth]{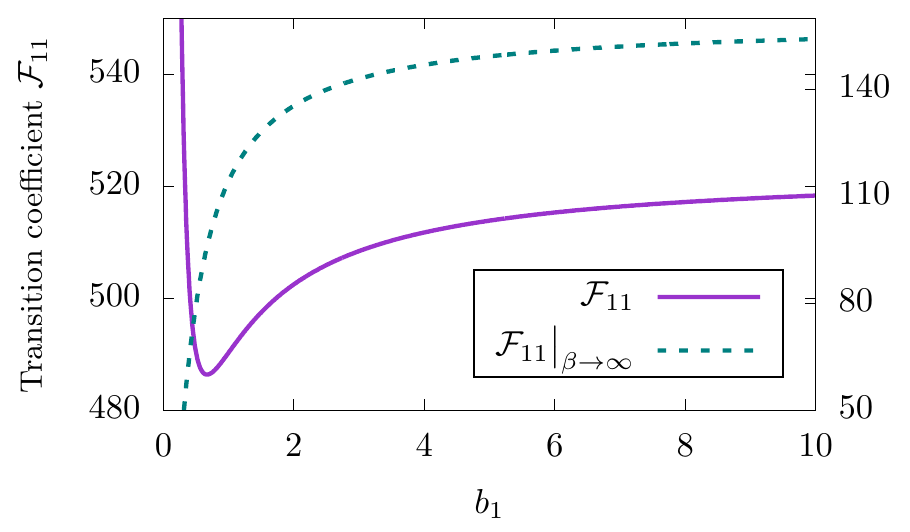}
 \caption{ The transition coefficient $\mathcal{F}_{11}(\Delta {E})$ with 
respect to varying $b_{1}$ for accelerated atoms in thermal background 
considering Minkowski modes. The solid violet line with the left vertical axis 
denotes this particular case. The transition coefficient 
$\mathcal{F}_{11}(\Delta {E})$ in the same scenario with the temperature of the 
thermal bath now zero denoted by the dotted green line with right vertical 
axis.}
 \label{fig:F11Dw}
\end{figure}

Now the question arises what is the origin of this anti-Unruh-like phenomena in 
these transition probabilities $\gamma_{se}$ and $\gamma_{ae}$? Is it happening 
solely because of the entanglement between the atoms? In this regard, we want to 
mention that this phenomena is not only visualized in $\gamma_{se}$ and 
$\gamma_{ae}$, but also in the transition coefficient $\mathcal{F}_{11}(\Delta 
{E})$, Fig. \ref{fig:F11Dw}. 
\noindent
It is to be noted that the transition coefficient $\mathcal{F}_{11}(\Delta {E})$ 
signifies the situation when a single detector is accelerated in a thermal bath. 
In that case the entanglement do not play any role into the picture and one can 
assert that the source of the anti-Unruh-like effect is not from entanglement, 
at least in this case. Then the attention is bound to be shifted towards the 
effects of the thermal bath as a possible origin of the anti-Unruh-like 
phenomena. In this regard, in the same Fig. \ref{fig:F11Dw} we have also plotted 
the $\mathcal{F}_{11}(\Delta {E})$ in the limit of $\beta\to\infty$ or for zero 
temperature of the thermal bath. Interestingly it shows no anti-Unruh-like 
effect, which suggests one reason behind the anti-Unruh-like effect to be the 
non zero temperature of the thermal bath in which the atoms are accelerating.

\begin{figure}[h]
\centering
 \includegraphics[width=0.8\linewidth]{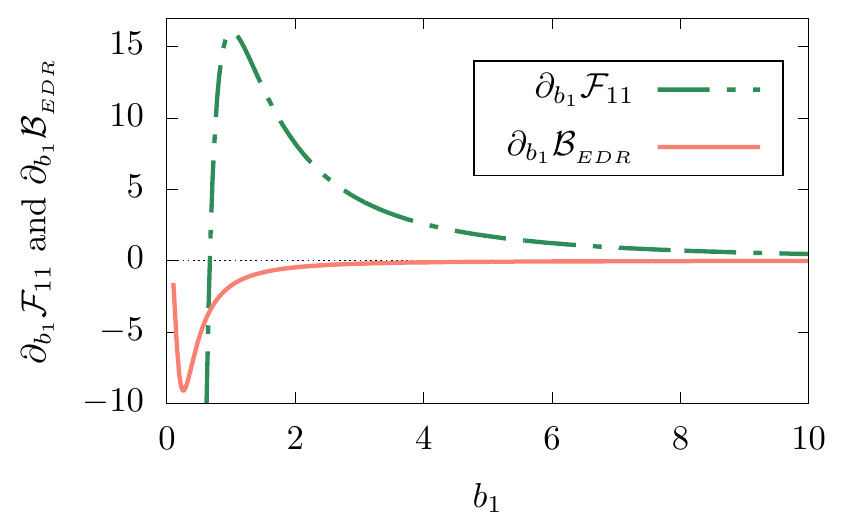}
 \caption{ Differentiation of the transition coefficient 
$\mathcal{F}_{11}(\Delta {E})$ with respect to $b_{1}$ plotted against varying 
$b_{1}$ for accelerated atoms in thermal background, denoted by the dash-dotted 
green line. The negative value of this quantity 
$\partial_{b_{1}}\mathcal{F}_{11}(\Delta {E})$ gives the condition for weak 
anti-Unruh effect. Here the quantity $\partial_{b_{1}} \beta_{\mathcal{F}_{11}}$ 
is also plotted with respect to varying $b_{1}$ and it is denoted by the solid 
violet line. The positive value of this quantity gives the strong condition for 
the anti-Unruh-like effect.}
 \label{fig:Anti-Unruh-F11}
\end{figure}

Subsequently, we have studied the nature of the anti-Unruh-like phenomenon 
arising in $\mathcal{F}_{11}(\Delta {E})$. We can observe that the transition 
coefficient $\mathcal{F}_{11}(\Delta {E})$ satisfies the condition for the weak 
anti-Unruh effect, but do not agree with the condition for the strong 
anti-Unruh effect (see Fig. \ref{fig:Anti-Unruh-F11}).
\noindent

\subsubsection{$(1+3)$-dimensions}
\begin{figure}[h]
\centering
 \includegraphics[width=0.8\linewidth]{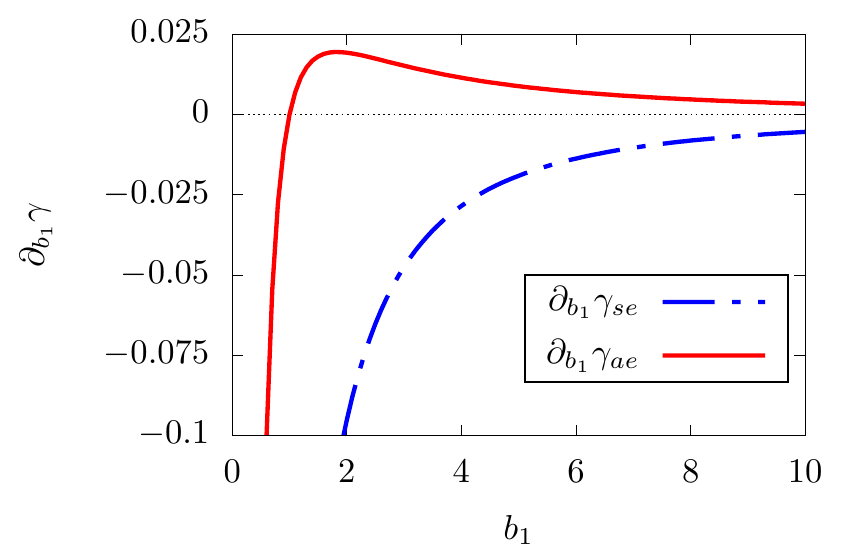}
 \caption{ The quantity $\partial_{b_{1}}\gamma_{se}$ is plotted with respect to 
varying $b_{1}$ for accelerated atoms in thermal background in $(1+3)$ 
dimensions, depicted by dash-dotted blue line. The quantity 
$\partial_{b_{1}}\gamma_{ae}$ is plotted with respect to varying $b_{1}$ for 
accelerated atoms in thermal background in $(1+3)$ dimensions, depicted by solid 
red line.}
 \label{fig:Anti-Unruh-Tsae-3DW}
\end{figure}

Next we consider the case for entangled atoms accelerated in a thermal bath as 
seen with respect to the Minkowski modes in a $(1+3)$ dimensional spacetime. For 
the same set of values of the parameters $\omega_{k}=0.1$, $\Delta {E}=0.1$, and 
$b_{2}=1$ as taken in Fig. \ref{fig:TsgTR-P3D} we have plotted, see Fig. 
\ref{fig:Anti-Unruh-Tsae-3DW},
\noindent
the $\partial_{b_{1}}\gamma_{se}$ and $\partial_{b_{1}}\gamma_{ae}$ with respect 
to $b_{1}$, which respectively correspond to transitions from the symmetric and 
anti-symmetric states to the collective excited state. For symmetric case, 
$\partial_{b_1} \gamma_{se}$ is negative for all selected values of $b_1$ while 
for other case it is negative till  $b_1=1$. These exactly comply with Fig. 
\ref{fig:TsgTR-P3D} and thereby provide confirmation in support of the 
satisfaction of the weak anti-Unruh condition. 

\begin{figure}[h]
\centering
 \includegraphics[width=0.8\linewidth]{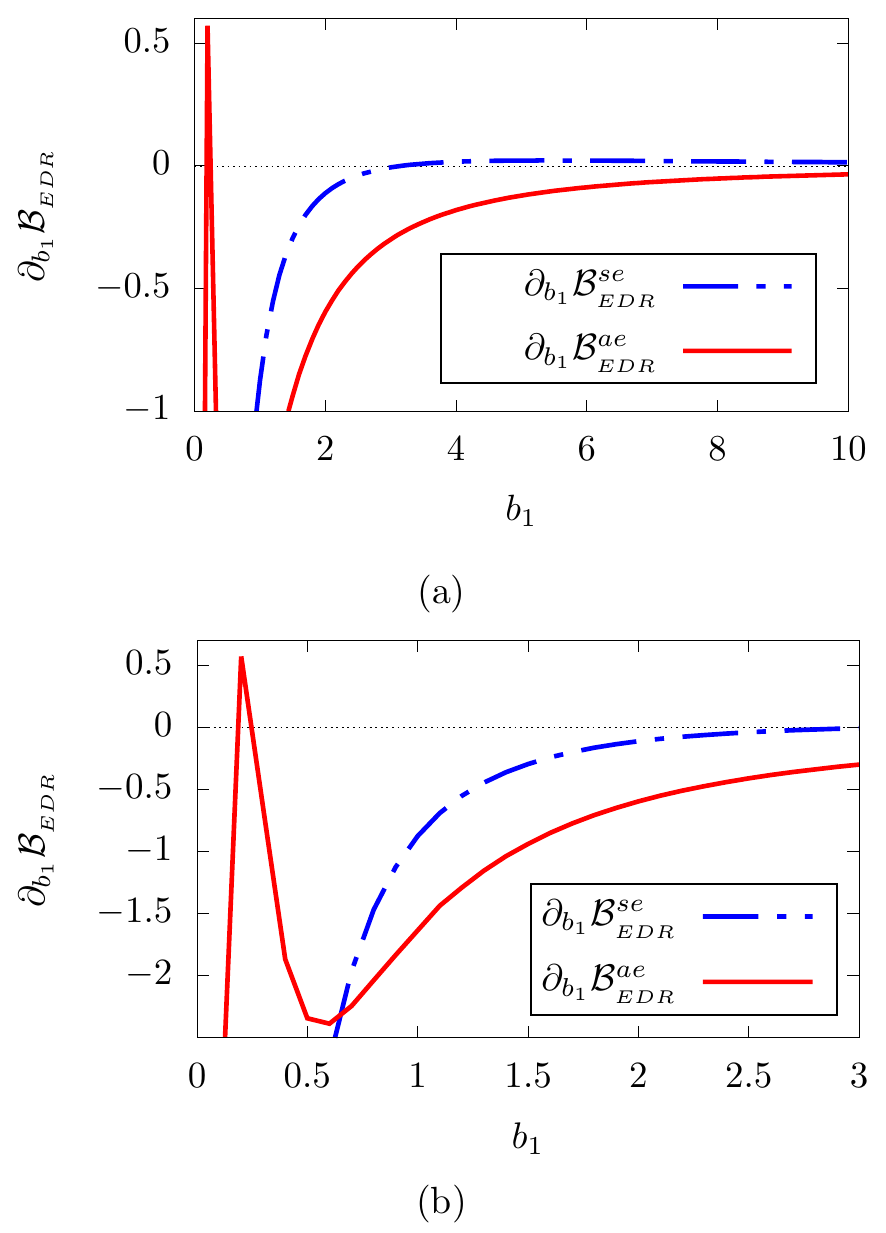}
 \caption{ (a) The quantity $\partial_{b_{1}}\mathcal{B}_{_{EDR}}^{se}$ is 
plotted with respect to varying $b_{1}$ for accelerated atoms in thermal 
background in $(1+3)$ dimensions, depicted by dash-dotted blue line. The 
quantity $\partial_{b_{1}}\mathcal{B}_{_{EDR}}^{ae}$ is plotted with respect to 
varying $b_{1}$ for accelerated atoms in thermal background in $(1+3)$ 
dimensions, depicted by solid red line. (b) The same plot as depicted in the 
previous sub-figure with the $x-$ range in the initial region now emphasized.}
 \label{fig:Anti-Unruh-Tsae-3DS}
\end{figure}

In Fig. \ref{fig:Anti-Unruh-Tsae-3DS} we have plotted the differentiation of EDR 
inverse temperature with respect to $b_{1}$ for the same two transitions, which 
provide the condition for the occurrence of the strong anti-Unruh effect for 
positive values.
We observed that for the transition from the symmetric entangled state to the 
collective excited state the quantity $\partial_{b_{1}} 
\mathcal{B}_{_{EDR}}^{se}$ has negative value up to around  $b_{1}=3$ and then it 
gets a small positive value and tends to decrease to zero for further increase 
in $b_{1}$. On the other hand, for the transition from the anti-symmetric 
entangled state to the collective excited state the quantity $\partial_{b_{1}} 
\mathcal{B}_{_{EDR}}^{ae}$ has positive value around  $b_{1}=0.25$ and then it 
becomes negative and remains so for further increase in $b_{1}$. These analysis 
suggest that while the condition for weak anti-Unruh effect is satisfied for a 
wide range of the parameter $b_{1}$, the condition for strong anti-Unruh effect 
is satisfied in a much smaller range residing inside that of the weak case. 
Therefore, for the parameter values of $b_{1}$ when strong anti-Unruh effect is 
satisfied the weak anti-Unruh effect is always satisfied, consistent with 
our previous assertions.

\begin{figure}[h]
\centering
 \includegraphics[width=0.8\linewidth]{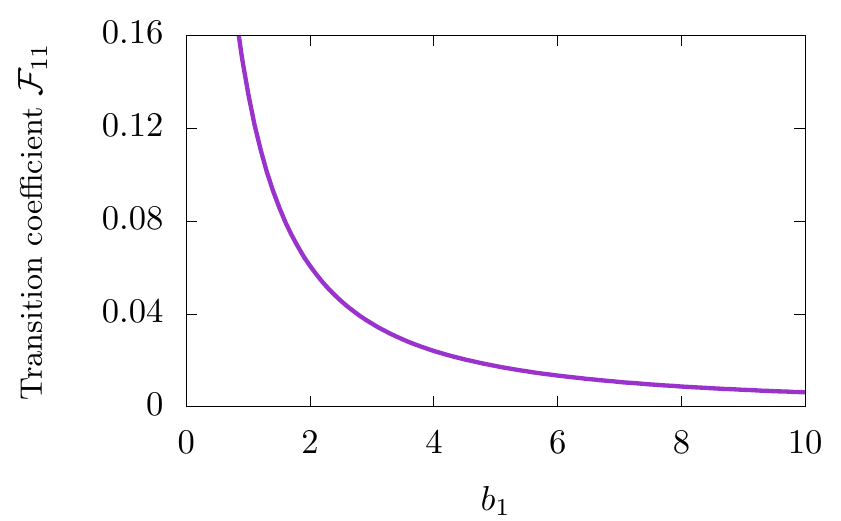}
 \caption{The transition coefficient $\mathcal{F}_{11}(\Delta {E})$, plotted 
with respect to varying $b_{1}$ for accelerated atoms in thermal background 
considering Minkowski modes and $(1+3)$ dimensions.}
 \label{fig:TR-F11-4D}
\end{figure}

\begin{figure}[h]
\centering
 \includegraphics[width=0.8\linewidth]{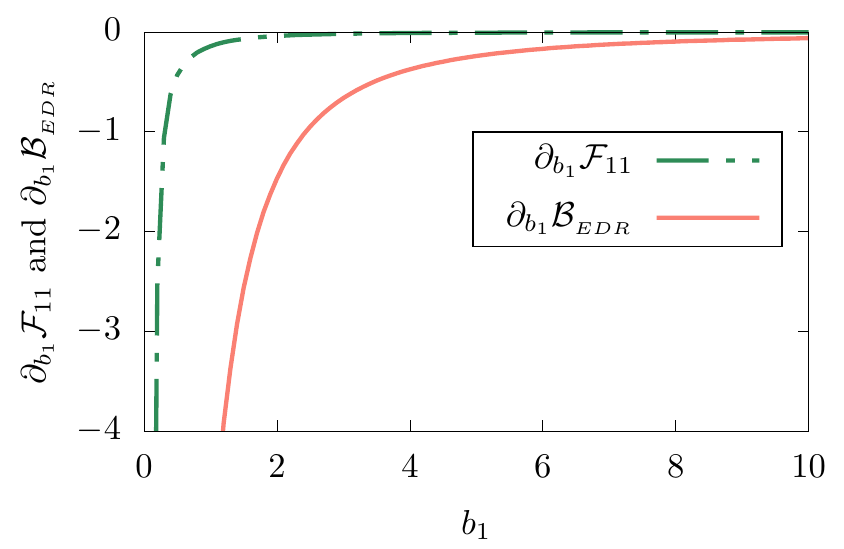}
 \caption{The derivative of the transition coefficient $\mathcal{F}_{11}(\Delta 
{E})$, plotted with respect to varying $b_{1}$ for accelerated atoms in thermal 
background considering Minkowski modes and $(1+3)$ dimensions, denoted by 
dash-dotted green line. In this figure the quantity $\partial_{b_{1}} 
\mathcal{B}_{_{EDR}}$ is also plotted, which is denoted by the solid line.}
 \label{fig:TR-F11-AU-4D}
\end{figure}

Here also like the previous $(1+1)$ dimensional case we have tried to understand 
the origin of the anti-Unruh-like effect. In this direction in Fig. 
\ref{fig:TR-F11-4D} we have plotted the transition coefficient 
$\mathcal{F}_{11}$ in $(1+3)$ dimensions considering the same sets of 
parameters. This particular transition coefficient signifies the situation if 
there were only one two-level atomic detector accelerating in the thermal bath. 
Like the $(1+1)$ dimensional case here also we have observed the anti-Unruh-like 
phenomenon, discarding any possibility of entanglement being the sole origin of 
this effect. Furthermore, in Fig. \ref{fig:TR-F11-AU-4D} we have studied whether 
the anti-Unruh-like effect for $\mathcal{F}_{11}$ is of weak or strong origin. 
From this figure we observed that in $(1+3)$ dimensions it satisfies the 
condition for weak anti-Unruh effect but not the strong one.

\subsection{Case II:  Rindler mode}\label{eq:AU-Unruh}

\subsubsection{$(1+1)$-dimensions}

\begin{figure}[h]
\centering
 \includegraphics[width=0.8\linewidth]{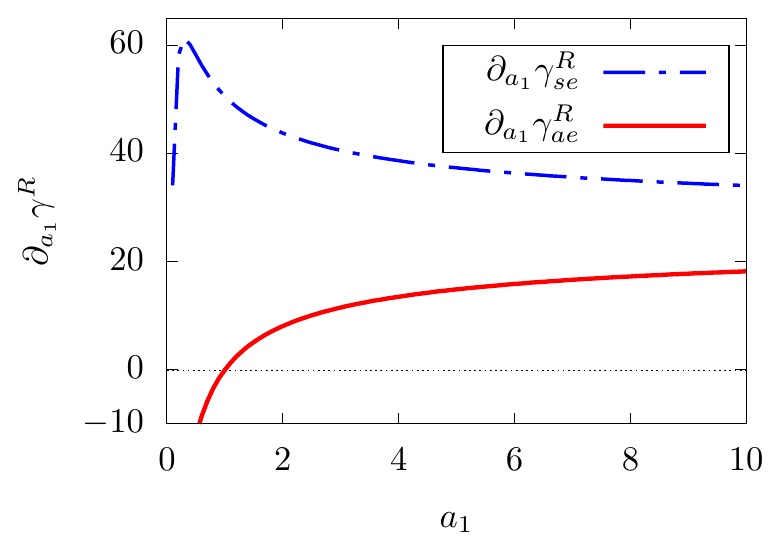} 
 \caption{ The quantity $\partial_{a_{1}}\gamma_{se}^{R}$ is plotted with 
respect to varying $a_{1}$ for accelerated atoms in thermal background 
considering Rindler modes, depicted by dash-dotted blue line. The quantity 
$\partial_{a_{1}}\gamma_{ae}^{R}$ is plotted with respect to varying $a_{1}$ for 
accelerated atoms in thermal background considering Rindler modes, depicted by 
solid red line.}
 \label{fig:Anti-Unruh-Tsae-TUW}
\end{figure}

To understand the anti-Unruh effect for the transition probabilities considering 
the Rindler modes, we have plotted $\partial_{a_{1}}\gamma_{se}^{R}$ and 
$\partial_{a_{1}}\gamma_{ae}^{R}$ with respect to varying $a_{1}$ in figure 
Fig. \ref{fig:Anti-Unruh-Tsae-TUW}. In obtaining this figure we have kept the 
other parameters fixed $\Delta {E}=0.1$, $a_{2}=1$, same as in the Fig. 
\ref{fig:Tsg-TU-1}. We observed that for transitions from the symmetric 
entangled state to the collective excited state there is no weak anti-Unruh 
effect, confirmed from Fig. \ref{fig:Anti-Unruh-Tsae-TUW}. On the other hand, 
for transition from the anti-symmetric state to the collective excited state 
there is weak anti-Unruh affect up to the value of $a_{1}=1$, complied from Fig. 
\ref{fig:Anti-Unruh-Tsae-TUW}.

\begin{figure}[h]
	\includegraphics[width=0.8\linewidth]{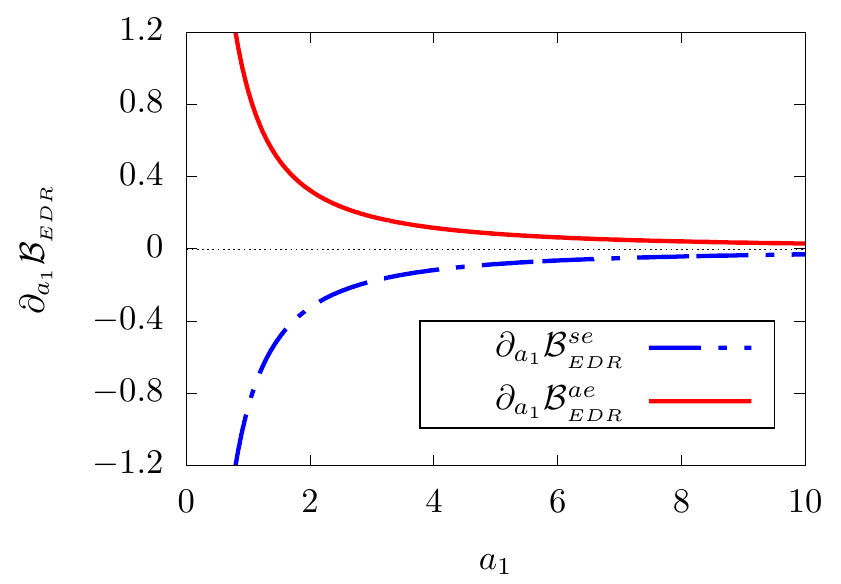}
	\caption{ The quantity $\partial_{a_{1}}\mathcal{B}_{_{EDR}}^{se}$ is plotted 
		with respect to varying $a_{1}$ for accelerated atoms in thermal background 
		considering Rindler modes, depicted by dash-dotted blue line. The 
quantity 
		$\partial_{a_{1}}\mathcal{B}_{_{EDR}}^{ae}$ is plotted with respect to varying 
		$a_{1}$ for accelerated atoms in thermal background considering Rindler 
modes, 
		depicted by solid red line.}
	\label{fig:Anti-Unruh-TsaeSTU}
\end{figure}

\begin{figure}[h]
\centering
 \includegraphics[width=0.8\linewidth]{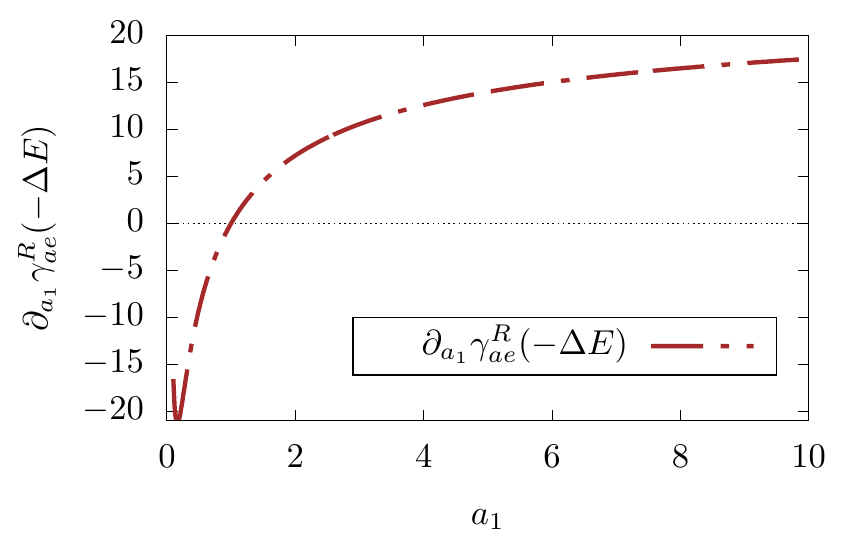} 
 \caption{The quantity $\partial_{a_{1}} \gamma_{ae}^{R} (-\Delta {E})$ 
corresponding to the transition from the anti-symmetric state to the collective 
excited state is plotted with respect to varying $a_{1}$ for accelerated atoms 
in thermal background with Rindler modes. This quantity is positive for values 
above $a_{1}=1$, like the $\partial_{a_{1}} \gamma_{ae}^{R} (\Delta {E})$ 
previously depicted in Fig. \ref{fig:Anti-Unruh-Tsae-TUW}.}
 \label{fig:TU-F11m-AU-2D}
\end{figure}

In Fig. \ref{fig:Anti-Unruh-TsaeSTU} we have plotted the quantity 
$\partial_{a_{1}}\mathcal{B}_{_{EDR}}^{se}$ and 
$\partial_{a_{1}}\mathcal{B}_{_{EDR}}^{ae}$ with respect to the acceleration 
$a_{1}$ considering the Rindler modes. These quantities correspond to the 
variation of the inverse EDR temperature for transitions from the symmetric and 
anti-symmetric states to the collective excited state with respect to $a_{1}$. 
These plots are meant to provide confirmation in support of occurrence of any 
strong anti-Unruh effect. We mention that while for the transition from the 
symmetric to excited state $\partial_{a_{1}}\mathcal{B}_{_{EDR}}^{se}$ is always 
negative and there is no sign of strong anti-Unruh effect, the case for 
anti-symmetric to excited state is different. In that case 
$\partial_{a_{1}}\mathcal{B}_{_{EDR}}^{ae}$ is positive in the whole region 
compared to a smaller region where weak condition is valid. From Fig. 
\ref{fig:TU-F11m-AU-2D} and Fig. \ref{fig:Anti-Unruh-Tsae-TUW} one can observe that 
$\partial_{a_{1}} \gamma_{ae}^{R} (-\Delta {E})$ and $\partial_{a_{1}} 
\gamma_{ae}^{R} (\Delta {E})$ are both positive for values of $a_{1}$ above 
$a_{1}=1$, and from Eq. (\ref{eq:Anti-Unruh-strongToweak}) it is to be noted 
that in that case strong anti-Unruh effect does not refer to the satisfaction of 
weak condition. Therefore, there is no shortcomings in the analysis. Then we 
note that the anti-Unruh effect occurring in this case is of both strong and 
weak nature below the value $a_{1}=1$, and above it there is no anti-Unruh 
effect.

Like the Minkowski mode case here also we have tried to understand the origin of 
the anti-Unruh effect. As discussed earlier the particular response function 
$\mathcal{R}_{11}(\Delta {E})$ signifies the contribution of a single two-level 
atomic detector, accelerated in a thermal background. Therefore, we consider 
the contribution from the response function $R_{11}$ for $\Delta {E}>0$ from 
Eq. (\ref{eq:response-TU-greater11}) and observe 
\begin{equation}
 \partial_{a_{1}}R_{11}(\Delta {E}) = 
\frac{\pi  \left(e^{\beta  \Delta {E}}+1\right) e^{\frac{2 \pi  
\Delta {E}}{a_{1}}}}{a_{1}^2 \left(e^{\beta \Delta {E}}-1\right) 
\left(e^{\frac{2 \pi  \Delta {E}}{a_{1}}}-1\right)^2}~,
\end{equation}
which is positive for all positive values of $\Delta {E}$, thus giving no 
anti-Unruh effect.

\begin{figure}[h]
	\centering
	\includegraphics[width=0.8\linewidth]{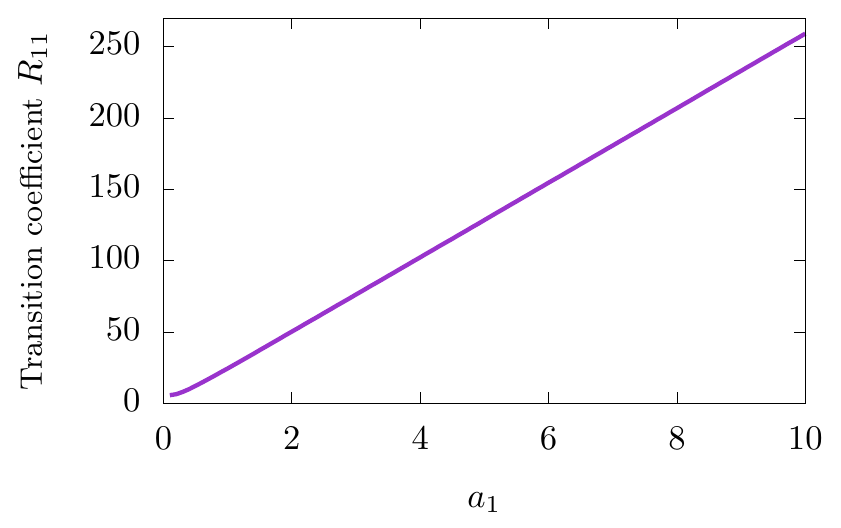}
	\caption{The response function $R_{11}(\Delta {E})$, plotted with respect to 
		varying $a_{1}$ for accelerated atoms in thermal background considering the 
		Rindler modes in $(1+1)$ dimensions.}
	\label{fig:TU-F11-2D}
\end{figure}

\begin{figure}[h]
	\centering
	\includegraphics[width=0.8\linewidth]{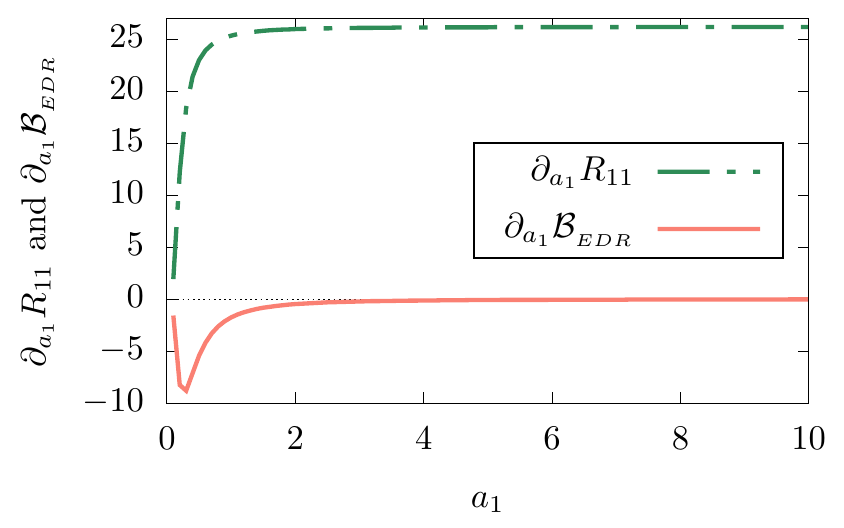}
	\caption{The differentiation of the response function $R_{11}(\Delta {E})$, 
		plotted with respect to varying $a_{1}$ for accelerated atoms in thermal 
		background considering the Rindler modes in $(1+1)$ dimensions, denoted 
by 
		dash-dotted green line. In this figure the quantity $\partial_{a_{1}} 
		\mathcal{B}_{_{EDR}}$ corresponding to the response function is also plotted, 
		which is denoted by the solid line.}
	\label{fig:TU-F11-AU-2D}
\end{figure}

This response function $R_{11}(\Delta {E})$ in $(1+1)$ dimensions considering 
the Rindler modes is plotted in Fig. \ref{fig:TU-F11-2D}, 
\noindent
where we observed that there is no visible case of anti-Unruh effect for the 
same set of values of the parameters. It should be noted that this is in 
contrary to the case considering the Minkowski modes. Therefore, in this case 
entanglement must have played a significant role to provide this outcome. In 
fact one may plot the differentiation of the transition probability 
$\partial_{a_{1}} \gamma_{ae}^{R}$ with respect to $a_{1}$ in the 
$\beta\to\infty$ limit, the zero temperature case, to check that the anti-Unruh 
effect is present there too. Furthermore, in Fig. \ref{fig:TU-F11-AU-2D} we have 
checked the conditions for weak and strong anti-Unruh effect for $R_{11}(\Delta 
{E})$ and we confirm the absence of any anti-Unruh effect in that parameter 
range. This reconfirms that entanglement is crucial for anti-Unruh phenomenon in case of Rindler mode analysis. 

\subsubsection{$(1+3)$-dimensions}

\begin{figure}[h]
\centering
 \includegraphics[width=0.8\linewidth]{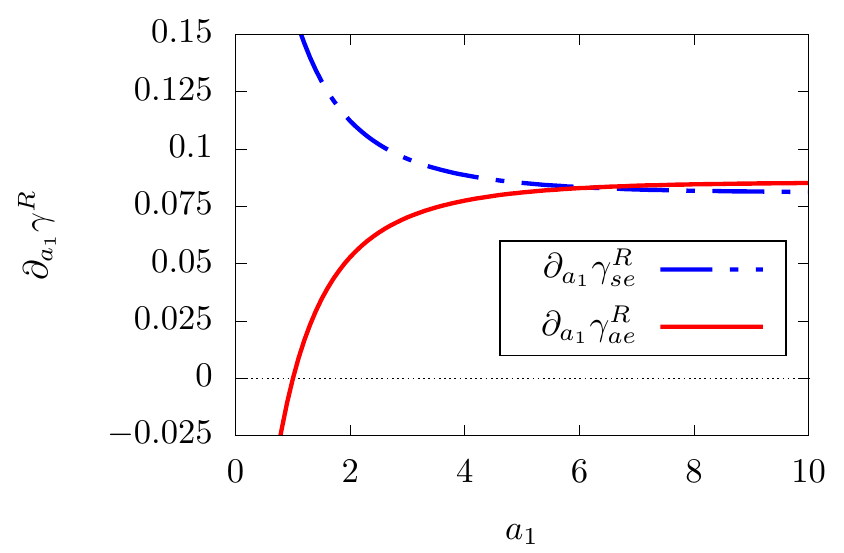}
 \caption{ The quantity $\partial_{a_{1}}\gamma_{se}^{R}$ is plotted with 
respect to varying $a_{1}$ for accelerated atoms in thermal background 
considering Rindler modes, depicted by dash-dotted blue line. The quantity 
$\partial_{a_{1}}\gamma_{ae}^{R}$ is plotted with respect to varying $a_{1}$ for 
accelerated atoms in thermal background considering Rindler modes in $(1+3)$ 
dimensions, depicted by solid red line.}
 \label{fig:AntiUnruh-3D-TUW}
\end{figure}

In Fig. \ref{fig:AntiUnruh-3D-TUW} we have plotted 
$\partial_{a_{1}}\gamma_{se}^{R}$ and $\partial_{a_{1}}\gamma_{ae}^{R}$, i.e., 
the derivatives of the transition probabilities corresponding to transitions 
from the symmetric and anti-symmetric states to the collective excited state 
considering the Rindler modes, with respect to varying $a_{1}$. The other 
parameters are kept fixed $\Delta {E}=0.1$, $a_{2}=1$, same as in the Fig. 
\ref{fig:TsgTU-P3D}. Like earlier these plots denote the conditions for weak 
anti-Unruh effect. In particular, we observed that for transition from the 
symmetric entangled state to the collective excited state there is no weak 
anti-Unruh effect for these particular parameter values, see Fig. 
\ref{fig:AntiUnruh-3D-TUW}. However, for the same parameter values for the 
transition from the anti-symmetric state to the collective excited state there 
is weak anti-Unruh affect up to the value of $a_{1}=1$ also seen from Fig. 
\ref{fig:AntiUnruh-3D-TUW}.
\begin{figure}[h]
\centering
 \includegraphics[width=0.8\linewidth]{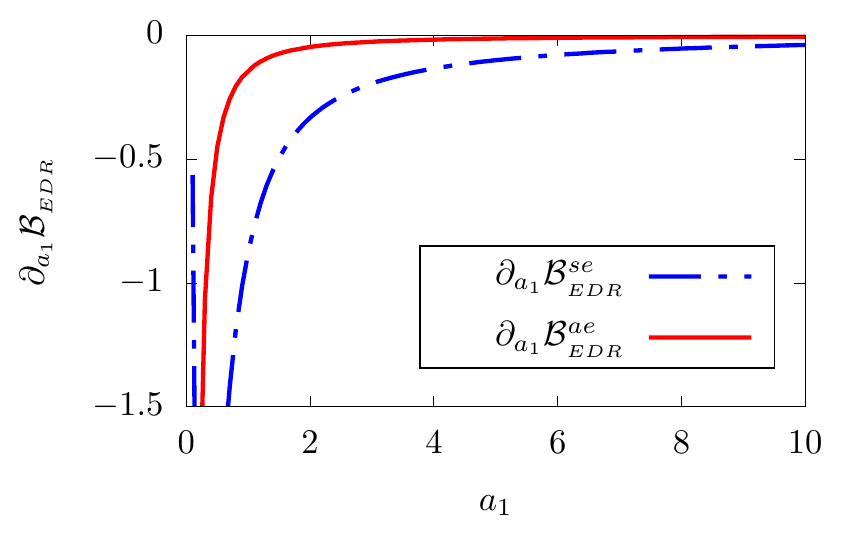}
 \caption{ The quantity $\partial_{a_{1}}\mathcal{B}_{_{EDR}}$ is plotted 
with respect to varying $a_{1}$ for accelerated atoms in thermal background 
considering Rindler modes in $(1+3)$ dimensions. For the transition from the 
symmetric state to the collective excited state the curve is given by the 
dash-dotted blue line. On the other hand, for the transition from the 
anti-symmetric state to the collective excited state the curve is given by the 
solid red line. The positivity of these curves are expected to provide the 
condition for strong anti-Unruh effect.}
 \label{fig:AntiUnruh-3D-TUS}
\end{figure}

In Fig. \ref{fig:AntiUnruh-3D-TUS} we have plotted the quantity 
$\partial_{a_{1}}\mathcal{B}_{_{EDR}}^{se}$ and 
$\partial_{a_{1}}\mathcal{B}_{_{EDR}}^{ae}$ with respect to the acceleration 
$a_{1}$ to understand the strong anti-Unruh effect in this case. These 
quantities correspond to the variation of the inverse EDR temperature for 
transitions from the symmetric and anti-symmetric states to the collective 
excited state with respect to $a_{1}$. These plots are meant to provide 
confirmation in support of the occurrence of any strong anti-Unruh effect. From 
this figure we observed that for both the transitions from the symmetric and 
anti-symmetric states to the collective excited state there is no sign of strong 
anti-Unruh effect.

\begin{figure}[h]
\centering
 \includegraphics[width=0.8\linewidth]{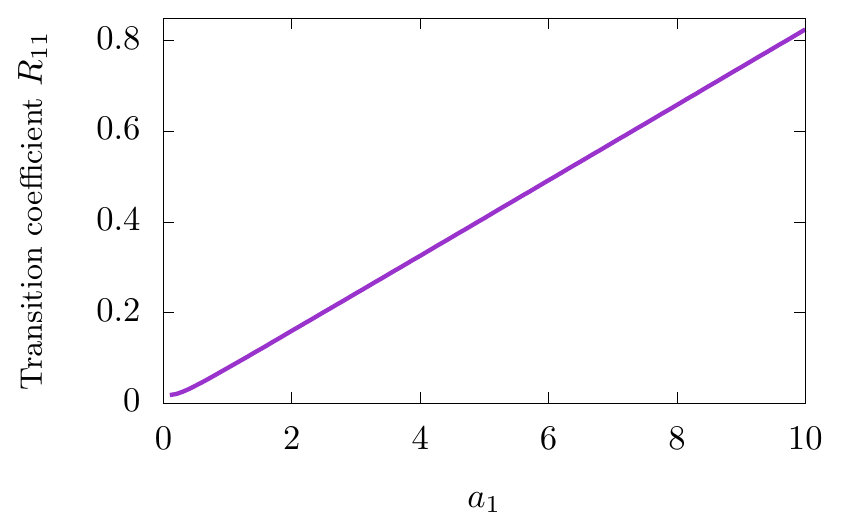}
 \caption{The response function $R_{11}(\Delta {E})$, plotted with respect to 
varying $a_{1}$ for accelerated atoms in thermal background considering the 
Rindler modes in $(1+3)$ dimensions.}
 \label{fig:TU-F11-4D}
\end{figure}

\begin{figure}[h]
\centering
 \includegraphics[width=0.8\linewidth]{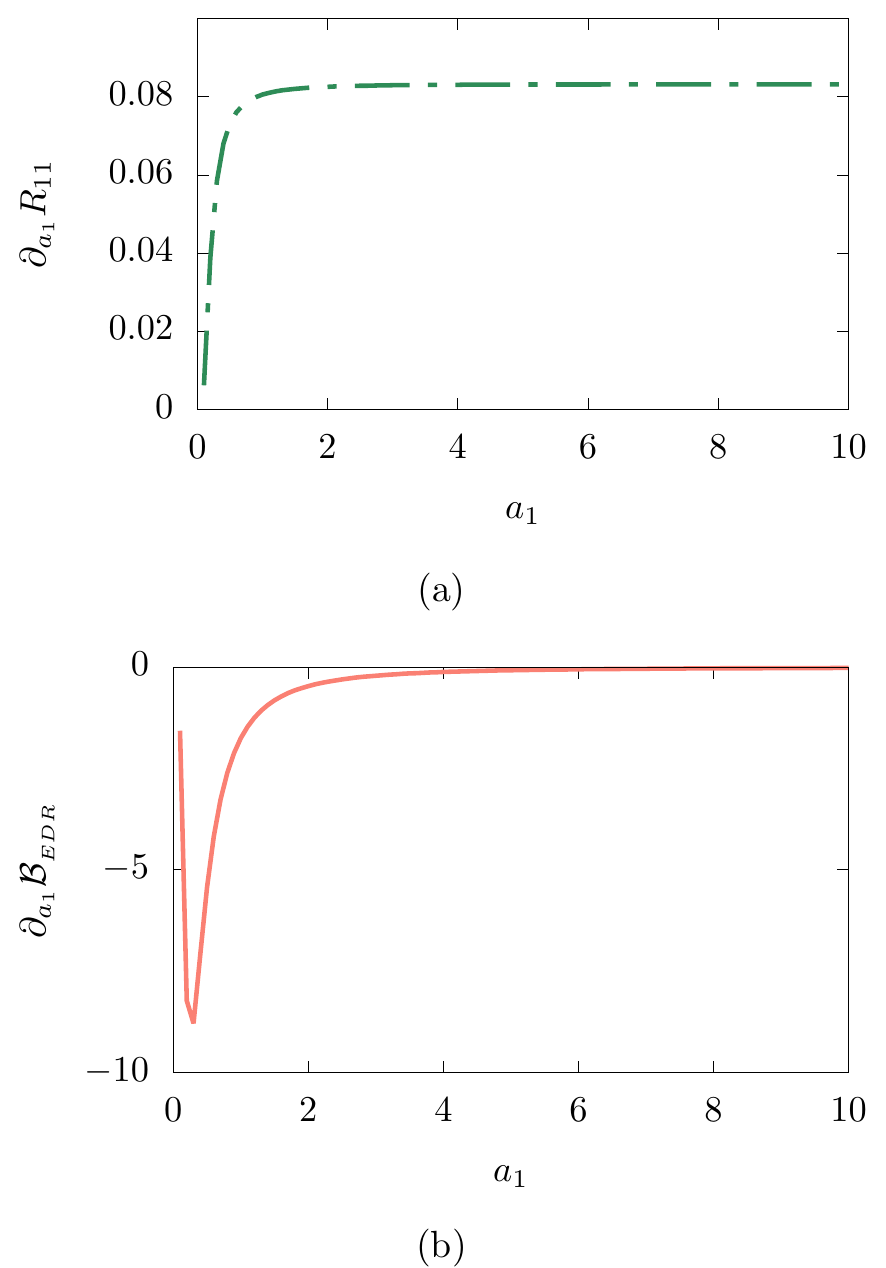}
 \caption{(a) The derivative of the response function $R_{11}(\Delta {E})$, 
plotted with respect to varying acceleration $a_{1}$ for accelerated atoms in 
thermal background considering the Rindler modes in $(1+3)$ dimensions, denoted 
by 
dash-dotted green line. (b) In this figure the quantity $\partial_{a_{1}} 
\mathcal{B}_{_{EDR}}$ corresponding to the response function is also plotted, 
which is denoted by the solid line.}
 \label{fig:TU-F11-AU-4D}
\end{figure}

Like the previous cases here also we have tried to understand the origin of 
these anti-Unruh effect. In this regard, we have  plotted the response function 
$R_{11}(\Delta {E})$ in $(1+3)$ dimensions considering the Rindler modes in 
Fig. 
\ref{fig:TU-F11-4D}. Here also we observed that there is no visible case of 
anti-Unruh effect for the selected set of values of the parameters, thus 
discarding the effects of the thermal bath in the occurrence of this anti-Unruh 
effect. It is in contrary to the case considering the Minkowski modes. 
Therefore, here also entanglement should be considered to be the significant 
contributor to the origin of the perceived  anti-Unruh phenomena, which can also 
be observed by plotting the $\partial_{a_{1}}\gamma_{ae}^{R}$ in the 
$\beta\to\infty$ limit. In Fig. \ref{fig:TU-F11-AU-4D} we have further checked 
the conditions for weak and strong anti-Unruh effect for $R_{11}(\Delta {E})$, 
and found the absence of the same in the considered parameter range.

\subsection{Summarizing the outcomes regarding the anti-Unruh(-like) effect}

In this part we summarize the results that we have arrived on, while studying 
the anti-Unruh effect considering two entangled atoms accelerated in a thermal 
bath. For the convenience of understanding the results will be tabulated below 
in a case by case manner. We shall first tabulate the results for the case 
with the Minkowski modes. Then we shall tabulate the results corresponding to 
the case with the Rindler modes.

In Table. \ref{tab:Minkowski-AU} we have tabulated the results corresponding to 
the case with the Minkowski modes. 
\begin{table}[h] \small%
	\centering
	\caption{The case with the Minkowski modes}\label{tab:Minkowski-AU}
	\begin{tabular}{|p{2.5cm}|p{1.6cm}|p{1.7cm}|p{2.1cm}|}
		\hline
		& Transitions & Anti-Unruh-like effect & Nature \\
		\hline
		$(1+1)$ dimensions & $\gamma_{se}$ & Yes & Entirely weak \\
		\cline{2-4}
		& $\gamma_{ae}$ & Yes & Entirely weak \\
		\cline{2-4}
		& $\mathcal{F}_{11}$ & Yes & Entirely weak \\
		\hline
		$(1+3)$ dimensions & $\gamma_{se}$ & Yes & Mostly weak, strong in some region \\
		\cline{2-4}
		& $\gamma_{ae}$ & Yes & Mostly weak, strong in some region \\
		\cline{2-4}
		& $\mathcal{F}_{11}$ & Yes & Entirely weak \\
		\hline
	\end{tabular}
\end{table}
On the other hand, in Table. 
\ref{tab:Unruh-AU} we have tabulated the results for the case with the Rindler 
modes. 
\begin{table}[h] \small%
	\centering
	\caption{The case with the Rindler modes}\label{tab:Unruh-AU}
	\begin{tabular}{|p{2.5cm}|p{1.6cm}|p{1.7cm}|p{2.1cm}|}
		\hline
		& Transitions & Anti-Unruh effect & Nature \\
		\hline
		$(1+1)$ dimensions & $\gamma_{se}^{R}$ & No & - \\
		\cline{2-4}
		& $\gamma_{ae}^{R}$ & Yes & Both strong and weak \\
		\cline{2-4}
		& $\mathcal{R}_{11}$ & No & - \\
		\hline
		$(1+3)$ dimensions & $\gamma_{se}^{R}$ & No & - \\
		\cline{2-4}
		& $\gamma_{ae}^{R}$ & Yes & Entirely weak \\
		\cline{2-4}
		& $\mathcal{R}_{11}$ & No & - \\
		\hline
	\end{tabular}
\end{table}

In both of the cases the set of parameters signify the similar scenarios 
and are considered to be in the same range.

\section{Discussion}\label{sec:discussion}

In this work we have attempted to understand the radiative process of two 
entangled accelerated atoms interacting with a massless scalar field in a 
thermal bath. In particular, the transitions form the symmetric and 
anti-symmetric entangled states to the collective excited or ground states are 
studied. It is to be noted that vacuum fluctuations effects acts as the cause 
for these transitions. In Sec. \ref{sec:Transition-prob-Mink} we have provided 
the estimations of the transition probabilities considering the Green's 
function, constructed from the Minkowski modes with a Rindler coordinate 
transformation, in both $(1+1)$ and $(1+3)$ dimensions. These transition 
probabilities correspond to certain frequencies of the field modes and they do 
not resemble the transition probabilities for unit time. In this case we 
observed that for both $(1+1)$ and $(1+3)$ dimensions there are visible cases of 
anti-Unruh-like effect in the transition probabilities. However, the $(1+1)$ 
dimensional results are qualitatively different from the $(1+3)$ dimensional 
one. For the transition from the symmetric state to the collective excited state 
we observed that in $(1+1)$ dimensions there is first anti-Unruh-like effect and 
then Unruh-like effect, see from Fig. \ref{fig:TsgTR-P1}. However, for the same 
transition in $(1+3)$ dimensions there is only anti-Unruh-like effect in the 
same parameter range, see Fig. \ref{fig:TsgTR-P3D}. Therefore, there is a bit of 
difference  between the $(1+1)$ and $(1+3)$ dimensional results. Another 
evidence of this mismatch is observed when the nature of the anti-Unruh-like 
effect is studied. In Sec. \ref{sec:Study-Anti-Unruh} we observed that in 
$(1+1)$ dimensions the anti-Unruh-like effect is of purely weak nature in the 
considered parameter range. On the other hand, in $(1+3)$ dimensions in the same 
parameter range we observed that there are also some regions where strong 
condition is satisfied. The regions where the strong condition is satisfied are 
always contained inside the regions for the weak condition. We have further 
plotted the quantities $\mathcal{F}_{11}$ and observed the anti-Unruh-like 
effect here also, suggesting that in the Minkowski mode case entanglement do not 
play a significant role in the occurrence of the anti-Unruh-like effect, i.e., 
thermal background plays the major role.

Furthermore, in Sec. \ref{sec:Transition-prob-Unruh} we have considered the 
Green's functions in terms of the Rindler modes, obtained using the Unruh 
operators in the Unruh Vacuum, for the estimation of the transition 
probabilities. These transition probabilities are time translation invariant and 
a unit time prescription can be provided for them. Unlike the Minkowski mode 
case here the $(1+1)$ and $(1+3)$ dimensional results are in agreement with each 
other, see Fig. \ref{fig:Tsg-TU-1} and \ref{fig:TsgTU-P3D}. For this case the 
occurrence of the anti-Unruh effect is confirmed only for the transition from 
the anti-symmetric state to the collective excited state, Sec. 
\ref{sec:Study-Anti-Unruh}. However, the quantities $R_{11}$, which signify the 
contribution if a single detector were accelerated in the thermal bath, do not 
show any anti-Unruh effect in the selected parameter range. Thus suggesting that 
here entanglement has a significant role in the occurrence of the anti-Unruh 
effect.

Subsequently, the transition coefficients $\mathcal{F}_{11}$ considering the 
Minkowski modes are not symmetric under the interchange between the temperature 
of the thermal bath and the Unruh temperature, i.e., under $\beta\leftrightarrow 
2\pi/b$. However, for the case with the Rindler modes with Unruh operators 
$R_{11}$ is symmetric under the interchange $\beta\leftrightarrow 2\pi/b$ (see 
Appendix. \ref{Apn:Diff-GreenFn-analogy} for qualitative difference in terms of 
time translation invariance, between the Green's functions considering the 
Minkowski and Rindler modes). Then the later case of Rindler modes with Unruh 
operators give a much more suitable representation for an accelerated observer, 
where the analogy with a thermal bath is concerned.

In summary through this work we have not only studied the radiative process of 
entangled atoms accelerated in a thermal bath but also provided understandings 
as to when an accelerated observer is invariably comparable to a static observer 
in thermal bath. These calculations and understandings motivate one to pursue 
other entanglement related studies \cite{Hu:2015lda, Zhou:2020oqa, 
Menezes:2015iva} such as entanglement dynamics, which includes rate of variation 
of the atomic energy, generation and decay of entangled states due to the 
contributions of vacuum fluctuations and radiation reaction, etc., for 
accelerated atoms in thermal bath.

\begin{acknowledgments}
S.B. would like to thank Indian Institute Technology Guwahati (IIT Guwahati) for 
supporting this work through a Post-Doctoral Fellowship.
\end{acknowledgments}

\appendix

\section{Different Green's functions and the consequence of their 
consideration}\label{Apn:Diff-GreenFn-analogy}

The fact that an accelerated observer in a Minkowski spacetime resembles an 
inertial observer in a thermal background is widely debated in literature. There 
are some studies \cite{Kolekar:2013hra, Kolekar:2013xua, Kolekar:2013aka} in 
favour of this resemblance and there also articles pointing out some very 
crucial contradictions \cite{Chowdhury:2019set}. The discourse is still open and 
here we are going to provide some insightful results  in this regard. Here we 
are going to point out the similarities and dissimilarities between between 
observers in a thermal bath or in a non-inertial motion with uniform 
acceleration at the Green's function level.

\subsubsection{Analogy between Green's functions of uniformly accelerated and 
static in thermal bath observers}

We mention that the momentum integral in Eq. (\ref{eq:Two-point-fn-thermal1}) 
can be explicitly carried out to provide a position space representation of the 
$(1+1)$ dimensional thermal Green's function as  
\begin{eqnarray}
 G_{\beta}^{+}(X_{2};X_{1}) &=& -\frac{1}{4\pi} 
\left(\ln{\left[1-e^{-\frac{2\pi}{\beta}(\Delta T-\Delta 
X)}\right]}\right. \nonumber\\
~&&~~\left. +~\ln{\left[1-e^{-\frac{2\pi}{\beta}(\Delta 
T+\Delta X)}\right]}\right).\label{eq:Greens-fn-thermal3}
\end{eqnarray}
Similarly in $(1+3)$ dimensions also the thermal Green's function of Eq. 
(\ref{eq:Greens-fn-thermal-1p3}) can be explicitly evaluated in position space, 
see \cite{Chowdhury:2019set}, as
\begin{eqnarray}
 G_{\beta}^{+}(X_{2};X_{1}) &=& \frac{1}{8\pi\beta|\Delta\mathbf{X}|} 
\left[\coth{\left(\frac{\pi}{\beta}(\Delta T+
|\Delta\mathbf{X}|)\right)}\right. \nonumber\\
~&&~~\left. ~-~\coth{\left(\frac{\pi}{\beta}(\Delta T-
|\Delta\mathbf{X}|)\right)}\right].\label{eq:Greens-fn-thermal4}
\end{eqnarray}
In the following discussions we are going to study these Green's 
functions in different limits and scenarios.\vspace{0.2cm}

\emph{$(1+1)$ dimensions.--}
One can take the expression of thermal Green's function in $(1+1)$ dimensions 
from Eq. (\ref{eq:Greens-fn-thermal3}) and represent it in a more suitable 
manner for the subsequent analysis as
\begin{eqnarray}
&&\scalebox{0.9}{$G_{\beta}^{+}(X_{2};X_{1})= -\frac{1}{4\pi} 
\ln{\left[\frac{\sinh{\left(\frac{\pi}{\beta}(\Delta T-\Delta 
X)\right)}\sinh{\left(\frac{\pi}{\beta}(\Delta T+\Delta 
X)\right)}}{\left(\frac{\pi}{\beta}\right)^2}\right]}$} \nonumber\\
~&& ~~~~~~~~~~~~~~~~~~~~~~~~~~-~ 
\scalebox{1}{$\frac{1}{2\pi}\ln{\left(\frac{2\pi}{\beta}e^{-\frac{\pi\Delta 
T}{\beta}}\right)}$}~.
\label{eq:GreenFn-thermal2-1p1}
\end{eqnarray}
From this expression of the Green's function one can arrive at the $(1+1)$ 
dimensional Minkowski Green's function as the limit $\beta\to\infty$ is taken. 
The corresponding expression is 
\begin{equation}
 G_{M}^{+}(X_{2};X_{1}) = \frac{1}{4\pi} 
\ln{\left[(\Delta T-\Delta X)(\Delta T+\Delta X)\right]}.
\label{eq:GrnFn-Mink2}
\end{equation}
However, it should be mentioned that this entire expression comes from the first 
term on the right hand side of Eq. (\ref{eq:GreenFn-thermal2-1p1}). The second 
term provides an infinite contribution, which is neglected out of convenience. 
Then neglecting this second term of Eq. (\ref{eq:GreenFn-thermal2-1p1}) one can 
observe that as one takes the limit $\Delta X\to0$ the $(1+1)$ dimensional 
thermal Green's function reduces to 
\begin{equation}
 G_{\beta}^{+}(X_{2};X_{1}) = -\frac{1}{2\pi} 
\ln{\left[\frac{\beta}{\pi}\sinh{\left(\frac{\pi~\Delta 
T}{\beta}\right)}\right]}~,
\label{eq:GrnFn-TB2}
\end{equation}
which gives the Green's function corresponding to a static observer in thermal 
bath in $(1+1)$ dimensions. On the other hand, from the expression of $(1+1)$ 
dimensional Minkowski Green's function from Eq. (\ref{eq:GrnFn-Mink2}), and 
using the Rindler transformation of Eq. (\ref{eq:Rindler1-trans2}) one can get 
the Green's function for an accelerated observer in $(1+1)$ dimensions as
\begin{eqnarray}
 G_{M}^{+}(X_{2};X_{1}) &=& -\frac{1}{2\pi} 
\ln{\left[\frac{2}{b}\sinh{\left(\frac{b~\Delta 
\tau}{2}\right)}\right]}~.
\label{eq:GrnFn-MinkRind2}
\end{eqnarray}
It can be observed that the Green's functions from Eq. (\ref{eq:GrnFn-TB2}) and 
Eq. (\ref{eq:GrnFn-MinkRind2}) are the same with $\Delta T$ and $\beta$ replaced 
by $\Delta \tau$ and $2\pi/b$. 
\vspace{0.2cm}

\emph{$(1+3)$ dimensions.--}
We note that in Eq. (\ref{eq:Greens-fn-thermal4}) as one takes $|\Delta 
\mathbf{X}|\to0$, one shall get
\begin{eqnarray}
 G_{\beta}^{+}(X_{2};X_{1}) &=& -\frac{1}{4~\beta^2} 
\frac{1}{\sinh^2{\left({\pi~\Delta T}/{\beta}\right)}}~.
\label{eq:GrnFn-TB4}
\end{eqnarray}
On the other hand, as one takes $1/\beta\to0$ in the expression of the Green's 
function from the same Eq. (\ref{eq:Greens-fn-thermal4}), one can obtain the 
Minkowski Green's function
\begin{eqnarray}
 G_{M}^{+}(X_{2};X_{1}) &=& \frac{1}{4\pi^2} 
\frac{1}{\left(-\Delta T^2+ |\Delta\mathbf{X}|^2\right)}~,
\label{eq:GrnFn-Mink4}
\end{eqnarray}
which for an accelerated observer in terms of the Rindler proper time 
(\ref{eq:Rindler1-trans2}) can be expressed as
\begin{eqnarray}
 G_{M}^{+}(X_{2};X_{1}) &=& -\frac{b^2}{16\pi^2} 
\frac{1}{\sinh^2{\left({b~\Delta\tau}/{2}\right)}}~.\label{ 
eq:Greens-fn-Minkowski-Rindler}
\end{eqnarray}
It should be noted that this Green's function and the thermal Green's function 
from Eq. (\ref{eq:GrnFn-TB4}) have the exact same expression with $\Delta T$ and 
$\beta$ now replaced by $\Delta \tau$ and $2\pi/b$.  We have tabulated 
characteristics of these different Green's functions corresponding to static 
observer in thermal bath or uniformly accelerated observer in non-thermal 
background in Table. \ref{tab:GrnFn-prop1}.

\begin{table} \small%
\centering
 \caption{Characteristics of different Green's 
functions (observers with uniform acceleration or static in thermal 
bath)}\label{tab:GrnFn-prop1}
\begin{tabular}{|p{1.4cm}|p{2.3cm}|p{1.7cm}|p{2.3cm}|}
\hline
 & Green's function & Time translation invariance & Analogous 
Green's function \\
\hline
$(1+1)$ dimensions & Thermal static & Yes & Uniformly accelerated with 
Minkowski modes \\
\cline{4-4}
                 &  &  & Uniformly accelerated with 
Rindler modes \\
\cline{2-4}
                & Uniformly accelerated with 
Minkowski modes & Yes & Uniformly accelerated with 
Rindler modes \\
\cline{4-4}
                 &  &  & Thermal static \\
\cline{2-4}
                & Uniformly accelerated with 
Rindler modes & Yes & Uniformly accelerated with 
Minkowski modes \\
\cline{4-4}
                 &  &  & Thermal static \\
\hline
$(1+3)$ dimensions & Thermal static & Yes & Uniformly accelerated with 
Minkowski modes \\
\cline{4-4}
                 &  &  & Uniformly accelerated with 
Rindler modes \\
\cline{2-4}
                & Uniformly accelerated with 
Minkowski modes & Yes & Uniformly accelerated with 
Rindler modes \\
\cline{4-4}
                 &  &  & Thermal static \\
\cline{2-4}
                & Uniformly accelerated with 
Rindler modes & Yes & Uniformly accelerated with 
Minkowski modes \\
\cline{4-4}
                 &  &  & Thermal static \\
\hline
\end{tabular}
\end{table}

\subsubsection{Analogy between Green's functions of accelerated atoms in 
thermal bath considering the Minkowski or Rindler modes, and atoms with double 
acceleration}

\emph{Accelerated observer in thermal bath.--} From Eq. 
(\ref{eq:Two-point-fn-thermal2}) with the coordinate transformation of Eq. 
(\ref{eq:TR-DT&DX}) one can find out the Green's function corresponding to an 
accelerated observer in a thermal bath in $(1+1)$ dimensions with respect to the 
Minkowski modes. Similarly, in $(1+3)$ dimensions the Green's function 
corresponding to an accelerated observer in thermal bath with respect to the 
Minkowski modes is given by the expression of Eq. (\ref{eq:Greens-fn-TR-4D}) 
with the coordinate transformation of Eq. Eq. (\ref{eq:TR-DT&DX-3D}). It should 
be noted that none of these Green's functions are time translation invariant 
with respect to the proper time of the accelerated observer.\vspace{0.1cm}

On the other hand, the $(1+1)$ and $(1+3)$ dimensional Green's functions 
corresponding to accelerated observers in thermal bath with respect to the 
Rindler modes are given by Eq. (\ref{eq:Greens-fn-TU}) and 
(\ref{eq:Greens-fn-TU-3D}). It is to be noted that these Green's functions are 
time translation invariant. We further briefly discuss about the Green's 
function in a Rindler-Rindler frame.

\begin{table} \small%
\centering
 \caption{Characteristics of different Green's 
functions (observers with uniform acceleration in thermal 
bath or with double acceleration)}\label{tab:GrnFn-prop2}
\begin{tabular}{|p{1.4cm}|p{2.3cm}|p{1.7cm}|p{2.3cm}|}
\hline
 & Green's function & Time translation invariance & Analogous 
Green's function \\
\hline
$(1+1)$ dimensions & Accelerated in thermal bath with Minkowski modes & No & 
- \\
\cline{2-4}
                & Accelerated in thermal bath with Rindler modes & Yes & 
- \\
\cline{2-4}
                & Double acceleration & - & - \\
\hline
$(1+3)$ dimensions & Accelerated in thermal bath with Minkowski modes & No & 
- \\
\cline{2-4}
                & Accelerated in thermal bath with Rindler modes & Yes & 
- \\
\cline{2-4}
                & Double acceleration & - & - \\
\hline
\end{tabular}
\end{table}

\emph{Rindler-Rindler.--} In the first Rindler spacetime defined by the 
coordinate transformation of Eq. (\ref{eq:Rindler1-trans}) if one considers 
another analogous coordinate transformation, one can form the so called 
Rindler-Rindler spacetime. This coordinate transformation is 
\begin{eqnarray}\label{eq:Rindler2-trans}
 \eta &=& \frac{e^{a' x}}{a'} \sinh{a' t}\nonumber\\
 \xi &=& \frac{e^{a' x}}{a'} \cosh{a' t}~,
\end{eqnarray}
which enables one to express the line element as
\begin{eqnarray}\label{eq:Rindler2-metric}
 ds^2 = e^{2a' x} 
\exp{\left\{\frac{2a}{a'}e^{a' x} \cosh{a' t} 
\right\}}\left[-dt^2+dx^2\right]~.\nonumber\\
\end{eqnarray}
It should be mentioned that here $a$ and $a'$ denote the acceleration parameters 
corresponding to the first and the second Rindler transformations. The relation 
between the Minkowski coordinates $(T,X)$ and thees Rindler-Rindler coordinates 
$(t,x)$ is 
\begin{eqnarray}\label{eq:Rindler3-trans}
 T &=& \frac{\tilde{X}}{a}  \sinh{\tilde{T}}\nonumber\\
~&=& \frac{e^{-a/a'}}{2a} \left\{\exp{\left[\frac{a}{a'}e^{a' (x+t)}\right]} - 
\exp{\left[\frac{a}{a'}e^{a' (x-t)}\right]}\right\}~\nonumber\\
 X &=& \frac{\tilde{X}}{a} \cosh{\tilde{T}}\nonumber\\
~&=& \frac{e^{-a/a'}}{2a} \left\{\exp{\left[\frac{a}{a'}e^{a' (x+t)}\right]} + 
\exp{\left[\frac{a}{a'}e^{a' (x-t)}\right]}\right\}~.\nonumber\\
\end{eqnarray}
Here $\tilde{X}=\exp{\left[\frac{a}{a'}e^{a' x} \cosh{a' t}\right]}$ and 
$\tilde{T}=\frac{a}{a'}e^{a' x} \sinh{a' t}$. Then we have the expression of 
\begin{eqnarray}\label{eq:Rindler3-trans}
 T+X &=& \frac{e^{-a/a'}}{a} \exp{\left[\frac{a}{a'}e^{a' (x+t)}\right]} 
~\nonumber\\
 T-X &=& -\frac{e^{-a/a'}}{a} \exp{\left[\frac{a}{a'}e^{a' 
(x-t)}\right]}~,
\end{eqnarray}
which can be used to obtain the $(1+1)$ and $(1+3)$ dimensional Green's 
functions from Eq. (\ref{eq:GrnFn-Mink2}) and (\ref{eq:GrnFn-Mink4}) 
corresponding to an observer in a Rindler-Rindler frame with respect to the 
Minkowski modes. It should be mentioned that in a Rindler-Rindler frame the 
exact expression of the proper time is not yet known up to our knowledge. Then 
it is not readily possible to comment about the time translational invariance 
for this Green's function. However, structure wise it can be observed that it is 
different than the thermal-Rindler case. We have tabulated characteristics of 
different Green's functions, corresponding to accelerated observers in thermal 
bath or observer in Rindler-Rindler frame, and the analogy between them in 
Table. \ref{tab:GrnFn-prop2}.

\section{Green's function of accelerated observer considering Rindler 
modes}\label{Apn:Unruh-Greens-fn}

To obtain the Green's function of an accelerated observer considering the 
Rindler modes we first take the definition of the Green's function 
\begin{equation}\label{eq:Green-fn-def-minkowski}
 G_{R}^{+} (X_{2},X_{1}) = \langle 
0_{M}|\Phi^{R}(X_{2})\Phi^{R}(X_{1})|0_{M}\rangle~.
\end{equation}
In this expression we put the expression of the scalar field from Eq. 
(\ref{eq:RRW-field-Unruh}), which denotes the scalar field decomposition in the 
right Rindler wedge. We also assume that for both of the spacetime points 
acceleration is the same $a$. Then the above Green's function 
(\ref{eq:Green-fn-def-minkowski}) becomes
\begin{eqnarray}\label{eq:Green-Unruh-der1}
 G_{R}^{+} &=& \langle 0_{M}|\sum_{k,k^{'}=-\infty}^{\infty} 
\frac{1}{2\sqrt{\sinh{\frac{\pi\omega_{k}}{a}}\sinh{\frac{\pi\omega_{k^{'}}}{a}}
} }\nonumber\\
~&& \left[\left(d^{1}_{k}e^{\frac{\pi\omega_{k}}{2a}}~ ^{R}u_{k} +  
d^{2}_{k}e^{-\frac{\pi\omega_{k}}{2a}}~ 
^{R}u^{*}_{-k}\right)\right.\nonumber\\
~&& \left.\left(d^{1^{\dagger}}_{k^{'}}e^{\frac{\pi\omega_{k^{'}}}{2a}}~ 
^{R}u^{*}_{k^{'}} +
d^{2^{\dagger}}_{k^{'}}e^{-\frac{\pi\omega_{k^{'}}}{2a}}~ 
^{R}u_{-k^{'}}\right)
\right]|0_{M}\rangle\nonumber\\
~&=& \sum_{k=-\infty}^{\infty} 
\frac{1}{2\sinh{\frac{\pi\omega_{k}}{a}}
} \left[e^{\frac{\pi\omega_{k}}{a}}~ ^{R}u_{k}~ 
^{R}u^{*}_{k}\right.\nonumber\\ 
~&& \left. ~~~~~+~~  
e^{-\frac{\pi\omega_{k}}{a}}~ 
^{R}u^{*}_{-k}~ 
^{R}u_{-k}\right]~,
\end{eqnarray}
where, we have used the commutation relation $[d^{j}_{k}, 
d^{j^{\dagger}}_{k^{'}}] =\delta_{k,k^{'}}$. Now by putting the expression of 
the modes $^{R}u_{k}$ from Eq. (\ref{eq:Rindler-modes}) one can obtain the 
Green's function to be
\begin{eqnarray}\label{eq:Green-Unruh-der2}
 G_{R}^{+} 
= \sum_{k=-\infty}^{\infty} 
\frac{1}{4\pi\omega_{k}
} \left[\frac{e^{ik\Delta\xi-i\omega_{k}\Delta\eta}}{1-e^{\frac{-2\pi\omega_{k}
} { a }}} + 
\frac{e^{ik\Delta\xi+i\omega_{k}\Delta\eta}}{e^{\frac{2\pi\omega_{k}}{ a 
}}-1}\right],
\end{eqnarray}
which, in the continuum momentum limit yields the desired expression of the 
Green's function for an accelerated observer in terms of the Rindler modes as 
given in Eq. (\ref{eq:Greens-fn-Unruh}). Note that this evaluation is done for 
$(1+1)$ dimensions. A similar evaluation can be done in the $(1+3)$ dimensions 
also considering the scalar field expansion in RRW from Eq. 
(\ref{eq:RRW-field-Unruh-1p3}) and putting it in Eq. 
(\ref{eq:Green-fn-def-minkowski}) and then using the expressions of the mode 
from Eq. (\ref{eq:Rindler-modes-3D}) to evaluate the Green's function 
corresponding to an accelerated observer in terms of the Rindler modes, the 
expression of which is given in Eq. (\ref{eq:Greens-fn-Unruh-3D}).

Next we construct the Green's function in the Minkowski vacuum considering the 
field decomposition given by Eq. (\ref{eq:Field-Unruh}), i.e., in terms of the 
Unruh modes and Unruh operators in $(1+1)$ dimensions. This Green's function is 
obtained as 
\begin{eqnarray}\label{eq:GreenFn-def-mink}
 \scalebox{0.9}{$G_{U}^{+} (X_{2},X_{1})$} &=& \langle 
0_{M}|\Phi(X_{2})\Phi(X_{1})|0_{M}\rangle\nonumber\\
~&=& \scalebox{0.9}{$G_{R}^{+} (X_{2},X_{1}) + G_{L}^{+} (X_{2},X_{1}) 
+G_{RL}^{+} (X_{2},X_{1})$}~,\nonumber\\
\end{eqnarray}
where the expressions $\scalebox{0.85}{$G_{R}^{+} (X_{2},X_{1})$}$ and 
$\scalebox{0.85}{$G_{L}^{+} (X_{2},X_{1})$}$ correspond to accelerated observers 
in the right and in left Rindler wedges respectively, and 
$\scalebox{0.85}{$G_{RL}^{+} (X_{2},X_{1})$}$ denotes the cross term. The 
expression of $\scalebox{0.85}{$G_{R}^{+} (X_{2},X_{1})$}$ is already given in 
Eq. (\ref{eq:Green-Unruh-der1}) and the other two quantities are given by 
\begin{eqnarray}\label{eq:GreenFn-Unruh-Diff}
 \scalebox{0.9}{$G_{L}^{+} (X_{2},X_{1})$} &=& \sum_{k=-\infty}^{\infty} 
\frac{1}{2\sinh{\frac{\pi\omega_{k}}{a}}
} \left[e^{\frac{\pi\omega_{k}}{a}}~ ^{L}u_{k}~ 
^{L}u^{*}_{k}\right.\nonumber\\ 
~&& \left. ~~~~~+~~  
e^{-\frac{\pi\omega_{k}}{a}}~ 
^{L}u^{*}_{-k}~ 
^{L}u_{-k}\right]~,\nonumber\\
 \scalebox{0.9}{$G_{RL}^{+} (X_{2},X_{1})$} &=& \sum_{k=-\infty}^{\infty} 
\frac{1}{2\sinh{\frac{\pi\omega_{k}}{a}}
} \left[ ^{R}u_{k} 
^{L}u_{-k} +  ^{L}u^{*}_{-k}~ 
^{R}u^{*}_{k}\right.\nonumber\\ 
~&& \left. ~~~~~+~~  
 ^{L}u_{k} 
^{R}u_{-k} +  ^{R}u^{*}_{-k}~ 
^{L}u^{*}_{k}\right]~.
\end{eqnarray}
One can use the explicit expressions of the Rindler field modes $^{R}u_{k}$ and 
$^{L}u_{k}$ from Eq. (\ref{eq:Rindler-modes}) and obtain the expression of 
$\scalebox{0.9}{$G_{R}^{+} (X_{2},X_{1})$}$ same as given in Eq. 
(\ref{eq:Green-Unruh-der2}) and other expressions of $\scalebox{0.9}{$G_{L}^{+} 
(X_{2},X_{1})$}$ and $\scalebox{0.9}{$G_{RL}^{+} (X_{2},X_{1})$}$ in a similar 
manner. A $(1+3)$ dimensional representation of this Green's function can be 
obtained in a similar manner.

\section{Green's function of accelerated observer in thermal bath considering 
Rindler modes}\label{Apn:TU-Greens-fn}

Considering Gibbs ensemble average definition from Eq. 
(\ref{eq:Greens-fn-thermal1}), and Rindler mode decomposition of the scalar 
field from Eq. (\ref{eq:RRW-field-Unruh}) we obtain the Green's function of an 
accelerated observer in a thermal bath as
\begin{eqnarray}\label{eq:Green-TU-der1}
 \scalebox{0.9}{$G_{\beta_R}^{+} (X_{j,2},X_{l,1})$} &=& 
\sum_{k,k^{'}=-\infty}^{\infty} 
\frac{1}{2\sqrt{\sinh{\frac{\pi\omega_{k}}{a_{j}}}\sinh{\frac{\pi\omega_{k^{'}}} 
{ a_{l} } } } }\nonumber\\
~&& 
\left\langle\left[d^{1}_{k}d^{1^{\dagger}}_{k^{'}}e^{\frac{\pi}{2}
\left(\frac{\omega_{k}}{a_{j}}+\frac{\omega_{k^{'}}}{a_{l}}\right) } ~ 
^{R}u^{j}_{k}~ ^{R}u^{l~*}_{k^{'}}\right.\right.\nonumber\\
~&+&  d^{1^{\dagger}}_{k}d^{1}_{k^{'}}e^{\frac{\pi}{2}
\left(\frac{\omega_{k}}{a_{j}}+\frac{\omega_{k^{'}}}{a_{l}}\right) } ~ 
^{R}u^{j~*}_{k}~ ^{R}u^{l}_{k^{'}}\nonumber\\
~&+&  d^{2}_{k}d^{2^{\dagger}}_{k^{'}}e^{-\frac{\pi}{2}
\left(\frac{\omega_{k}}{a_{j}}+\frac{\omega_{k^{'}}}{a_{l}}\right) } ~ 
^{R}u^{j~*}_{-k}~ ^{R}u^{l}_{-k^{'}}\nonumber\\
~&+& \left.\left.
 d^{2^{\dagger}}_{k}d^{2}_{k^{'}}e^{-\frac{\pi}{2}
\left(\frac{\omega_{k}}{a_{j}}+\frac{\omega_{k^{'}}}{a_{l}}\right) } ~ 
^{R}u^{j}_{-k}~ ^{R}u^{l~*}_{-k^{'}}
\right]\right\rangle_{\beta}~.\nonumber\\
\end{eqnarray}
Here $\langle\hat{O}\rangle_{\beta}$ denotes the Gibbs ensemble average and the 
superscript $j$(or $l$) denotes the $j^{th}$(or $l^{th}$ detector) which 
corresponds to the second(or first) spacetime point. We mention that the 
observer is considered to be confined in the right Rindler wedge. Furthermore, 
we consider the Hamiltonian corresponding to the $k^{th}$ excitation to be 
$\scalebox{0.9}{$H_{k}=(d^{1^{\dagger}}_{k} d^{1}_{k} + d^{2^{\dagger}}_{k} 
d^{2}_{k}) \omega_{k}$}$. Then the Gibbs ensemble average of the operators 
$\scalebox{0.9}{$\langle d^{j}_{k} d^{j^{\dagger}}_{k^{'}} \rangle_{\beta} = 
\delta_{k,k^{'}} /(1-e^{-\beta\omega_{k}})$}$ and $\scalebox{0.9}{$\langle 
d^{j^{\dagger}}_{k} d^{j}_{k^{'}}\rangle_{\beta} = 
\delta_{k,k^{'}}/(e^{\beta\omega_{k}}-1)$}$. These results can be used along 
with the expression of the modes $^{R}u_{k}$ from Eq. (\ref{eq:Rindler-modes}) 
to obtain the desired expression of the Green's function from Eq. 
(\ref{eq:Greens-fn-TU}). Here also this evaluation is provided for $(1+1)$ 
dimensions. One can evaluate the Green's function corresponding to accelerated 
detectors in thermal bath for $(1+3)$ dimensions considering the Rindler modes 
as given in Eq. (\ref{eq:Greens-fn-TU-3D}) in a similar manner. However, in the 
later case the scalar field expansion in RRW and the expression of the Rindler 
modes are taken from Eq. (\ref{eq:RRW-field-Unruh}) and 
(\ref{eq:Rindler-modes-3D}).

Next we consider the Gibbs ensemble average definition from Eq. 
(\ref{eq:Greens-fn-thermal1}) to obtain the Green's function in a thermal bath 
taking the field decomposition given by Eq. (\ref{eq:Field-Unruh}), i.e., in 
terms of the Unruh modes and Unruh operators in $(1+1)$ dimensions. This Green's 
function looks like
\begin{eqnarray}\label{eq:GreenFn-TUmodes}
 \scalebox{0.88}{$G_{\beta_U}^{+} (X_{j,2},X_{l,1})$} &=& 
\scalebox{0.88}{$G_{\beta_R}^{+} (X_{j,2},X_{l,1}) + G_{\beta_L}^{+} 
(X_{j,2},X_{l,1})$} \nonumber\\
~&&~~~~~~~~~+~~ 
\scalebox{0.88}{$G_{\beta_{RL}}^{+} (X_{j,2},X_{l,1}) $}~.
\end{eqnarray}
Here the expressions $\scalebox{0.85}{$G_{\beta_R}^{+} (X_{j,2},X_{l,1})$}$ and 
$\scalebox{0.85}{$G_{\beta_L}^{+} (X_{j,2},X_{l,1})$}$ correspond to accelerated 
observers in thermal bath in the right and in left Rindler wedges respectively, 
and $\scalebox{0.85}{$G_{\beta_{RL}}^{+} (X_{j,2},X_{l,1})$}$ denotes the cross 
term. The expression of $\scalebox{0.85}{$G_{\beta_R}^{+} (X_{j,2},X_{l,1})$}$ 
can be obtained from Eq. (\ref{eq:Green-TU-der1}) and the other two quantities 
are given by
%
\begin{eqnarray}
&& \scalebox{0.9}{$G_{\beta_L}^{+} (X_{j,2},X_{l,1})$} = 
\sum_{k=-\infty}^{\infty} 
\frac{1}{2\sqrt{\sinh{\frac{\pi\omega_{k}}{a_{j}}}\sinh{\frac{\pi\omega_{k}}
{ a_{l} } }
} }\nonumber\\
~&& ~~~~~~~~~~~~~
\left[\frac{1}{1-e^{-\beta\omega_{k}}} \left\{e^{\frac{\pi\omega_{k}}{2}
\left(\frac{1}{a_{j}}+\frac{1}{a_{l}}\right) } ~ 
^{L}u^{j}_{k}~ ^{L}u^{l~*}_{k}\right.\right.\nonumber\\
~&& ~~~~~~~~~~~~~~+ \left. e^{-\frac{\pi\omega_{k}}{2}
\left(\frac{1}{a_{j}}+\frac{1}{a_{l}}\right) } ~ 
^{L}u^{j~*}_{-k}~ ^{L}u^{l}_{-k} \right\}\nonumber\\
~&& ~~~~~~~~~~~~~~ +
\frac{1}{e^{\beta\omega_{k}}-1} \left\{ 
e^{\frac{\pi\omega_{k}}{2}
\left(\frac{1}{a_{j}}+\frac{1}{a_{l}}\right) } ~ 
^{L}u^{j~*}_{k}~ ^{L}u^{l}_{k} \right.\nonumber\\
~&& ~~~~~~~~~~~~~~ +\left.\left.
 e^{-\frac{\pi\omega_{k}}{2}
\left(\frac{1}{a_{j}}+\frac{1}{a_{l}}\right) } ~ 
^{L}u^{j}_{-k}~ ^{L}u^{l~*}_{-k} \right\}
\right]~,\nonumber
\end{eqnarray}
\begin{eqnarray}\label{eq:GreenFn-TherUnruh-Diff}
 && \scalebox{0.9}{$G_{\beta_{RL}}^{+} (X_{j,2},X_{l,1})$} = 
\sum_{k=-\infty}^{\infty} 
\frac{1}{2\sqrt{\sinh{\frac{\pi\omega_{k}}{a_{j}}}\sinh{\frac{\pi\omega_{k}}
{ a_{l} } }
} }\nonumber\\
~&& ~~~~~~~~~~~~~
\left[\frac{1}{1-e^{-\beta\omega_{k}}} \left\{e^{\frac{\pi\omega_{k}}{2}
\left(\frac{1}{a_{j}}-\frac{1}{a_{l}}\right) } ~ 
^{R}u^{j}_{k}~ ^{L}u^{l}_{-k}\right.\right.\nonumber\\
~&& ~~~~~~~~~~~~~~+  e^{-\frac{\pi\omega_{k}}{2}
\left(\frac{1}{a_{j}}-\frac{1}{a_{l}}\right) } ~ 
^{L}u^{j~*}_{-k}~ ^{R}u^{l~*}_{k}\nonumber\\
~&& ~~~~~~~~~~~~~~+  e^{\frac{\pi\omega_{k}}{2}
\left(\frac{1}{a_{j}}-\frac{1}{a_{l}}\right) } ~ 
^{L}u^{j}_{k}~ ^{R}u^{l}_{-k}\nonumber\\
~&& ~~~~~~~~~~~~~~+ \left. e^{-\frac{\pi\omega_{k}}{2}
\left(\frac{1}{a_{j}}-\frac{1}{a_{l}}\right) } ~ 
^{R}u^{j~*}_{-k}~ ^{L}u^{l~*}_{k} \right\}\nonumber\\
~&& ~~~~~~~~~~~~~~ +
\frac{1}{e^{\beta\omega_{k}}-1} \left\{ 
e^{\frac{\pi\omega_{k}}{2}
\left(\frac{1}{a_{j}}-\frac{1}{a_{l}}\right) } ~ 
^{R}u^{j~*}_{k}~ ^{L}u^{l~*}_{-k} \right.\nonumber\\
~&& ~~~~~~~~~~~~~~ +
 e^{-\frac{\pi\omega_{k}}{2}\left(\frac{1}{a_{j}}-\frac{1}{a_{l}}\right) } ~ 
^{L}u^{j}_{-k}~ ^{R}u^{l}_{k}\nonumber\\
~&& ~~~~~~~~~~~~~~ +
 e^{\frac{\pi\omega_{k}}{2} \left(\frac{1}{a_{j}}-\frac{1}{a_{l}}\right) } ~ 
^{L}u^{j~*}_{k}~ ^{R}u^{l~*}_{-k}\nonumber\\
~&& ~~~~~~~~~~~~~~ +\left.\left.
 e^{-\frac{\pi\omega_{k}}{2}\left(\frac{1}{a_{j}}-\frac{1}{a_{l}}\right) } ~ 
^{R}u^{j}_{-k}~ ^{L}u^{l}_{k} \right\}
\right]~.
\end{eqnarray}
One can use the explicit expressions of the Rindler field modes $^{R}u_{k}$ and 
$^{L}u_{k}$ from Eq. (\ref{eq:Rindler-modes}) and further express the quantities 
$\scalebox{0.9}{$ G_{\beta_R}^{+} (X_{2},X_{1})$}$, 
$\scalebox{0.9}{$G_{\beta_L}^{+} (X_{2},X_{1})$}$, and 
$\scalebox{0.9}{$G_{\beta_{RL}}^{+} (X_{2},X_{1})$}$ in terms of Rindler 
coordinates. Here also a $(1+3)$ dimensional representation of this Green's 
function can be provided in a similar manner.

\section{Relation between the detector proper times}\label{Apn:rel-proper-time}

We take the Minkowski to Rindler coordinate transformation from Eq. 
(\ref{eq:Rindler1-trans}) to express the two accelerated observers. In 
particular the coordinate transformation corresponding to our first accelerated 
observer is
\begin{eqnarray}\label{eq:Rindler1-trans-trel1}
 T_{1} &=& \frac{e^{a_{1}\xi_{1}}}{a_{1}} \sinh{a_{1}\eta_{1}}\nonumber\\
 X_{1} &=& \frac{e^{a_{1}\xi_{1}}}{a_{1}} \cosh{a_{1}\eta_{1}}~,
\end{eqnarray}
and a similar coordinate transformation corresponding to the second 
accelerated observer is
\begin{eqnarray}\label{eq:Rindler1-trans-trel2}
 T_{2} &=& \frac{e^{a_{2}\xi_{2}}}{a_{2}} \sinh{a_{2}\eta_{2}}\nonumber\\
 X_{2} &=& \frac{e^{a_{2}\xi}}{a_{2}} \cosh{a_{2}\eta_{2}}~.
\end{eqnarray}
%
\begin{figure}[h]
\centering
\includegraphics[width=0.9\linewidth]{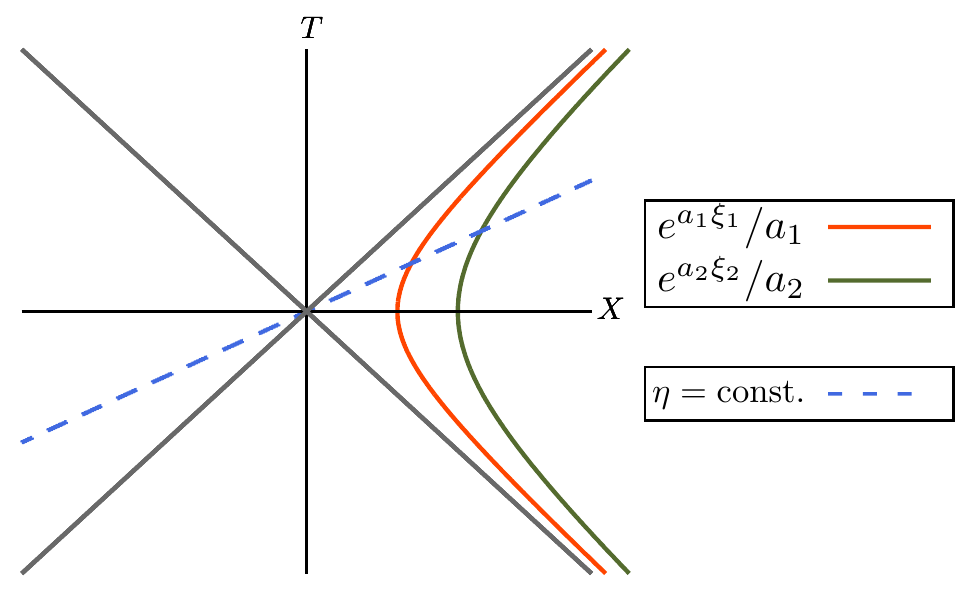}
 \caption{Here the trajectories of two observers with two different 
proper accelerations are depicted in the right Rindler wedge.}
 \label{fig:Accelerated-observers}
\end{figure}
From these coordinate transformations one can observe that $\eta_{j} = (1/a_{j}) 
\tanh^{-1}(T_{j}/X_{j})$ and ${e^{a_{j}\xi_{j}}}/{a_{j}} = 
(X_{j}^2-T_{j}^2)^{-1/2}$. It means constant Rindler times signify straight 
lines in the Minkowski $T-X$ plane, see Fig. \ref{fig:Accelerated-observers}. On 
the other hand observers with constant proper accelerations 
$b_{j}=a_{j}e^{-a_{j}\xi_{j}}$ follow the hyperbolic trajectories, also depicted 
in the Fig. \ref{fig:Accelerated-observers}. Now the scenario of constant proper 
acceleration can be achieved in a few different ways. One can take two 
accelerated observers with the same Rindler parameter $a_{1}=a_{2}$ and keep 
them in two different Rindler spatial points $\xi_{1}\neq\xi_{2}$. On the other, 
hand one can also take the $a_{1}\neq a_{2}$ from the beginning with the 
consideration of $\xi_{1}=\xi_{2}$. In both of the cases the observers have 
different proper acceleration, i.e., they signify two different hyperbolic 
trajectories in the Minkowski $T-X$ plane. Interestingly in both of the cases 
these trajectories can be cut by a single $\eta=const.$ line. It is noticed that 
if one considers both of the observers to be described by equal Rindler time 
$\eta$ then a relation between the proper times of the two different observers 
can be obtained.

\bibliographystyle{apsrev}

\bibliography{bibtexfile}

\begin{thebibliography}{64}
\expandafter\ifx\csname natexlab\endcsname\relax\def\natexlab#1{#1}\fi
\expandafter\ifx\csname bibnamefont\endcsname\relax
  \def\bibnamefont#1{#1}\fi
\expandafter\ifx\csname bibfnamefont\endcsname\relax
  \def\bibfnamefont#1{#1}\fi
\expandafter\ifx\csname citenamefont\endcsname\relax
  \def\citenamefont#1{#1}\fi
\expandafter\ifx\csname url\endcsname\relax
  \def\url#1{\texttt{#1}}\fi
\expandafter\ifx\csname urlprefix\endcsname\relax\def\urlprefix{URL }\fi
\providecommand{\bibinfo}[2]{#2}
\providecommand{\eprint}[2][]{\url{#2}}

\bibitem[{\citenamefont{Kirby and Franson}(2013)}]{PhysRevA.87.053822}
\bibinfo{author}{\bibfnamefont{B.~T.} \bibnamefont{Kirby}} \bibnamefont{and}
  \bibinfo{author}{\bibfnamefont{J.~D.} \bibnamefont{Franson}},
  \bibinfo{journal}{Phys. Rev. A} \textbf{\bibinfo{volume}{87}},
  \bibinfo{pages}{053822} (\bibinfo{year}{2013}).

\bibitem[{\citenamefont{Hensen et~al.}(2015)}]{Hensen:2015ccp}
\bibinfo{author}{\bibfnamefont{B.}~\bibnamefont{Hensen}} \bibnamefont{et~al.},
  \bibinfo{journal}{Nature} \textbf{\bibinfo{volume}{526}},
  \bibinfo{pages}{682} (\bibinfo{year}{2015}), \eprint{arXiv:1508.05949}.

\bibitem[{\citenamefont{Tittel et~al.}(1998)\citenamefont{Tittel, Brendel,
  Zbinden, and Gisin}}]{Tittel:1998ja}
\bibinfo{author}{\bibfnamefont{W.}~\bibnamefont{Tittel}},
  \bibinfo{author}{\bibfnamefont{J.}~\bibnamefont{Brendel}},
  \bibinfo{author}{\bibfnamefont{H.}~\bibnamefont{Zbinden}}, \bibnamefont{and}
  \bibinfo{author}{\bibfnamefont{N.}~\bibnamefont{Gisin}},
  \bibinfo{journal}{Phys. Rev. Lett.} \textbf{\bibinfo{volume}{81}},
  \bibinfo{pages}{3563} (\bibinfo{year}{1998}),
  \eprint{arXiv:quant-ph/9806043}.

\bibitem[{Sal()}]{Salart-2008}
\bibinfo{note}{Salart, D., Baas, A., Branciard, C. et al. Testing the speed of
  ‘spooky action at a distance’. Nature 454, 861–864 (2008).}

\bibitem[{\citenamefont{Fuentes-Schuller and
  Mann}(2005)}]{FuentesSchuller:2004xp}
\bibinfo{author}{\bibfnamefont{I.}~\bibnamefont{Fuentes-Schuller}}
  \bibnamefont{and} \bibinfo{author}{\bibfnamefont{R.~B.} \bibnamefont{Mann}},
  \bibinfo{journal}{Phys. Rev. Lett.} \textbf{\bibinfo{volume}{95}},
  \bibinfo{pages}{120404} (\bibinfo{year}{2005}),
  \eprint{arXiv:quant-ph/0410172}.

\bibitem[{\citenamefont{Reznik}(2003)}]{Reznik:2002fz}
\bibinfo{author}{\bibfnamefont{B.}~\bibnamefont{Reznik}},
  \bibinfo{journal}{Found. Phys.} \textbf{\bibinfo{volume}{33}},
  \bibinfo{pages}{167} (\bibinfo{year}{2003}), \eprint{arXiv:quant-ph/0212044}.

\bibitem[{\citenamefont{Lin and Hu}(2010)}]{Lin:2010zzb}
\bibinfo{author}{\bibfnamefont{S.-Y.} \bibnamefont{Lin}} \bibnamefont{and}
  \bibinfo{author}{\bibfnamefont{B.}~\bibnamefont{Hu}}, \bibinfo{journal}{Phys.
  Rev. D} \textbf{\bibinfo{volume}{81}}, \bibinfo{pages}{045019}
  (\bibinfo{year}{2010}), \eprint{arXiv:0910.5858}.

\bibitem[{\citenamefont{Ball et~al.}(2006)\citenamefont{Ball, Fuentes-Schuller,
  and Schuller}}]{Ball:2005xa}
\bibinfo{author}{\bibfnamefont{J.~L.} \bibnamefont{Ball}},
  \bibinfo{author}{\bibfnamefont{I.}~\bibnamefont{Fuentes-Schuller}},
  \bibnamefont{and} \bibinfo{author}{\bibfnamefont{F.~P.}
  \bibnamefont{Schuller}}, \bibinfo{journal}{Phys. Lett. A}
  \textbf{\bibinfo{volume}{359}}, \bibinfo{pages}{550} (\bibinfo{year}{2006}),
  \eprint{arXiv:quant-ph/0506113}.

\bibitem[{\citenamefont{Cliche and Kempf}(2010)}]{Cliche:2009fma}
\bibinfo{author}{\bibfnamefont{M.}~\bibnamefont{Cliche}} \bibnamefont{and}
  \bibinfo{author}{\bibfnamefont{A.}~\bibnamefont{Kempf}},
  \bibinfo{journal}{Phys. Rev. A} \textbf{\bibinfo{volume}{81}},
  \bibinfo{pages}{012330} (\bibinfo{year}{2010}), \eprint{arXiv:0908.3144}.

\bibitem[{\citenamefont{Martin-Martinez and
  Menicucci}(2012)}]{MartinMartinez:2012sg}
\bibinfo{author}{\bibfnamefont{E.}~\bibnamefont{Martin-Martinez}}
  \bibnamefont{and} \bibinfo{author}{\bibfnamefont{N.~C.}
  \bibnamefont{Menicucci}}, \bibinfo{journal}{Class. Quant. Grav.}
  \textbf{\bibinfo{volume}{29}}, \bibinfo{pages}{224003}
  (\bibinfo{year}{2012}), \eprint{arXiv:1204.4918}.

\bibitem[{\citenamefont{Salton et~al.}(2015)\citenamefont{Salton, Mann, and
  Menicucci}}]{Salton:2014jaa}
\bibinfo{author}{\bibfnamefont{G.}~\bibnamefont{Salton}},
  \bibinfo{author}{\bibfnamefont{R.~B.} \bibnamefont{Mann}}, \bibnamefont{and}
  \bibinfo{author}{\bibfnamefont{N.~C.} \bibnamefont{Menicucci}},
  \bibinfo{journal}{New J. Phys.} \textbf{\bibinfo{volume}{17}},
  \bibinfo{pages}{035001} (\bibinfo{year}{2015}), \eprint{arXiv:1408.1395}.

\bibitem[{\citenamefont{Martin-Martinez
  et~al.}(2016)\citenamefont{Martin-Martinez, Smith, and
  Terno}}]{Martin-Martinez:2015qwa}
\bibinfo{author}{\bibfnamefont{E.}~\bibnamefont{Martin-Martinez}},
  \bibinfo{author}{\bibfnamefont{A.~R.~H.} \bibnamefont{Smith}},
  \bibnamefont{and} \bibinfo{author}{\bibfnamefont{D.~R.} \bibnamefont{Terno}},
  \bibinfo{journal}{Phys. Rev. D} \textbf{\bibinfo{volume}{93}},
  \bibinfo{pages}{044001} (\bibinfo{year}{2016}), \eprint{arXiv:1507.02688}.

\bibitem[{\citenamefont{Cai and Ren}(2018{\natexlab{a}})}]{Cai:2018xuo}
\bibinfo{author}{\bibfnamefont{H.}~\bibnamefont{Cai}} \bibnamefont{and}
  \bibinfo{author}{\bibfnamefont{Z.}~\bibnamefont{Ren}}, \bibinfo{journal}{Sci.
  Rep.} \textbf{\bibinfo{volume}{8}}, \bibinfo{pages}{11802}
  (\bibinfo{year}{2018}{\natexlab{a}}).

\bibitem[{\citenamefont{Menezes}(2018)}]{Menezes:2017oeb}
\bibinfo{author}{\bibfnamefont{G.}~\bibnamefont{Menezes}},
  \bibinfo{journal}{Phys. Rev.} \textbf{\bibinfo{volume}{D97}},
  \bibinfo{pages}{085021} (\bibinfo{year}{2018}), \eprint{arXiv:1712.07151}.

\bibitem[{\citenamefont{Menezes et~al.}(2017)\citenamefont{Menezes, Svaiter,
  and Zarro}}]{Menezes:2017rby}
\bibinfo{author}{\bibfnamefont{G.}~\bibnamefont{Menezes}},
  \bibinfo{author}{\bibfnamefont{N.}~\bibnamefont{Svaiter}}, \bibnamefont{and}
  \bibinfo{author}{\bibfnamefont{C.}~\bibnamefont{Zarro}},
  \bibinfo{journal}{Phys. Rev. A} \textbf{\bibinfo{volume}{96}},
  \bibinfo{pages}{062119} (\bibinfo{year}{2017}), \eprint{arXiv:1709.08702}.

\bibitem[{\citenamefont{Zhou and Yu}(2017)}]{Zhou:2017axh}
\bibinfo{author}{\bibfnamefont{W.}~\bibnamefont{Zhou}} \bibnamefont{and}
  \bibinfo{author}{\bibfnamefont{H.}~\bibnamefont{Yu}}, \bibinfo{journal}{Phys.
  Rev. D} \textbf{\bibinfo{volume}{96}}, \bibinfo{pages}{045018}
  (\bibinfo{year}{2017}).

\bibitem[{\citenamefont{Henderson et~al.}(2018)\citenamefont{Henderson,
  Hennigar, Mann, Smith, and Zhang}}]{Henderson:2017yuv}
\bibinfo{author}{\bibfnamefont{L.~J.} \bibnamefont{Henderson}},
  \bibinfo{author}{\bibfnamefont{R.~A.} \bibnamefont{Hennigar}},
  \bibinfo{author}{\bibfnamefont{R.~B.} \bibnamefont{Mann}},
  \bibinfo{author}{\bibfnamefont{A.~R.} \bibnamefont{Smith}}, \bibnamefont{and}
  \bibinfo{author}{\bibfnamefont{J.}~\bibnamefont{Zhang}},
  \bibinfo{journal}{Class. Quant. Grav.} \textbf{\bibinfo{volume}{35}},
  \bibinfo{pages}{21LT02} (\bibinfo{year}{2018}), \eprint{arXiv:1712.10018}.

\bibitem[{\citenamefont{Henderson and Menicucci}(2020)}]{Henderson:2020ucx}
\bibinfo{author}{\bibfnamefont{L.~J.} \bibnamefont{Henderson}}
  \bibnamefont{and} \bibinfo{author}{\bibfnamefont{N.~C.}
  \bibnamefont{Menicucci}}, \bibinfo{journal}{Phys. Rev. D}
  \textbf{\bibinfo{volume}{102}}, \bibinfo{pages}{125026}
  (\bibinfo{year}{2020}), \eprint{arXiv:2005.05330}.

\bibitem[{\citenamefont{Stritzelberger
  et~al.}(2020)\citenamefont{Stritzelberger, Henderson, Baccetti, Menicucci,
  and Kempf}}]{Stritzelberger:2020hde}
\bibinfo{author}{\bibfnamefont{N.}~\bibnamefont{Stritzelberger}},
  \bibinfo{author}{\bibfnamefont{L.~J.} \bibnamefont{Henderson}},
  \bibinfo{author}{\bibfnamefont{V.}~\bibnamefont{Baccetti}},
  \bibinfo{author}{\bibfnamefont{N.~C.} \bibnamefont{Menicucci}},
  \bibnamefont{and} \bibinfo{author}{\bibfnamefont{A.}~\bibnamefont{Kempf}}
  (\bibinfo{year}{2020}), \eprint{arXiv:2006.11291}.

\bibitem[{\citenamefont{Rodr\'\i{}guez-Camargo
  et~al.}(2018)\citenamefont{Rodr\'\i{}guez-Camargo, Svaiter, and
  Menezes}}]{Rodriguez-Camargo:2016fbq}
\bibinfo{author}{\bibfnamefont{C.}~\bibnamefont{Rodr\'\i{}guez-Camargo}},
  \bibinfo{author}{\bibfnamefont{N.}~\bibnamefont{Svaiter}}, \bibnamefont{and}
  \bibinfo{author}{\bibfnamefont{G.}~\bibnamefont{Menezes}},
  \bibinfo{journal}{Annals Phys.} \textbf{\bibinfo{volume}{396}},
  \bibinfo{pages}{266} (\bibinfo{year}{2018}), \eprint{arXiv:1608.03365}.

\bibitem[{\citenamefont{Menezes and Svaiter}(2016)}]{Menezes:2015iva}
\bibinfo{author}{\bibfnamefont{G.}~\bibnamefont{Menezes}} \bibnamefont{and}
  \bibinfo{author}{\bibfnamefont{N.}~\bibnamefont{Svaiter}},
  \bibinfo{journal}{Phys. Rev. A} \textbf{\bibinfo{volume}{93}},
  \bibinfo{pages}{052117} (\bibinfo{year}{2016}), \eprint{arXiv:1512.02886}.

\bibitem[{\citenamefont{Hu and Yu}(2015)}]{Hu:2015lda}
\bibinfo{author}{\bibfnamefont{J.}~\bibnamefont{Hu}} \bibnamefont{and}
  \bibinfo{author}{\bibfnamefont{H.}~\bibnamefont{Yu}}, \bibinfo{journal}{Phys.
  Rev. A} \textbf{\bibinfo{volume}{91}}, \bibinfo{pages}{012327}
  (\bibinfo{year}{2015}), \eprint{arXiv:1501.03321}.

\bibitem[{\citenamefont{Rizzuto et~al.}(2016)\citenamefont{Rizzuto, Lattuca,
  Marino, Noto, Spagnolo, Passante, and Zhou}}]{Rizzuto:2016ijj}
\bibinfo{author}{\bibfnamefont{L.}~\bibnamefont{Rizzuto}},
  \bibinfo{author}{\bibfnamefont{M.}~\bibnamefont{Lattuca}},
  \bibinfo{author}{\bibfnamefont{J.}~\bibnamefont{Marino}},
  \bibinfo{author}{\bibfnamefont{A.}~\bibnamefont{Noto}},
  \bibinfo{author}{\bibfnamefont{S.}~\bibnamefont{Spagnolo}},
  \bibinfo{author}{\bibfnamefont{R.}~\bibnamefont{Passante}}, \bibnamefont{and}
  \bibinfo{author}{\bibfnamefont{W.}~\bibnamefont{Zhou}},
  \bibinfo{journal}{Phys. Rev. A} \textbf{\bibinfo{volume}{94}},
  \bibinfo{pages}{012121} (\bibinfo{year}{2016}), \eprint{arXiv:1601.04502}.

\bibitem[{\citenamefont{Arias et~al.}(2016)\citenamefont{Arias, Due\~nas,
  Menezes, and Svaiter}}]{Arias:2015moa}
\bibinfo{author}{\bibfnamefont{E.}~\bibnamefont{Arias}},
  \bibinfo{author}{\bibfnamefont{J.}~\bibnamefont{Due\~nas}},
  \bibinfo{author}{\bibfnamefont{G.}~\bibnamefont{Menezes}}, \bibnamefont{and}
  \bibinfo{author}{\bibfnamefont{N.}~\bibnamefont{Svaiter}},
  \bibinfo{journal}{JHEP} \textbf{\bibinfo{volume}{07}}, \bibinfo{pages}{147}
  (\bibinfo{year}{2016}), \eprint{arXiv:1510.00047}.

\bibitem[{\citenamefont{Pican\c{c}o et~al.}(2020)\citenamefont{Pican\c{c}o,
  Svaiter, and Zarro}}]{Costa:2020aqa}
\bibinfo{author}{\bibfnamefont{G.}~\bibnamefont{Pican\c{c}o}},
  \bibinfo{author}{\bibfnamefont{N.~F.} \bibnamefont{Svaiter}},
  \bibnamefont{and} \bibinfo{author}{\bibfnamefont{C.~A.} \bibnamefont{Zarro}},
  \bibinfo{journal}{JHEP} \textbf{\bibinfo{volume}{08}}, \bibinfo{pages}{025}
  (\bibinfo{year}{2020}), \eprint{arXiv:2002.06085}.

\bibitem[{\citenamefont{Zhou and Yu}(2020)}]{Zhou:2020oqa}
\bibinfo{author}{\bibfnamefont{W.}~\bibnamefont{Zhou}} \bibnamefont{and}
  \bibinfo{author}{\bibfnamefont{H.}~\bibnamefont{Yu}}, \bibinfo{journal}{Phys.
  Rev. D} \textbf{\bibinfo{volume}{101}}, \bibinfo{pages}{025009}
  (\bibinfo{year}{2020}), \eprint{arXiv:2001.00750}.

\bibitem[{\citenamefont{Cai and Ren}(2019)}]{Cai:2019pnw}
\bibinfo{author}{\bibfnamefont{H.}~\bibnamefont{Cai}} \bibnamefont{and}
  \bibinfo{author}{\bibfnamefont{Z.}~\bibnamefont{Ren}},
  \bibinfo{journal}{Class. Quant. Grav.} \textbf{\bibinfo{volume}{36}},
  \bibinfo{pages}{165001} (\bibinfo{year}{2019}).

\bibitem[{\citenamefont{Lima et~al.}(2019)\citenamefont{Lima, Lima, and
  Lyra}}]{Lima:2019mbt}
\bibinfo{author}{\bibfnamefont{F.}~\bibnamefont{Lima}},
  \bibinfo{author}{\bibfnamefont{R.}~\bibnamefont{Lima}}, \bibnamefont{and}
  \bibinfo{author}{\bibfnamefont{M.}~\bibnamefont{Lyra}},
  \bibinfo{journal}{Braz. J. Phys.} \textbf{\bibinfo{volume}{49}},
  \bibinfo{pages}{423} (\bibinfo{year}{2019}).

\bibitem[{\citenamefont{Liu et~al.}(2018)\citenamefont{Liu, Tian, Wang, and
  Jing}}]{Liu:2018zod}
\bibinfo{author}{\bibfnamefont{X.}~\bibnamefont{Liu}},
  \bibinfo{author}{\bibfnamefont{Z.}~\bibnamefont{Tian}},
  \bibinfo{author}{\bibfnamefont{J.}~\bibnamefont{Wang}}, \bibnamefont{and}
  \bibinfo{author}{\bibfnamefont{J.}~\bibnamefont{Jing}},
  \bibinfo{journal}{Phys. Rev. D} \textbf{\bibinfo{volume}{97}},
  \bibinfo{pages}{105030} (\bibinfo{year}{2018}), \eprint{arXiv:1805.04470}.

\bibitem[{\citenamefont{Cai and Ren}(2018{\natexlab{b}})}]{Cai:2017jan}
\bibinfo{author}{\bibfnamefont{H.}~\bibnamefont{Cai}} \bibnamefont{and}
  \bibinfo{author}{\bibfnamefont{Z.}~\bibnamefont{Ren}},
  \bibinfo{journal}{Class. Quant. Grav.} \textbf{\bibinfo{volume}{35}},
  \bibinfo{pages}{025016} (\bibinfo{year}{2018}{\natexlab{b}}).

\bibitem[{\citenamefont{Zhou et~al.}(2018)\citenamefont{Zhou, Rizzuto, and
  Passante}}]{Zhou:2017fmu}
\bibinfo{author}{\bibfnamefont{W.}~\bibnamefont{Zhou}},
  \bibinfo{author}{\bibfnamefont{L.}~\bibnamefont{Rizzuto}}, \bibnamefont{and}
  \bibinfo{author}{\bibfnamefont{R.}~\bibnamefont{Passante}},
  \bibinfo{journal}{Phys. Rev. A} \textbf{\bibinfo{volume}{97}},
  \bibinfo{pages}{042503} (\bibinfo{year}{2018}), \eprint{arXiv:1711.08249}.

\bibitem[{\citenamefont{Menezes}(2016)}]{Menezes:2015veo}
\bibinfo{author}{\bibfnamefont{G.}~\bibnamefont{Menezes}},
  \bibinfo{journal}{Phys. Rev.} \textbf{\bibinfo{volume}{D94}},
  \bibinfo{pages}{105008} (\bibinfo{year}{2016}), \eprint{arXiv:1512.03636}.

\bibitem[{\citenamefont{Flores-Hidalgo
  et~al.}(2015)\citenamefont{Flores-Hidalgo, Rojas, and
  Rojas}}]{Flores-Hidalgo:2015urj}
\bibinfo{author}{\bibfnamefont{G.}~\bibnamefont{Flores-Hidalgo}},
  \bibinfo{author}{\bibfnamefont{M.}~\bibnamefont{Rojas}}, \bibnamefont{and}
  \bibinfo{author}{\bibfnamefont{O.}~\bibnamefont{Rojas}}
  (\bibinfo{year}{2015}), \eprint{arXiv:1511.01416}.

\bibitem[{\citenamefont{Menezes and Svaiter}(2015)}]{Menezes:2015uaa}
\bibinfo{author}{\bibfnamefont{G.}~\bibnamefont{Menezes}} \bibnamefont{and}
  \bibinfo{author}{\bibfnamefont{N.}~\bibnamefont{Svaiter}},
  \bibinfo{journal}{Phys. Rev. A} \textbf{\bibinfo{volume}{92}},
  \bibinfo{pages}{062131} (\bibinfo{year}{2015}), \eprint{arXiv:1508.04513}.

\bibitem[{\citenamefont{Zhou et~al.}(2016)\citenamefont{Zhou, Passante, and
  Rizzuto}}]{Zhou:2016urt}
\bibinfo{author}{\bibfnamefont{W.}~\bibnamefont{Zhou}},
  \bibinfo{author}{\bibfnamefont{R.}~\bibnamefont{Passante}}, \bibnamefont{and}
  \bibinfo{author}{\bibfnamefont{L.}~\bibnamefont{Rizzuto}},
  \bibinfo{journal}{Phys. Rev. D} \textbf{\bibinfo{volume}{94}},
  \bibinfo{pages}{105025} (\bibinfo{year}{2016}), \eprint{arXiv:1609.06931}.

\bibitem[{\citenamefont{Unruh}(1976)}]{Unruh:1976db}
\bibinfo{author}{\bibfnamefont{W.}~\bibnamefont{Unruh}},
  \bibinfo{journal}{Phys.Rev.} \textbf{\bibinfo{volume}{D14}},
  \bibinfo{pages}{870} (\bibinfo{year}{1976}).

\bibitem[{\citenamefont{Unruh and Wald}(1984)}]{Unruh:1983ms}
\bibinfo{author}{\bibfnamefont{W.~G.} \bibnamefont{Unruh}} \bibnamefont{and}
  \bibinfo{author}{\bibfnamefont{R.~M.} \bibnamefont{Wald}},
  \bibinfo{journal}{Phys. Rev.} \textbf{\bibinfo{volume}{D29}},
  \bibinfo{pages}{1047} (\bibinfo{year}{1984}).

\bibitem[{\citenamefont{Scully et~al.}(2018)\citenamefont{Scully, Fulling, Lee,
  Page, Schleich, and Svidzinsky}}]{Scully:2017utk}
\bibinfo{author}{\bibfnamefont{M.~O.} \bibnamefont{Scully}},
  \bibinfo{author}{\bibfnamefont{S.}~\bibnamefont{Fulling}},
  \bibinfo{author}{\bibfnamefont{D.}~\bibnamefont{Lee}},
  \bibinfo{author}{\bibfnamefont{D.~N.} \bibnamefont{Page}},
  \bibinfo{author}{\bibfnamefont{W.}~\bibnamefont{Schleich}}, \bibnamefont{and}
  \bibinfo{author}{\bibfnamefont{A.}~\bibnamefont{Svidzinsky}},
  \bibinfo{journal}{Proc. Nat. Acad. Sci.} \textbf{\bibinfo{volume}{115}},
  \bibinfo{pages}{8131} (\bibinfo{year}{2018}), \eprint{arXiv:1709.00481}.

\bibitem[{\citenamefont{Chakraborty and Majhi}(2019)}]{Chakraborty:2019ltu}
\bibinfo{author}{\bibfnamefont{K.}~\bibnamefont{Chakraborty}} \bibnamefont{and}
  \bibinfo{author}{\bibfnamefont{B.~R.} \bibnamefont{Majhi}},
  \bibinfo{journal}{Phys. Rev. D} \textbf{\bibinfo{volume}{100}},
  \bibinfo{pages}{045004} (\bibinfo{year}{2019}), \eprint{arXiv:1905.10554}.

\bibitem[{\citenamefont{Majhi}(2020)}]{Majhi:2020pps}
\bibinfo{author}{\bibfnamefont{B.~R.} \bibnamefont{Majhi}},
  \bibinfo{journal}{Phys. Lett. B} \textbf{\bibinfo{volume}{808}},
  \bibinfo{pages}{135640} (\bibinfo{year}{2020}), \eprint{arXiv:2006.04486}.

\bibitem[{\citenamefont{Comp\`ere et~al.}(2019)\citenamefont{Comp\`ere, Long,
  and Riegler}}]{Compere:2019rof}
\bibinfo{author}{\bibfnamefont{G.}~\bibnamefont{Comp\`ere}},
  \bibinfo{author}{\bibfnamefont{J.}~\bibnamefont{Long}}, \bibnamefont{and}
  \bibinfo{author}{\bibfnamefont{M.}~\bibnamefont{Riegler}},
  \bibinfo{journal}{JHEP} \textbf{\bibinfo{volume}{05}}, \bibinfo{pages}{053}
  (\bibinfo{year}{2019}), \eprint{arXiv:1903.01812}.

\bibitem[{\citenamefont{Plenio et~al.}(1999)\citenamefont{Plenio, Huelga,
  Beige, and Knight}}]{Plenio:1998wq}
\bibinfo{author}{\bibfnamefont{M.}~\bibnamefont{Plenio}},
  \bibinfo{author}{\bibfnamefont{S.}~\bibnamefont{Huelga}},
  \bibinfo{author}{\bibfnamefont{A.}~\bibnamefont{Beige}}, \bibnamefont{and}
  \bibinfo{author}{\bibfnamefont{P.}~\bibnamefont{Knight}},
  \bibinfo{journal}{Phys. Rev. A} \textbf{\bibinfo{volume}{59}},
  \bibinfo{pages}{2468} (\bibinfo{year}{1999}),
  \eprint{arXiv:quant-ph/9811003}.

\bibitem[{\citenamefont{Ficek and Tana}(2003)}]{Ficek2003}
\bibinfo{author}{\bibfnamefont{Z.}~\bibnamefont{Ficek}} \bibnamefont{and}
  \bibinfo{author}{\bibfnamefont{R.}~\bibnamefont{Tana}},
  \bibinfo{journal}{Journal of Modern Optics} \textbf{\bibinfo{volume}{50}},
  \bibinfo{pages}{2765} (\bibinfo{year}{2003}).

\bibitem[{\citenamefont{Tana and Ficek}(2004)}]{Tana_2004}
\bibinfo{author}{\bibfnamefont{R.}~\bibnamefont{Tana}} \bibnamefont{and}
  \bibinfo{author}{\bibfnamefont{Z.}~\bibnamefont{Ficek}},
  \bibinfo{journal}{Journal of Optics B: Quantum and Semiclassical Optics}
  \textbf{\bibinfo{volume}{6}}, \bibinfo{pages}{S90} (\bibinfo{year}{2004}).

\bibitem[{\citenamefont{Costa and Matsas}(1995)}]{Costa:1994yx}
\bibinfo{author}{\bibfnamefont{S.~S.} \bibnamefont{Costa}} \bibnamefont{and}
  \bibinfo{author}{\bibfnamefont{G.~E.~A.} \bibnamefont{Matsas}},
  \bibinfo{journal}{Phys. Rev. D} \textbf{\bibinfo{volume}{52}},
  \bibinfo{pages}{3466} (\bibinfo{year}{1995}), \eprint{arXiv:gr-qc/9412030}.

\bibitem[{\citenamefont{Kolekar and Padmanabhan}(2014)}]{Kolekar:2013hra}
\bibinfo{author}{\bibfnamefont{S.}~\bibnamefont{Kolekar}} \bibnamefont{and}
  \bibinfo{author}{\bibfnamefont{T.}~\bibnamefont{Padmanabhan}},
  \bibinfo{journal}{Phys. Rev. D} \textbf{\bibinfo{volume}{89}},
  \bibinfo{pages}{064055} (\bibinfo{year}{2014}), \eprint{arXiv:1309.4424}.

\bibitem[{\citenamefont{Hodgkinson et~al.}(2014)\citenamefont{Hodgkinson,
  Louko, and Ottewill}}]{Hodgkinson:2014iua}
\bibinfo{author}{\bibfnamefont{L.}~\bibnamefont{Hodgkinson}},
  \bibinfo{author}{\bibfnamefont{J.}~\bibnamefont{Louko}}, \bibnamefont{and}
  \bibinfo{author}{\bibfnamefont{A.~C.} \bibnamefont{Ottewill}},
  \bibinfo{journal}{Phys. Rev. D} \textbf{\bibinfo{volume}{89}},
  \bibinfo{pages}{104002} (\bibinfo{year}{2014}), \eprint{arXiv:1401.2667}.

\bibitem[{\citenamefont{Brenna et~al.}(2016)\citenamefont{Brenna, Mann, and
  Martin-Martinez}}]{Brenna:2015fga}
\bibinfo{author}{\bibfnamefont{W.~G.} \bibnamefont{Brenna}},
  \bibinfo{author}{\bibfnamefont{R.~B.} \bibnamefont{Mann}}, \bibnamefont{and}
  \bibinfo{author}{\bibfnamefont{E.}~\bibnamefont{Martin-Martinez}},
  \bibinfo{journal}{Phys. Lett. B} \textbf{\bibinfo{volume}{757}},
  \bibinfo{pages}{307} (\bibinfo{year}{2016}), \eprint{arXiv:1504.02468}.

\bibitem[{\citenamefont{Garay et~al.}(2016)\citenamefont{Garay,
  Martin-Martinez, and de~Ramon}}]{Garay:2016cpf}
\bibinfo{author}{\bibfnamefont{L.~J.} \bibnamefont{Garay}},
  \bibinfo{author}{\bibfnamefont{E.}~\bibnamefont{Martin-Martinez}},
  \bibnamefont{and} \bibinfo{author}{\bibfnamefont{J.}~\bibnamefont{de~Ramon}},
  \bibinfo{journal}{Phys. Rev. D} \textbf{\bibinfo{volume}{94}},
  \bibinfo{pages}{104048} (\bibinfo{year}{2016}), \eprint{arXiv:1607.05287}.

\bibitem[{\citenamefont{Kolekar}(2014)}]{Kolekar:2013xua}
\bibinfo{author}{\bibfnamefont{S.}~\bibnamefont{Kolekar}},
  \bibinfo{journal}{Phys. Rev. D} \textbf{\bibinfo{volume}{89}},
  \bibinfo{pages}{044036} (\bibinfo{year}{2014}), \eprint{arXiv:1309.3261}.

\bibitem[{\citenamefont{Kolekar and Padmanabhan}(2015)}]{Kolekar:2013aka}
\bibinfo{author}{\bibfnamefont{S.}~\bibnamefont{Kolekar}} \bibnamefont{and}
  \bibinfo{author}{\bibfnamefont{T.}~\bibnamefont{Padmanabhan}},
  \bibinfo{journal}{Class. Quant. Grav.} \textbf{\bibinfo{volume}{32}},
  \bibinfo{pages}{202001} (\bibinfo{year}{2015}), \eprint{arXiv:1308.6289}.

\bibitem[{\citenamefont{Adhikari et~al.}(2018)\citenamefont{Adhikari,
  Bhattacharya, Chowdhury, and Majhi}}]{Adhikari:2017gyb}
\bibinfo{author}{\bibfnamefont{A.}~\bibnamefont{Adhikari}},
  \bibinfo{author}{\bibfnamefont{K.}~\bibnamefont{Bhattacharya}},
  \bibinfo{author}{\bibfnamefont{C.}~\bibnamefont{Chowdhury}},
  \bibnamefont{and} \bibinfo{author}{\bibfnamefont{B.~R.} \bibnamefont{Majhi}},
  \bibinfo{journal}{Phys. Rev. D} \textbf{\bibinfo{volume}{97}},
  \bibinfo{pages}{045003} (\bibinfo{year}{2018}), \eprint{arXiv:1707.01333}.

\bibitem[{\citenamefont{Das et~al.}(2019)\citenamefont{Das, Dalui, Chowdhury,
  and Majhi}}]{Das:2019aii}
\bibinfo{author}{\bibfnamefont{A.}~\bibnamefont{Das}},
  \bibinfo{author}{\bibfnamefont{S.}~\bibnamefont{Dalui}},
  \bibinfo{author}{\bibfnamefont{C.}~\bibnamefont{Chowdhury}},
  \bibnamefont{and} \bibinfo{author}{\bibfnamefont{B.~R.} \bibnamefont{Majhi}},
  \bibinfo{journal}{Phys. Rev. D} \textbf{\bibinfo{volume}{100}},
  \bibinfo{pages}{085002} (\bibinfo{year}{2019}), \eprint{arXiv:1902.03735}.

\bibitem[{\citenamefont{Chowdhury et~al.}(2019)\citenamefont{Chowdhury, Das,
  Dalui, and Majhi}}]{Chowdhury:2019set}
\bibinfo{author}{\bibfnamefont{C.}~\bibnamefont{Chowdhury}},
  \bibinfo{author}{\bibfnamefont{S.}~\bibnamefont{Das}},
  \bibinfo{author}{\bibfnamefont{S.}~\bibnamefont{Dalui}}, \bibnamefont{and}
  \bibinfo{author}{\bibfnamefont{B.~R.} \bibnamefont{Majhi}},
  \bibinfo{journal}{Phys. Rev. D} \textbf{\bibinfo{volume}{99}},
  \bibinfo{pages}{045021} (\bibinfo{year}{2019}), \eprint{arXiv:1902.06900}.

\bibitem[{\citenamefont{Lima et~al.}(2020)\citenamefont{Lima, Alencar, and
  Landim}}]{Lima:2020czr}
\bibinfo{author}{\bibfnamefont{A.~P. C.~M.} \bibnamefont{Lima}},
  \bibinfo{author}{\bibfnamefont{G.}~\bibnamefont{Alencar}}, \bibnamefont{and}
  \bibinfo{author}{\bibfnamefont{R.~R.} \bibnamefont{Landim}},
  \bibinfo{journal}{Phys. Rev. D} \textbf{\bibinfo{volume}{101}},
  \bibinfo{pages}{125008} (\bibinfo{year}{2020}), \eprint{arXiv:2002.02020}.

\bibitem[{\citenamefont{Banerjee and Majhi}(2020)}]{Banerjee:2019tbr}
\bibinfo{author}{\bibfnamefont{R.}~\bibnamefont{Banerjee}} \bibnamefont{and}
  \bibinfo{author}{\bibfnamefont{B.~R.} \bibnamefont{Majhi}},
  \bibinfo{journal}{Eur. Phys. J. C} \textbf{\bibinfo{volume}{80}},
  \bibinfo{pages}{435} (\bibinfo{year}{2020}), \eprint{arXiv:1909.03760}.

\bibitem[{\citenamefont{Dicke}(1954)}]{Dicke:1954}
\bibinfo{author}{\bibfnamefont{R.~H.} \bibnamefont{Dicke}},
  \bibinfo{journal}{Phys. Rev.} \textbf{\bibinfo{volume}{93}},
  \bibinfo{pages}{99} (\bibinfo{year}{1954}).

\bibitem[{\citenamefont{Mijic}(1993)}]{Mijic:1993wm}
\bibinfo{author}{\bibfnamefont{M.}~\bibnamefont{Mijic}}, in
  \emph{\bibinfo{booktitle}{{Belgrade Workshop (Danube 93)}}}
  (\bibinfo{year}{1993}), \eprint{arXiv:hep-th/9311030}.

\bibitem[{\citenamefont{Weldon}(2000)}]{Weldon:2000pe}
\bibinfo{author}{\bibfnamefont{H.~A.} \bibnamefont{Weldon}},
  \bibinfo{journal}{Phys. Rev. D} \textbf{\bibinfo{volume}{62}},
  \bibinfo{pages}{056010} (\bibinfo{year}{2000}),
  \eprint{arXiv:hep-ph/0007138}.

\bibitem[{\citenamefont{Crispino et~al.}(2008)\citenamefont{Crispino, Higuchi,
  and Matsas}}]{Crispino:2007eb}
\bibinfo{author}{\bibfnamefont{L.~C.} \bibnamefont{Crispino}},
  \bibinfo{author}{\bibfnamefont{A.}~\bibnamefont{Higuchi}}, \bibnamefont{and}
  \bibinfo{author}{\bibfnamefont{G.~E.} \bibnamefont{Matsas}},
  \bibinfo{journal}{Rev.Mod.Phys.} \textbf{\bibinfo{volume}{80}},
  \bibinfo{pages}{787} (\bibinfo{year}{2008}), \eprint{arXiv:0710.5373}.

\bibitem[{\citenamefont{Birrell and Davies}(1984)}]{book:Birrell}
\bibinfo{author}{\bibfnamefont{N.~D.} \bibnamefont{Birrell}} \bibnamefont{and}
  \bibinfo{author}{\bibfnamefont{P.~C.~W.} \bibnamefont{Davies}},
  \emph{\bibinfo{title}{Quantum fields in curved space}}, Cambridge Monographs
  on Mathematical Physics (\bibinfo{publisher}{Cambridge University Press},
  \bibinfo{year}{1984}).

\bibitem[{\citenamefont{Carroll}(2004)}]{book:carroll}
\bibinfo{author}{\bibfnamefont{S.}~\bibnamefont{Carroll}},
  \emph{\bibinfo{title}{Spacetime and geometry. An introduction to general
  relativity}} (\bibinfo{publisher}{AW}, \bibinfo{year}{2004}).

\bibitem[{\citenamefont{Higuchi et~al.}(2017)\citenamefont{Higuchi, Iso, Ueda,
  and Yamamoto}}]{Higuchi:2017gcd}
\bibinfo{author}{\bibfnamefont{A.}~\bibnamefont{Higuchi}},
  \bibinfo{author}{\bibfnamefont{S.}~\bibnamefont{Iso}},
  \bibinfo{author}{\bibfnamefont{K.}~\bibnamefont{Ueda}}, \bibnamefont{and}
  \bibinfo{author}{\bibfnamefont{K.}~\bibnamefont{Yamamoto}},
  \bibinfo{journal}{Phys. Rev. D} \textbf{\bibinfo{volume}{96}},
  \bibinfo{pages}{083531} (\bibinfo{year}{2017}), \eprint{arXiv:1709.05757}.

\bibitem[{\citenamefont{Padmanabhan}(2010)}]{book:PadmanabhanGrav}
\bibinfo{author}{\bibfnamefont{T.}~\bibnamefont{Padmanabhan}},
  \emph{\bibinfo{title}{Gravitation: Foundations and Frontiers}}
  (\bibinfo{publisher}{Cambridge University Press}, \bibinfo{year}{2010}),
  \bibinfo{edition}{1st} ed.

\end{thebibliography}

\end{document}